\documentclass[12pt,english]{article}
\usepackage{times}
\usepackage[T1]{fontenc}
\usepackage{geometry}
\usepackage{array}
\usepackage{color}
\usepackage{graphicx}
\usepackage{setspace}
\usepackage{amssymb}

\makeatletter

\providecommand{\tabularnewline}{\\}

\usepackage{bm}

\usepackage{babel}
\makeatother
\begin{document}

\section*{\noindent On the Design of Modular Reflecting \emph{EM} Skins for
Enhanced Urban Wireless Coverage}

\noindent P. Rocca,\textcolor{black}{$^{(1)(2)}$} \emph{Senior Member,
IEEE}, P. Da R\'u,$^{(1)}$ N. Anselmi,$^{(1)}$ \emph{Member, IEEE},
M. Salucci,$^{(1)}$ \emph{Member, IEEE}, G. Oliveri,$^{(1)}$ \emph{Senior
Member, IEEE}, D. Erricolo,$^{(4)}$ \emph{Fellow, IEEE}, and A. Massa,$^{(1)(3)(5)}$
\emph{Fellow, IEEE}

\noindent ~

\noindent {\footnotesize $^{(1)}$} \emph{\footnotesize ELEDIA Research
Center} {\footnotesize (}\emph{\footnotesize ELEDIA}{\footnotesize @}\emph{\footnotesize UniTN}
{\footnotesize - University of Trento)}{\footnotesize \par}

\noindent {\footnotesize DICAM - Department of Civil, Environmental,
and Mechanical Engineering}{\footnotesize \par}

\noindent {\footnotesize Via Mesiano 77, 38123 Trento - Italy}{\footnotesize \par}

\noindent \textit{\emph{\footnotesize E-mail:}} {\footnotesize \{}\emph{\footnotesize paolo.rocca}{\footnotesize ,}
\emph{\footnotesize pietro.daru}{\footnotesize ,} \emph{\footnotesize nicola.anselmi}{\footnotesize ,}
\emph{\footnotesize marco.salucci}{\footnotesize ,} \emph{\footnotesize giacomo.oliveri}{\footnotesize ,}
\emph{\footnotesize andrea.massa}{\footnotesize \}@}\emph{\footnotesize unitn.it}{\footnotesize \par}

\noindent {\footnotesize Website:} \emph{\footnotesize www.eledia.org/eledia-unitn}{\footnotesize \par}

\noindent {\footnotesize ~}{\footnotesize \par}

\noindent {\footnotesize $^{(2)}$} \emph{\footnotesize ELEDIA Research
Center} {\footnotesize (}\emph{\footnotesize ELEDIA@XIDIAN} {\footnotesize -
Xidian University)}{\footnotesize \par}

\noindent {\footnotesize P.O. Box 191, No.2 South Tabai Road, 710071
Xi'an, Shaanxi Province - China}{\footnotesize \par}

\noindent {\footnotesize E-mail:} \emph{\footnotesize paolo.rocca@xidian.edu.cn}{\footnotesize \par}

\noindent {\footnotesize Website:} \emph{\footnotesize www.eledia.org/eledia-xidian}{\footnotesize \par}

\noindent ~

\noindent {\footnotesize $^{(3)}$} \emph{\footnotesize ELEDIA Research
Center} {\footnotesize (}\emph{\footnotesize ELEDIA}{\footnotesize @}\emph{\footnotesize UESTC}
{\footnotesize - UESTC)}{\footnotesize \par}

\noindent {\footnotesize School of Electronic Engineering, Chengdu
611731 - China}{\footnotesize \par}

\noindent \textit{\emph{\footnotesize E-mail:}} \emph{\footnotesize andrea.massa@uestc.edu.cn}{\footnotesize \par}

\noindent {\footnotesize Website:} \emph{\footnotesize www.eledia.org/eledia}{\footnotesize -}\emph{\footnotesize uestc}{\footnotesize \par}

\noindent {\footnotesize ~}{\footnotesize \par}

\noindent {\footnotesize $^{(4)}$} \emph{\footnotesize Andrew Electromagnetics
Laboratory} {\footnotesize - University of Illinois Chicago}{\footnotesize \par}

\noindent {\footnotesize Department of Electrical and Computer Engineering,
851 South Morgan Street, Chicago, IL 60607-7053 - USA}{\footnotesize \par}

\noindent {\footnotesize E-mail:} \emph{\footnotesize derric1@uic.edu}{\footnotesize \par}

\noindent {\footnotesize Website:} \emph{\footnotesize https://andrew.lab.uic.edu/}{\footnotesize \par}

\noindent ~

\noindent {\footnotesize $^{(5)}$} \emph{\footnotesize ELEDIA Research
Center} {\footnotesize (}\emph{\footnotesize ELEDIA@TSINGHUA} {\footnotesize -
Tsinghua University)}{\footnotesize \par}

\noindent {\footnotesize 30 Shuangqing Rd, 100084 Haidian, Beijing
- China}{\footnotesize \par}

\noindent {\footnotesize E-mail:} \emph{\footnotesize andrea.massa@tsinghua.edu.cn}{\footnotesize \par}

\noindent {\footnotesize Website:} \emph{\footnotesize www.eledia.org/eledia-tsinghua}{\footnotesize \par}

\noindent \emph{\footnotesize This work has been submitted to the
IEEE for possible publication. Copyright may be transferred without
notice, after which this version may no longer be accessible.}{\footnotesize \par}

\newpage
\section*{On the Design of Modular Reflecting \emph{EM} Skins for Enhanced
Urban Wireless Coverage}

~

~

~

\begin{flushleft}P. Rocca, P. Da R\'u, N. Anselmi, M. Salucci, G.
Oliveri, D. Erricolo, and A. Massa\end{flushleft}

\vfill

\begin{abstract}
\noindent The design of modular, passive, and static artificial metasurfaces
to be used as electromagnetic skins (\emph{EMS}s) of buildings for
improving the coverage in urban millimeter-wave communication scenarios
is addressed. Towards this end, an \emph{ad-hoc} design strategy is
presented to determine optimal trade-off implementative solutions
that assure a suitable coverage of the areas of interest, where the
signal from the base station is too weak, with the minimum complexity.
More specifically, the admissible surface in the building facade is
first partitioned into tiles, which are the minimum-size elements
of the artificial coating (i.e., the building block of an \emph{EMS}).
Then, the search for the optimal \emph{EMS} layout (i.e., the minimum
number and the positions of the tiles to be installed) is carried
out with a binary multi-objective optimization method. Representative
numerical results are reported and discussed to point out the features
and the potentialities of the \emph{EMS} solution in the smart electromagnetic
environment (\emph{SEME}) as well as the effectiveness of the proposed
design method.

\vfill
\end{abstract}
\noindent \textbf{Key words}: Smart EM Environment, Artificial Materials,
Mobile Communications, Millimeter-wave, Multi-Objective Optimization.

\newpage
\section{Introduction}

\noindent Starting from the first generation of cellular networks,
back when the communications were analog and the portable devices
heavy and cumbersome, there has always been a continuous technological
push towards higher data rates, which are a mandatory requirement
with the introduction of smart-phones and the sharing of multimedia
contents now enabled by the fast and reliable data streams of 4G and
5G communication networks \cite{Haupt 2020}-\cite{Cox 2021}. Regardless
of the throughput of modern communication networks, the data traffic
is expected to further increase in the next years because of the massive
proliferation of wireless devices and systems (e.g., smart-phones,
tablets, internet-of-things (\emph{IoT}) sensors, and robots) as well
as the introduction of novel applications including, for instance,
the autonomous driving \cite{Yu 2021}, the tactile internet \cite{Li 2019},
and the remote control of robots \cite{Acemoglu 2020}. All these
applications will unavoidably require improved coverage (i.e., a higher
level of the \emph{EM} signal in the coverage area) and better quality-of-service
(\emph{QoS}) for mobile users/devices as well as wireless links characterized
by lower latency and higher throughput/resiliency \cite{Bennis 2018}\cite{Wang 2019a}.
Towards this end, future mobile communication networks will have to
assure more and more reliable and ubiquitous connections, everywhere
and anytime, as never seen before. However, the standard solutions
chosen by the operators (i.e., installing more base stations (\emph{BS}s),
transmitting more power, or using new frequency bands) are no longer
applicable because of the too high power consumption due to the foreseen
explosion of the traffic needs as well as the spectrum congestion
\cite{Puglielli 2016}. Moreover, the obstacles/scatterers in the
environment cannot be neglected due to the increase in the operation
frequencies (e.g., millimeter-waves in 5G \cite{Hong 2021}) so that
the no line-of-sight (\emph{NLOS}) condition has to be taken into
account since the design of the wireless network architecture.

\noindent A possible countermeasure to these issues and challenges
is the {}``implementation'' of the so-called smart electromagnetic
environment (\emph{SEME}) \cite{IEEE-TAP SI 2021}, where the objects
and the scatterers within the environment are considered, unlike the
past, as enablers \cite{Salucci 2020} of the electromagnetic (\emph{EM})
propagation and not impairments. The environment is thus exploited
as an additional degree-of-freedom (\emph{DoF}) for tailoring the
propagation of the \emph{EM} waves and to enhance the signal strength
in the {}``blind spots'', namely the zones where the signal from
the base station is too weak to support a desired throughput for users'
applications \cite{Salucci 2020}. By using artificial materials (e.g.,
engineered materials or metamaterials \cite{Yang 2019}) on the building
facades or integrated within panels along the streets, the propagation
of the \emph{EM} waves in complex urban scenarios is controlled to
fit \emph{QoS} requirements and coverage targets. Passive (i.e., without
active amplifiers) reconfigurable metasurfaces, which behave like
intelligent reflecting surfaces (\emph{IRS}s) \cite{Di Renzo 2019}-\cite{Wu 2021}
thanks to simple electronic devices such as radio-frequency switches
\cite{Wang 2019b}, have been successfully installed, but they will
not be massively deployed until their cost is significantly reduced.

\noindent An opposite strategy still profitably exploits the objects
in the environment, without introducing additional materials, but
optimizing the excitations of the array elements of the \emph{BS}
to generate a desired \emph{EM} field distribution in the desired
spots \cite{Massa 2021}. Although there are no additional costs and
an installation of new hardware is not required, the opportunistic
use of the \emph{BS} array offers to the designer a limited number
of \emph{DoF}s and an accurate knowledge of the surrounding scenario
is also needed to assure a suitable/stable \emph{QoS}.

\noindent Differently, this paper is concerned with the instance of
the \emph{SEME} vision aimed at providing a doable large-scale solution
suitable for the mass production. More specifically, it proposes a
method for the design of modular, passive, and static metasurfaces
to build effective and low-cost \emph{EMS}s. In telecommunication
engineering, the term \emph{EMS} refers to a device conformal to the
external surface of the object, where it is installed, that offers
a set of functionalities related to the sensing and the manipulation
of the \emph{EM} waves \cite{Zhou 2019}. A typical deployment for
an \emph{EMS} is over the facades of a building, which are strategic
assets in a urban scenario for redirecting the impinging \emph{EM}
field towards areas where the signal would otherwise be weak. Starting
from the selection/definition of the maximal area available on a building
facade, the support of the \emph{EMS} is discretized into tiles, which
are the elementary building blocks of an \emph{EMS}. The arrangement
of the tiles on the admissible surface of the facade is then optimized
with a multi-objective global optimization strategy based on a binary
implementation of the Non-Dominated Sorting Genetic Algorithm II (\emph{NSGA-II})
\cite{Deb 2002} to yield the optimal trade-off solutions between
the best coverage of the zones of interest and the minimum number
of tiles. 

\noindent To the best of the authors' knowledge, the main innovative
contributions of this research work include (\emph{i}) the description,
the statement, and the mathematical formalization of a novel design
problem within the \emph{SEME} framework, (\emph{ii}) the development
of a customized design strategy for selecting the minimum number of
tiles of the \emph{EMS} that assures the coverage of the regions of
interest, and (\emph{iii}) the synthesis of innovative tiled \emph{EMS}s
to be embedded in the facade of buildings for improving the \emph{EM}
coverage within urban millimeter-wave communication scenarios.

\noindent The rest of the paper is organized as follows. The synthesis
problem of the modular, passive, and static \emph{EMS} is mathematically
described and formulated in Sect. 2, while the optimization-based
design method is presented in Sect. 3. Section 4 deals with the validation
and the numerical assessment of the proposed concepts and synthesis
method by considering realistic urban scenarios served by a millimeter-wave
band of 5G systems. Eventually, some conclusions and final remarks
are drawn (Sect. 5).

\section{\noindent Mathematical Formulation}

\noindent Let us consider an urban scenario {[}Fig. 1(\emph{a}){]}
where a \emph{BS} serves the terminals and the devices in the surrounding
environment to implement the communication network by providing wireless
services to the users with a suitable \emph{QoS}. Because of the presence
and the configuration of the buildings, the signal from the \emph{BS}
is absent or too weak to guarantee a connection to the network or
a sufficient throughput in some areas, \{$\Omega_{b}$; $b=1,...,B$\},
of the scenario at hand. In order to increase the signal strength
in these {}``blind spots'', the installation of artificial \emph{EMS}s
is considered. 

\noindent For the sake of formulation simplicity, the case of a single
{}``blind spot'' (i.e., $B=1$, $\Omega\leftarrow\Omega_{b}$),
where there is neither direct nor reflected signal from the \emph{BS},
and the design of a single \emph{EMS} on a facade of a building is
considered in this work. However, it is worthwhile pointing out that
both the theoretical description and the proposed design method can
be straightforwardly extended to the case with multiple skins over
multiple buildings to cover multiple blind spots. With reference to
such a benchmark, an area $\mathcal{S}$ on the external wall of a
selected building is assumed to be located on the $y-z$ plane, subdivided
into $N$ square tiles, \{$\mathcal{S}^{\left(n\right)}$; $n=1,...,N$\}
($\bigcup_{n=1}^{N}\mathcal{S}^{\left(n\right)}=\mathcal{S}$), of
equal size $\Delta\mathcal{S}$ ($\Delta\mathcal{S}=L\times L$, $L$
being the side length of each tile), and centered at the positions
$\mathbf{r}_{\mathcal{S}}^{\left(n\right)}=y_{\mathcal{S}}^{\left(n\right)}\widehat{\mathbf{y}}+z_{\mathcal{S}}^{\left(n\right)}\widehat{\mathbf{z}}$
($n=1,...,N$) {[}Fig. 1(\emph{b}){]}, while the \emph{BS} is located
at $\mathbf{r}_{BS}=x_{BS}\widehat{\mathbf{x}}+y_{BS}\widehat{\mathbf{y}}+z_{BS}\widehat{\mathbf{z}}$
in the far-field of $\mathcal{S}$.

\noindent Without loss of generality, the \emph{EM} wave generated
from the \emph{BS} and impinging on the \emph{EMS} is modeled as a
monochromatic plane wave at the working frequency $f$ with electric
field \cite{Osipov 2017}\cite{Lindell 2019}\begin{equation}
\mathbf{E}_{\Im}\left(\mathbf{r}\right)\triangleq E_{\Im}\,\widehat{\mathbf{e}}_{\Im}\, e^{-j\mathbf{k}_{\Im}\cdot\left(\mathbf{r}-\mathbf{r}_{\mathcal{S}}^{\left(0\right)}\right)}\label{eq:_incident.wave}\end{equation}
where $E_{\Im}$ is the complex-valued wave amplitude, $\widehat{\mathbf{e}}_{\Im}$
is the complex polarization vector, and $\mathbf{k}_{\Im}$ ($\mathbf{k}_{\Im}\triangleq-k\left[\sin\theta_{\Im}\cos\phi_{\Im}\widehat{\mathbf{x}}+\sin\theta_{\Im}\sin\phi_{\Im}\widehat{\mathbf{y}}+\cos\theta_{\Im}\widehat{\mathbf{z}}\right]$)
is the incident wave vector, $k$ being the free-space wavenumber
($k\triangleq\frac{2\pi}{\lambda}$, $\lambda$ being the wavelength
at $f$), while $\left(\theta_{\Im}^{\left(0\right)},\phi_{\Im}^{\left(0\right)}\right)$
is the direction of arrival of the incident wave from the \emph{BS}
to the center of $\mathcal{S}$, $\mathbf{r}_{\mathcal{S}}^{\left(0\right)}$
($\mathbf{r}_{\mathcal{S}}^{\left(0\right)}=y_{\mathcal{S}}^{\left(0\right)}\widehat{\mathbf{y}}+z_{\mathcal{S}}^{\left(0\right)}\widehat{\mathbf{z}}$)
{[}Fig. 1(\emph{b}){]}, being $\left.\theta_{\Im}^{\left(n\right)}\right\rfloor _{n=0}=\arccos\left.\left(\frac{z_{BS}-z_{\mathcal{S}}^{\left(n\right)}}{\left|\mathbf{r}_{BS}-\mathbf{r}_{\mathcal{S}}^{\left(n\right)}\right|}\right)\right\rfloor _{n=0}$
and $\left.\phi_{\Im}^{\left(n\right)}\right\rfloor _{n=0}=\arctan\left.\left(\frac{y_{BS}-y_{\mathcal{S}}^{\left(n\right)}}{x_{BS}-x_{\mathcal{S}}^{\left(n\right)}}\right)\right\rfloor _{n=0}=\arctan\left.\left(\frac{y_{BS}-y_{\mathcal{S}}^{\left(n\right)}}{x_{BS}}\right)\right\rfloor _{n=0}$
since $x_{\mathcal{S}}^{\left(0\right)}=0$.

\noindent Each tile $\mathcal{S}^{\left(n\right)}$ ($n=1,...,N$)
of the \emph{EMS} is illuminated from the direction $\left(\theta_{\Im}^{\left(n\right)},\phi_{\Im}^{\left(n\right)}\right)$
and it reflects the impinging wave towards the direction $\left(\theta_{\Re}^{\left(n\right)},\phi_{\Re}^{\left(n\right)}\right)$
{[}$\theta_{\Re}^{\left(n\right)}=\arccos\left(\frac{z_{\Omega}^{\left(n\right)}-z_{\mathcal{S}}^{\left(0\right)}}{\left|\mathbf{r}_{\Omega}^{\left(n\right)}-\mathbf{r}_{\mathcal{S}}^{\left(0\right)}\right|}\right)$,
$\phi_{\Re}^{\left(n\right)}=\arctan\left(\frac{y_{\Omega}^{\left(n\right)}-y_{\mathcal{S}}^{\left(0\right)}}{x_{\Omega}^{\left(n\right)}}\right)${]}
by focusing the reflected beam in the point $\mathbf{r}_{\Omega}^{\left(n\right)}$
($\mathbf{r}_{\Omega}^{\left(n\right)}=x_{\Omega}^{\left(n\right)}\widehat{\mathbf{x}}+y_{\Omega}^{\left(n\right)}\widehat{\mathbf{y}}+z_{\Omega}^{\left(n\right)}\widehat{\mathbf{z}}$)
of the coverage region $\Omega$, which is also called area of interest
(\emph{AoI}). More in detail, the \emph{AoI} is assumed parallel to
the $x-y$ plane (i.e., $z_{\Omega}^{\left(n\right)}=z_{\Omega}$,
$\forall n$), centered at $\mathbf{r}_{\Omega}^{\left(0\right)}$,
and discretized into $N$ partitions, \{$\Omega^{\left(n\right)}$;
$n=1,...,N$\} (i.e., $\bigcup_{n=1}^{N}\Omega^{\left(n\right)}=\Omega$),
located at $\mathbf{r}_{\Omega}^{\left(n\right)}$ ($n=1,...,N$)
with equal dimensions $\Delta\Omega^{\left(n\right)}=\Delta\Omega$
{[}Fig. 1(\emph{b}){]}.

\noindent By neglecting the polarization and the reflection losses,
the far-field expression of the electric field reflected from the
$n$-th ($n=1,...,N$) tile $\mathcal{S}^{\left(n\right)}$ in a generic
point $\tilde{\mathbf{r}}$ of the local coordinate system%
\footnote{\noindent The following relationships between the reference and the
local coordinate systems hold true: $\tilde{x}=y$, $\tilde{y}=z-z_{\mathcal{S}}^{\left(n\right)},$and
$\tilde{z}=x$.%
} (Fig. 2), $\mathbf{E}_{\Re}^{\left(n\right)}\left(\tilde{\mathbf{r}}\right)=E_{\Re}^{\left(n\right)}\left(\tilde{\mathbf{r}}\right)\,\widehat{\mathbf{e}}_{\Re}$,
is given by the following closed-form expression \cite{Danufane 2021}\begin{equation}
E_{\Re}^{\left(n\right)}\left(\tilde{\mathbf{r}}\right)\simeq-jk\eta\frac{e^{-jk\left(d_{\Im}^{\left(n\right)}+d_{\Re}^{\left(n\right)}\right)}}{4\pi d_{\Im}^{\left(n\right)}d_{\Re}^{\left(n\right)}}L^{2}\left(\cos\tilde{\theta}_{\Im}^{\left(n\right)}+\cos\tilde{\theta}_{\Re}^{\left(n\right)}\right)\textnormal{sinc}\left(kL\mathcal{D}_{\tilde{x}}\right)\textnormal{sinc}\left(kL\mathcal{D}_{\tilde{y}}\right)e^{-j\left(\varphi_{\Im}+\varphi_{\Re}\right)}\label{eq:_Danufane}\end{equation}
where $\eta$ is the free-space impedance, while $d_{\Im}^{\left(n\right)}$
and $d_{\Re}^{\left(n\right)}$ are the distances travelled by the
incident wave from the \emph{BS} to the barycenter of the $n$-th
($n=1,...,N$) tile ($d_{\Im}^{\left(n\right)}=\left|\mathbf{r}_{BS}-\mathbf{r}_{\mathcal{S}}^{\left(n\right)}\right|$)
and by the reflected wave from the barycenter of the $n$-th ($n=1,...,N$)
tile to the point $\mathbf{r}_{\Omega}^{\left(n\right)}$ where the
peak of the reflected beam, generated from the same $n$-th tile,
is directed ($d_{\Re}^{\left(n\right)}=\left|\mathbf{r}_{\mathcal{S}}^{\left(n\right)}-\mathbf{r}_{\Omega}^{\left(n\right)}\right|$),
respectively. Moreover, $\varphi_{\Im}$ is the phase associated to
the modulation of the signal generated by the \emph{BS}, $\varphi_{\Re}$
is the phase term that can be engineered with the design of the surface,
while $\mathcal{D}_{\tilde{x}}=\sin\tilde{\theta}\cos\tilde{\phi}-\sin\tilde{\theta}_{\Re}^{\left(n\right)}\cos\tilde{\phi}_{\Re}^{\left(n\right)}$
and $\mathcal{D}_{\tilde{y}}=\sin\tilde{\theta}\sin\tilde{\phi}-\sin\tilde{\theta}_{\Re}^{\left(n\right)}\sin\tilde{\phi}_{\Re}^{\left(n\right)}.$

\noindent To improve the coverage within the \emph{AoI} by guaranteeing
a suitable intensity of the signal from the \emph{BS} by means of
the reflection from the \emph{EMS}, the following synthesis problem
is formulated:

\begin{quotation}
\noindent \emph{Modular Reflecting EM Skin} (\emph{MREMS}) \emph{Design
Problem} - Given an admissible skin surface $\mathcal{S}$ discretized
into $N$ square tiles, \{$\mathcal{S}^{\left(n\right)}$; $n=1,...,N$\},
and an \emph{AoI} $\Omega$, select the minimum number of tiles $M$
(i.e., $M\leq N$), which reflect the \emph{EM} wave from the \emph{BS}
towards the corresponding focusing points, \{$\mathbf{r}_{\Omega}^{\left(n\right)}$;
$n=1,...,M$\}, within the \emph{AoI} ($\mathbf{r}_{\Omega}^{\left(n\right)}\in\Omega$),
so that the power collected by a receiver at the position $\mathbf{r}_{u}$
of the \emph{AoI} ($\mathbf{r}_{u}\in\Omega$), $\mathcal{P}_{\Re}\left(\mathbf{r}_{u}\right)$
($\mathcal{P}_{\Re}\left(\mathbf{r}\right)\triangleq\sum_{m=1}^{M}\left|E_{\Re}^{\left(m\right)}\left(\mathbf{r}\right)\right|^{2}$),
fulfils the condition\begin{equation}
\mathcal{P}_{\Re}\left(\mathbf{r}_{u}\right)\geq\mathcal{P}_{th}\label{eq:_coverage.condition}\end{equation}
where $\mathcal{P}_{th}$ is a user-defined coverage threshold ($\mathcal{P}_{th}\ge\mathcal{P}_{bls}$,
$\mathcal{P}_{bls}$ being the minimum level for a wireless connection).
\end{quotation}

\section{\emph{MREMS} Synthesis Method}

\noindent The \emph{MREMS} Design Problem is addressed through a multi-objective
optimization strategy based on a binary implementation of the \emph{NSGA-II}
\cite{Deb 2002}. Towards this end, the presence/absence of a tile
on the final \emph{EMS} layout $\mathcal{S}_{opt}$ is mathematically
modeled by means of a binary variable, $t_{n}$ ($t_{n}\in\left\{ 0,\,1\right\} $)
($n=1,...,N$). If $t_{n}=1$, then the $n$-th ($n=1,...,N$) tile
$\mathcal{S}^{\left(n\right)}$ is present on the facade of the building
and it contributes to the reflection of the \emph{EM} wave towards
the \emph{AoI}. Otherwise (i.e., $t_{n}=0$), the $n$-th ($n=1,...,N$)
tile $\mathcal{S}^{\left(n\right)}$ is not installed on the external
wall of the building, which maintains its original scattering properties
without contributing to the enhancement of the signal in $\Omega$.
Accordingly, an admissible layout of the \emph{EMS} (i.e., an arrangement
of tiles over the available surface $\mathcal{S}$) is univocally
described by the binary vector $\mathbf{T}$ ($\mathbf{T}\triangleq\left\{ t_{n};\, n=1,...,N\right\} $.

\noindent In order to determine the final structure of the \emph{EMS},
$\mathcal{S}_{opt}$, namely the best subset of $M$ tiles among the
$N$ admissible ones to be installed in $\mathcal{S}$, the problem
is formulated as an optimization one by defining suitable optimization
objectives aimed at quantifying the mismatch between the desired coverage
(\ref{eq:_coverage.condition}) with that afforded by a trial \emph{EMS}
arrangement, $\mathbf{T}$, as well as the complexity of the final
layout of the \emph{EMS}, $\mathbf{T}_{opt}$. More specifically,
the following two cost functions are defined. The former is the {}``\emph{coverage
term}'', $\Phi_{1}\left(\mathbf{T}\right)$, given by\begin{equation}
\Phi_{1}\left(\mathbf{T}\right)\triangleq\frac{1}{U}\sum_{u=1}^{U}\frac{\left|\sum_{n=1}^{N}t_{n}\mathcal{P}_{\Re}^{\left(n\right)}\left(\mathbf{r}_{u};\,\mathbf{T}\right)-\mathcal{P}_{th}\right|}{\mathcal{P}_{th}}\mathcal{H}\left\{ \mathcal{P}_{th}-\sum_{n=1}^{N}t_{n}\mathcal{P}_{\Re}^{\left(n\right)}\left(\mathbf{r}_{u};\,\mathbf{T}\right)\right\} ,\label{eq:_cost.function.coverage}\end{equation}
while the latter is the {}``\emph{complexity term}'' defined as\begin{equation}
\Phi_{2}\left(\mathbf{T}\right)\triangleq\frac{M}{N}\label{eq:_cost.function.complexity}\end{equation}
where $U$ is the number of receivers, which are uniformly distributed
inside the \emph{AoI} $\Omega$, and $M$ ($M=\sum_{n=1}^{N}t_{n}$)
is the number of tiles composing the \emph{EMS}. Moreover, $\mathcal{H}\left(\cdot\right)$
is the Heaviside function equal to $1$ when the argument is positive
(i.e., $\mathcal{P}_{th}>\mathcal{P}_{\Re}\left(\mathbf{r}_{u};\,\mathbf{T}\right)$
$\to$ the power strength at $\mathbf{r}_{u}$ is below the desired
value) and $0$ otherwise {[}i.e., $\mathcal{P}_{th}\le\mathcal{P}_{\Re}\left(\mathbf{r}_{u};\,\mathbf{T}\right)$
$\to$ the coverage condition (\ref{eq:_coverage.condition}) is fulfilled{]}. 

\noindent Since the two cost functions to be minimized impose, on
the one hand, the fitting of the coverage condition (\ref{eq:_cost.function.coverage}),
while, on the other hand, the reduction of the number of tiles to
minimize the area/cost of the skin (\ref{eq:_cost.function.complexity}),
they are by definition conflicting. Indeed, a larger number of tiles
leads to a stronger electric field in $\Omega$ and vice versa. Thus,
the optimization problem turns out to be natively multi-objective
and a natural solution strategy is that of defining a Pareto front
of multiple optimal solutions, each being a valid trade-off to be
considered for the final implementation of the \emph{EMS} on the building
facade. Such a multiplicity of solutions gives to the designer the
possibility of choosing the \emph{EMS} layout to be implemented according
to its feeling and other non-functional constraints (e.g., architectural
and landscaping restrictions, costs, etc ...). Following such a guideline,
the binary \emph{NSGA-II} \cite{Deb 2002} is chosen as optimization
algorithm because of the binary nature of the design problem at hand
($\mathbf{T}$ being a binary vector) and the need of synthesizing
multiple trade-off solutions among the conflicting objectives. Moreover,
thanks to its hill-climbing features \cite{Rocca 2009}, the Genetic
Algorithm (\emph{GA}) has global optimization features, which are
here compulsory due to the non-convex behaviors of the cost functions.
Indeed, $\Phi_{1}$ and $\Phi_{2}$ are non-continuous functions also
characterized by the presence of local minima (i.e., sub-optimal solutions
of the corresponding \emph{EMS} design problem).

\noindent More in detail, the following implementation of the binary
\emph{NSGA-II} is taken into account:

\begin{itemize}
\item \emph{Step 0} - \textbf{\emph{NSGA-II}} \textbf{Setup}. Select the
number of $P$ individuals (i.e., trial layouts of the \emph{EMS})
of the \emph{GA} population and set the control parameters of the
\emph{NSGA}-II, namely the crossover rate, $\wp_{c}$, the polynomial
mutation rate, $\wp_{m}$, the distribution index for both the crossover,
$\aleph_{c}$, and the mutation rate, $\aleph_{m}$, and the maximum
number of iterations, $I$, $i$ being the iteration index ($i=0,...,I$);
\item \emph{Step 1} - \textbf{Population Initialization} ($i=0$). Randomly
set the initial trial solutions, $\left\{ \mathbf{T}_{i}^{\left(p\right)};\, p=1,...,P\right\} $,
and compute the cost function terms, $\Phi_{1,i}^{\left(p\right)}=\Phi_{1}\left(\mathbf{T}_{i}^{\left(p\right)}\right)$
and $\Phi_{2,i}^{\left(p\right)}=\Phi_{2}\left(\mathbf{T}_{i}^{\left(p\right)}\right)$,
for each individual of the population ($p=1,...,P$);
\item \emph{Step 2 -} \textbf{\emph{EMS}} \textbf{Optimization} ($i=1,...,I$).
Apply the evolutionary operators of the \emph{NSGA-II} to iteratively
($i\leftarrow i+1$) generate the offsprings, $\left\{ \mathbf{T}_{i}^{\left(p\right)};\, p=1,...,P\right\} $,
from the current population of parents, $\left\{ \mathbf{T}_{i-1}^{\left(p\right)};\, p=1,...,P\right\} $,
and compute their fitness values (\ref{eq:_cost.function.coverage})
and (\ref{eq:_cost.function.complexity}). Stop the iterative process
when the maximum number of iterations ($i=I$) is reached;
\item \emph{Step 3 -} \textbf{Final Trade-Off} \textbf{\emph{EMS}} \textbf{Design}.
Select the set of $O$ trial solutions that are non-dominated and
belonging to the optimized Pareto front \cite{Deb 2002}, \{$\mathbf{T}_{opt}^{\left(o\right)}$;
$o=1,...,O$\}. Such solutions are ordered according to the $\Phi_{2}$
cost function value, namely $\left.\mathbf{T}_{opt}^{\left(o\right)}\right\rfloor _{o=1}=\min_{o=1,...,O}\left\{ \Phi_{2}\left(\mathbf{T}_{opt}^{\left(o\right)}\right)\right\} $
and $\left.\mathbf{T}_{opt}^{\left(o\right)}\right\rfloor _{o=O}=\max_{o=1,...,O}\left\{ \Phi_{2}\left(\mathbf{T}_{opt}^{(o)}\right)\right\} $.
For a given $o$-th ($o=1,...,O$) Pareto optimal solution, $\mathbf{T}_{opt}^{\left(o\right)}$,
place the $n$-th ($n=1,...,N$) tile $\mathcal{S}^{\left(n\right)}$
of the skin at the position $\mathbf{r}_{\mathcal{S}}^{\left(n\right)}$
on the wall of the building if $t_{n,opt}^{\left(o\right)}=1$.
\end{itemize}

\section{\noindent Numerical Results}

\noindent The objective of this section is twofold. On the one hand,
the critical evaluation of the impact on the wireless coverage of
using modular \emph{EMS}s. On the other, the assessment of the effectiveness
of the proposed design method by considering different scenarios and
tiles, while validating (\ref{eq:_Danufane}) through full-wave simulations
with \emph{ANSYS HFSS} \cite{HFSS 2019}.

\noindent As for this latter, the validation benchmark consists of
a single ($N=1$) tile located on the $y-z$ plane at the center of
a Cartesian coordinate system and illuminated by a millimeter ($f=27$
{[}GHz{]}) plane wave generated from the \emph{BS} (\ref{eq:_incident.wave})
that impinges with an orthogonal incidence $\left(\theta_{\Im}^{\left(1\right)},\phi_{\Im}^{\left(1\right)}\right)=\left(0,\,0\right)$
{[}deg{]} on the tile with a vertical polarized (i.e., $\widehat{\mathbf{e}}_{r}=\widehat{\mathbf{y}}$)
electric field. The distance between the \emph{BS} and the tile has
been chosen equal to $d_{\Im}^{\left(1\right)}=100$ {[}m{]} ($\to$
$9\times10^{3}$ $\lambda$), while the side and the square area of
the tile has been set to $L^{\left(1\right)}=25$ $\lambda$ and $\Delta\mathcal{S}^{\left(1\right)}\simeq0.277\times0.277$
{[}$\textnormal{m}^{2}${]}, respectively.

\noindent The realistic \emph{EMS} tile has been modeled with the
\emph{ANSYS HFSS} software as a metasurface defined by a lattice of
unit cells uniformly-space along the $x$- and the $y$-axis by $\frac{\lambda}{2}$.
Each unit-cell is composed by a square metallic patch printed on a
single-layer Rogers 3003 dielectric substrate with thickness $5.08\times10^{-4}$
{[}m{]} {[}Fig. 3(\emph{a}){]}. The single-tile \emph{EMS} has been
then designed by shaping the metallic patches of the metasurface so
that it reflects the impinging wave towards the direction $\left(\theta_{\Re}^{\left(1\right)},\phi_{\Re}^{\left(1\right)}\right)=\left(40,\,-20\right)$
{[}deg{]} \cite{Oliveri 2020}.

\noindent For comparison purposes, the power, $\mathcal{P}_{\Re}\left(\mathbf{r}\right)$,
reflected from the realistic {[}Fig. 3(\emph{b}){]} and the ideal
{[}Fig. 3(\emph{c}){]} tiles on a sphere at a distance of $5$ {[}m{]}
from the skin barycenter $\mathbf{r}_{\mathcal{S}}^{\left(0\right)}$
is shown in Fig. 3. As for the reflected electric field computed in
\emph{HFSS}, only the co-polar component is shown for the sake of
comparison with the field reflected from ideal skin, which is not
affected by the polarization loss \cite{Balanis 2005}. Albeit the
presence of undesired sidelobes, which are generally unavoidable when
dealing with a real implementation of a low-cost artificial metasurface,
the main lobes have a similar shape in both cases and they are steered
towards the same desired angular direction {[}Figs. 3(\emph{b})-3(\emph{c}){]}.
Since the pattern generated from the ideal tile according to (\ref{eq:_Danufane})
is free of undesired sidelobes and polarization losses, it seems reasonable
to infer that the modular \emph{EMS}s synthesized in this work will
represent reference ideal solutions. From an operative viewpoint,
this means that the number $M$ of tiles composing the synthesized
\emph{EMS} layout has to be considered as a lower bound (i.e., the
minimum number of tiles for approximating a project target) to be
probably increased when going to the implementation of the \emph{EMS}
in a real scenario.

\noindent Moving to the design of modular reflecting \emph{EMS}s for
an enhanced wireless coverage, the first test case (\emph{Simple Skin
Layout} - \emph{Orthogonal Incidence}) refers to the urban scenario
depicted in Fig. 4(\emph{a}) where the \emph{BS} is located at $\mathbf{r}_{BS}$
with Cartesian coordinates $(x_{BS},\, y_{BS},\, z_{BS})=(100,\,0,\,10)$
{[}m{]}. The height of the \emph{BS} from the ground (i.e., $z_{BS}=10$
{[}m{]}) has been set according to the \emph{3GPP} guidelines for
the \emph{Urban-Micro} (\emph{UMi}) cell scenario \cite{3GPP 36.873}.
Moreover, the \emph{BS} has been assumed to radiate a plane wave at
$f=27$ {[}GHz{]} having the electric field vertically-polarized ($\widehat{\mathbf{e}}_{r}=\widehat{\mathbf{z}}$)
with unitary amplitude $E_{\Im}=1.0$ {[}V/m{]} that impinges from
the $\phi$-normal direction ($\phi_{\Im}^{\left(0\right)}=0.0$ {[}deg{]})
on the \emph{EMS} placed on the $y-z$ plane {[}i.e., $x_{\mathcal{S}}^{\left(n\right)}=0$
($n=1,...,N$){]} {[}Fig. 4(\emph{b}){]} at a distance of $d_{\Im}^{\left(0\right)}=100$
{[}m{]} from the \emph{BS}. The admissible surface $\mathcal{S}$
for the deployment of the artificial \emph{EMS} on the building facade
{[}Fig. 4(\emph{b}){]} has been chosen with an area $\Delta\mathcal{S}=15$
{[}$\textnormal{m}^{2}${]} and it extends within the range $-2.5$
{[}m{]} $\le$ $y_{\mathcal{S}}$ $\le$ $2.5$ {[}m{]} and $5.0$
{[}m{]} $\le$ $z_{\mathcal{S}}$ $\le$ $8.0$ {[}m{]} along the
$y$- and the $z$-axis, respectively. Such an area $\mathcal{S}$
has been partitioned into $N=60$ square sub-domains of size $\Delta\mathcal{S}^{\left(n\right)}=L^{\left(n\right)}\times L^{\left(n\right)}$
($n=1,...,N$) being $L^{\left(n\right)}=0.5$ {[}m{]}, so that there
are $N_{y}=10$ and $N_{z}=6$ partitions along the $y$- and the
$z$-axis, respectively (i.e., $N=N_{y}\times N_{z}$). By enumerating
the admissible locations of the \emph{EMS} tiles in a raster scan
way, starting from the top left corner of $\mathcal{S}$, the barycenter
of the $n$-th ($n=1,...,N$) sub-domain of $\mathcal{S}$ has the
following coordinates\begin{eqnarray}
y_{\mathcal{S}}^{\left(n\right)} & = & y_{\mathcal{S}}^{\left(1\right)}+\left(n-1-\left\lfloor \frac{n-1}{N_{y}}\right\rfloor N_{y}\right)\times L^{\left(n\right)}\nonumber \\
z_{\mathcal{S}}^{\left(n\right)} & = & z_{\mathcal{S}}^{\left(1\right)}-\left\lfloor \frac{n-1}{N_{y}}\right\rfloor \times L^{\left(n\right)}\label{eq:_TC1.tile.barycenters}\end{eqnarray}
($n=2,...,N$), $y_{\mathcal{S}}^{\left(1\right)}=-2.25$ {[}m{]}
and $z_{\mathcal{S}}^{\left(1\right)}=7.75$ {[}m{]} being the coordinates
of the barycenter of the first ($n=1$) location admissible for a
tile.

\noindent The goal of the \emph{EMS} design it that of enhancing the
power strength in the \emph{AoI} $\Omega$ of size $\Delta\Omega=10\times50$
{[}$\textnormal{m}^{2}${]} located in $x_{\Omega}^{\left(0\right)}=80.35$
{[}m{]} and $y_{\Omega}^{\left(0\right)}=95.75$ {[}m{]} {[}Fig. 4(\emph{a}){]}
along the azimuth direction $\phi_{\Re}^{\left(0\right)}=50.0$ {[}deg{]}
with respect to the \emph{EMS}, $\mathcal{P}_{th}=-70$ {[}dB{]} being
the threshold on the desired coverage in (\ref{eq:_cost.function.coverage}),
while the {}``no connection'' power level has been set to $\mathcal{P}_{bls}\approx-100$
{[}dB{]}. In order to assess the {}``coverage'' condition within
the \emph{AoI}, $U=500$ ideal receivers have been uniformly-distributed
within $\Omega$, that is one receiver every $\Delta\Omega=1$ {[}$\textnormal{m}^{2}${]},
at the height $z_{u}=1.5$ {[}m{]} to emulate users on the ground
\cite{3GPP 36.873}.

\noindent Concerning the \emph{NSGA-II} algorithm, the following setup
of the control parameters has been used: $P=2\times N$, $I=1000$,
$\wp_{c}=1.0$, $\wp_{m}=\frac{1}{N}$, $\aleph_{c}=15$, and $\aleph_{m}=20$.
Moreover, each simulation has been repeated $50$ times with different
random seeds to statistically validate the results from the stochastic
optimization. However, since all simulations led to similar Pareto
fronts at convergence ($i=I$), only the solutions of a representative
run will be reported and discussed in the following. 

\noindent Figure 5 shows the population of trial solutions, $\left\{ \mathbf{T}_{i}^{\left(p\right)};\, p=1,...,P\right\} $,
at the iterations $i=100$, $i=500$, and $i=I$ in the space of the
objectives along with the Pareto front of $O=12$ non-dominated solutions
at the convergence ($i=I$), \{$\mathbf{T}_{opt}^{\left(o\right)}$;
$o=1,...,O$\}. Let us now analyze the \emph{EMS} solution that fully
fits the coverage requirements {[}i.e., $\Phi_{1}\left(\mathbf{T}_{opt}^{\left(o\right)}\right)=0${]},
which corresponds to the $O$-th representative point of the Pareto
front, whose chromosome has $M=12$ bits at one (i.e., $\left.t_{n,opt}^{\left(o\right)}\right\rfloor _{o=12}=1$,
$n$ $=$ \{$3$, $4$, $5$, $6$, $8$, $12$, $30$, $32$, $43$,
$44$, $45$, $46$\}) so that the \emph{EMS} layout turns out being
composed by $M=12$ tiles as shown in Fig. 6(\emph{a}). The support
of such an \emph{EMS} amounts to $\left.\Delta\mathcal{S}_{opt}^{\left(o\right)}\right\rfloor _{o=12}=3$
{[}$\textnormal{m}^{2}${]}, which is one fifth of the whole admissible
\emph{EMS} surface (i.e., $\frac{\left.\Delta\mathcal{S}_{opt}^{\left(o\right)}\right\rfloor _{o=12}}{\Delta\mathcal{S}}=20$
\%). The coverage improvement enabled by the installation of such
an artificial skin on the building facade is pointed out in Fig. 6(\emph{b})
where the map of the reflected power $\mathcal{P}_{\Re}\left(\mathbf{r}\right)$
at $z=1.5$ {[}m{]} from the ground is shown in a region $\Psi$,
around the \emph{AoI,} of extension $\Delta\Psi=200\times200$ {[}$\textnormal{m}^{2}${]}.
As it can be observed, the power intensity along the direction of
the street passing through $\Omega$ has been significantly increased.
As a matter of fact, the signal turns out to be stronger not only
in $\Omega$, but also before and after {[}Fig. 6(\emph{b}){]} since
the skin tiles generate simple pencil beams (\ref{eq:_Danufane})
with an elongated footprint on the ground, the dots in Figs. 6(\emph{b})-6(\emph{c})
being the $M=12$ points $\mathbf{r}_{\Omega}^{\left(n\right)}$,
$n=\left\{ 3,\,4,\,5,\,6,\,8,\,12,\,30,\,32,\,43,\,44,\,45,\,46\right\} $
where the peaks of the beams reflected from the $M$ \emph{EMS} tiles
are directed. Regardless of the simplicity of the beam afforded by
a single \emph{EMS} tile, the combined use of multiple/modular tiles
has allowed to reach the desired average power threshold $\mathcal{P}_{th}=-70$
{[}dB{]} (i.e., $\Phi_{1}\left(\left.\mathbf{T}_{opt}^{\left(o\right)}\right\rfloor _{o=12}\right)=0$)
in the whole \emph{AoI} as shown in Fig. 6(\emph{c}), $\gamma-\xi$
being the $\Omega$ local coordinate system {[}see Fig. 6(\emph{b}){]}.
Indeed, the statistics of the power reflected in $\Omega$ are: $\min_{\mathbf{r}\in\Omega}\left\{ \mathcal{P}_{\Re}\left(\mathbf{r}_{u};\,\left.\mathbf{T}_{opt}^{(o)}\right\rfloor _{o=12}\right)\right\} =-69.9$
{[}dB{]}, $\max_{\mathbf{r}\in\Omega}\left\{ \mathcal{P}_{\Re}\left(\mathbf{r}_{u};\,\left.\mathbf{T}_{opt}^{(o)}\right\rfloor _{o=12}\right)\right\} =-63.0$
{[}dB{]}, and $\textnormal{avg}_{\mathbf{r}\in\Omega}\left\{ \mathcal{P}_{\Re}\left(\mathbf{r}_{u};\,\left.\mathbf{T}_{opt}^{(o)}\right\rfloor _{o=12}\right)\right\} =-66.8$
{[}dB{]}, respectively (Tab. I).

\noindent For the sake of completeness, other two representative solutions
of the Pareto front in Fig. 5 are analyzed. The \emph{EMS} layouts
and the maps of the power reflected in $\Psi$ of the solution with
minimum complexity ($o=1$ - Fig. 5) and the one having $\left.\Phi_{1}\left(\mathbf{T}_{opt}^{\left(o\right)}\right)\right\rfloor _{o=4}\approx0.1$
(i.e., $\textnormal{avg}_{\mathbf{r}\in\Omega}\left\{ \mathcal{P}_{\Re}\left(\mathbf{r}_{u};\,\left.\mathbf{T}_{opt}^{(o)}\right\rfloor _{o=12}\right)\right\} =-77$
{[}dB{]}) are reported in Fig. 7.

\noindent The minimum complexity \emph{EMS} (i.e., $\Phi_{2}\left(\left.\mathbf{T}_{opt}^{\left(o\right)}\right\rfloor _{o=1}\right)=\frac{1}{60}$)
needs only one ($M=1$) tile {[}Fig. 7(\emph{a}){]}, but the average
power level in $\Omega$ reduces of $\left.\delta\Phi_{1}\right]_{o=1}^{o=12}=17.9$
{[}dB{]} ($\left.\delta\Phi\right]_{o'}^{o}\triangleq\Phi\left(\mathbf{T}_{opt}^{\left(o\right)}\right)-\Phi\left(\mathbf{T}_{opt}^{\left(o'\right)}\right)$;
$o$, $o'$ $\in$ $\left[1,\, O\right]$) with respect to that in
Fig. 6(\emph{a}). Owing to the presence of a single tile, the map
of the \emph{EM} power in Fig. 7(\emph{c}) shows the classical footprint
of a pencil beam characterized by a mainlobe focused in the point
$\left.\mathbf{r}_{\Omega}^{\left(n\right)}\right\rfloor _{n=28}$
within $\Omega$ {[}Fig. 7(\emph{c}){]} along the central line of
the \emph{AoI} ($\xi=0$) {[}Fig. 7(\emph{e}){]}, while there are
portions of $\Omega$ close to $\xi=\pm5$ {[}m{]} where the power
strength is very low {[}Fig. 7(\emph{e}){]}. Quantitatively, it turns
out that the condition $\mathcal{P}_{th}\le\mathcal{P}_{\Re}\left(\mathbf{r}_{u};\,\left.\mathbf{T}_{opt}^{\left(o\right)}\right\rfloor _{o=1}\right)$
never holds true {[}Fig. 7(\emph{e}){]} since also the power peak
is below the desired \emph{QoS} threshold ($\mathcal{P}_{\Re}\left(\mathbf{r}_{u};\,\left.\mathbf{T}_{opt}^{\left(o\right)}\right\rfloor _{o=1}\right)=-78.1$
{[}dB{]}). Moreover, the minimum level of the power reflected within
the \emph{AoI} by such a single-tile \emph{EMS} of side $L^{\left(1\right)}=0.5$
{[}m{]} is equal to $\min_{\mathbf{r}\in\Omega}\left\{ \mathcal{P}_{\Re}\left(\mathbf{r}_{u};\,\left.\mathbf{T}_{opt}^{\left(o\right)}\right\rfloor _{o=1}\right)\right\} =-173.9$
{[}dB{]} (Tab. I), that is (well) below the {}``connectivity'' threshold
of $\mathcal{P}_{bls}=-100$ {[}dB{]}. Such an undesired condition
verifies in other portions of $\Omega$ where there is not enough
signal for assuring the users' connections {[}$\mathcal{P}_{\Re}\left(\mathbf{r}_{u};\,\left.\mathbf{T}_{opt}^{\left(o\right)}\right\rfloor _{o=1}\right)<\mathcal{P}_{bls}$
- Fig. 7(\emph{g}){]}.

\noindent By using three more tiles {[}i.e., $M=4$ - Fig. 7(\emph{b}){]},
the power level reflected on $\Omega$ turns out significantly higher
{[}Fig. 7(\emph{d}){]}, the average power being increased of almost
ten times (i.e. $\left.\delta\Phi_{1}\right]_{o=1}^{o=4}=9.7$ {[}dB{]}),
and there are no more ''no connection'' zones within $\Omega$ {[}Fig.
7(\emph{h}){]} since $\min_{\mathbf{r}\in\Omega}$\{ $\mathcal{P}_{\Re}$$\left(\mathbf{r}_{u};\,\left.\mathbf{T}_{opt}^{\left(o\right)}\right\rfloor _{o=4}\right)$\}
$=$ $-90.9$ {[}dB{]} (Tab. I), even though the power peak is still
slightly lower than the \emph{QoS} threshold ($\max_{\mathbf{r}\in\Omega}\left\{ \mathcal{P}_{\Re}\left(\mathbf{r}_{u};\,\left.\mathbf{T}_{opt}^{\left(o\right)}\right\rfloor _{o=4}\right)\right\} =-71.7$
{[}dB{]}).

\noindent In the second test case (\emph{Simple Skin Layout} - \emph{Oblique
Incidence}), the field generated from the \emph{BS} has been assumed
impinging on the \emph{EMS} with an oblique incidence on the azimuth
plane (i.e., $\phi_{\Im}^{\left(0\right)}=20$ {[}deg{]}), being $(x_{BS},\, y_{BS},\, z_{BS})=(93.9,\,34.2,\,10)$
{[}m{]} such that $d_{\Im}^{\left(0\right)}=100$ {[}m{]} as in the
previous example {[}Fig. 8(\emph{a}){]}. All other features concerned
with the \emph{EM} field generated from the \emph{BS} (i.e., frequency
and polarization), the area and the discretization of $\mathcal{S}$,
and the coverage area under analysis $\Omega$ have been kept unaltered
from the first test case.

\noindent The Pareto front of the $O$ optimal trade-off solutions
determined by the proposed \emph{NSGA-II} based approach is compared
in Fig. 8(\emph{b}) with that obtained in the {}``normal incidence''
case (Fig. 5). The reader can observe that the Pareto front of the
{}``oblique incidence'' scenario consists of $O=14$ \emph{EMS}
designs (vs. $O=12$ - {}``normal incidence'') and it turns out
that the oblique incidence from the \emph{BS} needs a higher number
of tiles to yield the same coverage of the {}``normal incidence''
solutions {[}Fig. 8(\emph{b}){]}. As expected, a wider area is now
required because of the reduction of the effective area of the \emph{EMS}
(\ref{eq:_Danufane}) since a lower amount of power is intercepted
from the incident wave with the same area of the {}``normal incidence''
\emph{EMS}.

\noindent Figure 9 summarizes the characteristics of the $O$-th ($O=14$)
solution that fits the coverage requirement {[}i.e., $\mathcal{P}_{\Re}\left(\mathbf{r}_{u};\,\mathbf{T}_{opt}^{\left(O\right)}\right)\ge\mathcal{P}_{th}$
$\to$ $\Phi_{1}\left(\mathbf{T}_{opt}^{\left(O\right)}\right)=0${]}.
More in detail, the layout of the corresponding \emph{EMS} is composed
by $M=14$ tiles {[}Fig. 9(\emph{a}){]}, that is two more than those
of the $O$-th \emph{EMS} for the normal incidence, and the arising
tiles arrangement is also quite different {[}Fig. 9(\emph{a}) vs.
Fig. 6(\emph{a}){]}. On the contrary, the power distributions are
quite similar as pictorially shown by comparison of the maps in Figs.
9(\emph{b})-9(\emph{c}) with those in Figs. 6(\emph{b})-6(\emph{c})
and also confirmed by the statistics of the power reflected by the
\emph{EMS} within the \emph{AoI} (Tab. I). Indeed, the differences
among the values of the statistical indices are null or negligible
(i.e., $\left.\delta\mathcal{P}_{\Re}^{min}\right]_{Normal}^{Oblique}=0.0$
{[}dB{]}, $\left.\delta\mathcal{P}_{\Re}^{max}\right]_{Normal}^{Oblique}=0.2$
{[}dB{]}, and $\left.\delta\mathcal{P}_{\Re}^{av}\right]_{Normal}^{Oblique}$
$=$ $\left.\delta\Phi_{1}\right]_{Normal}^{Oblique}$ $=$$-0.1$
{[}dB{]}, being $\left.\delta\mathcal{P}_{\Re}^{stat}\right]_{Normal}^{Oblique}$
$\triangleq$ $\textnormal{stat}_{\mathbf{r}\in\Omega}\left\{ \mathcal{P}_{\Re}\left(\mathbf{r}_{u};\,\left.\mathbf{T}_{opt}^{\left(o\right)}\right\rfloor _{o=O}^{Oblique}\right)\right\} $
$-$ $\textnormal{stat}_{\mathbf{r}\in\Omega}\left\{ \mathcal{P}_{\Re}\left(\mathbf{r}_{u};\,\left.\mathbf{T}_{opt}^{\left(o\right)}\right\rfloor _{o=O}^{Normal}\right)\right\} $
- Tab. I).

\noindent The third design experiment (\emph{Simple Skin Layout} -
\emph{Varying Tiles Size}) is concerned with a surface $\mathcal{S}$
still discretized with uniform tiles, but considering different tile
sizes: $L^{\left(n\right)}=1.0$ {[}m{]} {[}$\to$ $N=N_{y}\times N_{z}=5\times3=15$
- Fig. 10(\emph{a}){]} or $L^{\left(n\right)}=0.25$ {[}m{]} {[}$\to$
$N=N_{y}\times N_{z}=20\times12=240$ - Fig. 10(\emph{b}){]} ($n=1,...,N$).
In the former case, there are few admissible tiles reflecting a narrow
beam towards the \emph{AoI}, while the number of tiles and \emph{DoF}s
is $16$ times larger in the latter case where the beam reflected
by each $m$-th ($m=1,...,M$) installed tile has a broader coverage.

\noindent The \emph{NSGA-II} optimization has been run for both tile
sizes and the Pareto fronts obtained at the convergence ($i=I$) are
shown in Fig. 11(\emph{a}) along with that of Fig. 5, which is related
to the tile size $L^{\left(n\right)}=0.5$ {[}m{]} ($n=1,...,N$).
The plots in Fig. 11(\emph{a}) indicate that, the wider the tile size,
the higher is the value of the complexity index $\Phi_{2}$ to fulfil
(\ref{eq:_coverage.condition}). Furthermore, it is worth highlighting
that when $L^{\left(n\right)}=1.0$ {[}m{]}, no solution of the Pareto
front satisfies the coverage requirement since $\Phi_{1}\left(\mathbf{T}_{opt}^{\left(O\right)}\right)>0$,
while the coverage condition holds true using smaller tiles. 

\noindent The same conclusions arise when extending the coverage area
$\Omega$ from $\Delta\Omega=10\times50$ {[}$\textnormal{m}^{2}${]}
{[}Fig. 11(\emph{a}){]} up to $\Delta\Omega=10\times100$ {[}$\textnormal{m}^{2}${]}
{[}Fig. 11(\emph{b}){]}. For this latter case, the layouts and the
coverage maps of the solutions providing the best coverage (i.e.,
the $O$-th of the Pareto front) for each tile size are shown in Fig.
12. By analyzing the power distributions in Figs. 12(\emph{d})-12(\emph{f}),
it turns out that the main advantage of using larger tiles, which
reflect narrower beams, is the capability of better focusing the reflected
field only along the direction of the \emph{AoI} {[}Fig. 12(\emph{f})
vs. Fig. 12(\emph{d}){]}. This is not for free and the cost to pay
is that of having a very large \emph{EMS} composed by $M=13$ tiles,
each of $\Delta\mathcal{S}^{\left(m\right)}=1$ {[}$\textnormal{m}^{2}${]}
($m=1,...,M$), for a total surface of $\left.\Delta\mathcal{S}_{opt}^{\left(O\right)}\right\rfloor _{L=1.0\,[\textnormal{m}]}=13$
{[}$\textnormal{m}^{2}${]}, while the area occupied by the \emph{EMS}
when using square tiles of size $L^{\left(n\right)}=0.25$ {[}m{]}
{[}Fig. 12(\emph{a}){]} and $L^{\left(n\right)}=0.5$ {[}m{]} {[}Fig.
12(\emph{b}){]} amounts to $\left.\Delta\mathcal{S}_{opt}^{\left(O\right)}\right\rfloor _{L=0.25\,[\textnormal{m}]}\simeq1.69$
{[}$\textnormal{m}^{2}${]} and $\left.\Delta\mathcal{S}_{opt}^{\left(O\right)}\right\rfloor _{L=0.5\,[\textnormal{m}]}=5$
{[}$\textnormal{m}^{2}${]}, respectively. This means that the limited
focusing capability of smaller tiles is balanced by a reduction of
the required \emph{EMS} extension, thus a lower cost of the \emph{EMS}.

\noindent In the last design example (\emph{Complex Skin Layout} -
\emph{Orthogonal Incidence}), the admissible region $\mathcal{S}$
on the building facade is more complex {[}Fig. 13(\emph{a}){]} since
the area dedicated to the \emph{EMS} deployment is smaller and there
are more architectural constraints (e.g., misaligned windows and open
window shutters) as in historical buildings. As for the descriptive
parameters of the scenario at hand, they have been defined as in the
{}``\emph{Simple Skin Layout} - \emph{Orthogonal Incidence}'' case,
but a larger \emph{AoI} (i.e., $\Delta\Omega=10\times100$ {[}$\textnormal{m}^{2}${]})
has been considered. 

\noindent The evolution ($i=100$, $i=500$, and $i=I$) of the population
of \emph{EMS} trial solutions, $\left\{ \mathbf{T}_{i}^{\left(p\right)};\, p=1,...,P\right\} $,
in the space of the objectives is shown in Fig. 13(\emph{b}) together
with the Pareto front at convergence ($i=I$), which includes $O=31$
non-dominated solutions, \{$\mathbf{T}_{opt}^{\left(o\right)}$; $o=1,...,O$\}.
As it can be noticed {[}Fig. 13(\emph{b}){]}, $15$ \emph{EMS} of
the Pareto front have values of the coverage index smaller than $\Phi_{1}\left(\left.\mathbf{T}_{opt}^{\left(o\right)}\right\rfloor _{o}\right)<10^{-2}$.

\noindent The $O$-th solution, which fully satisfies the coverage
requirements, is done by $M=32$ tiles of size $\Delta\mathcal{S}^{\left(m\right)}=0.5\times0.5$
{[}$\textnormal{m}^{2}${]} ($m=1,...,M$) and it covers a surface
area of $\Delta\mathcal{S}_{opt}^{\left(O\right)}=8$ {[}$\textnormal{m}^{2}${]}
{[}Fig. 14(\emph{a}){]}. Despite the irregularity of the \emph{EMS}
layout, the coverage maps in Figs. 14(\emph{b})-14(\emph{d}) confirm
that the proposed \emph{EMS} design method properly selects, from
the admissible pool, a subset of tiles that guarantees the required
power level within the \emph{AoI} $\Omega$ (Tab. I).

\section{\noindent Conclusions}

\noindent Within the \emph{SEME} vision, this paper has proposed a
novel strategy to improve the signal strength in urban areas where
the power radiated by the \emph{BS} is too strongly attenuated. More
specifically, such an approach proposes the use of modular, passive,
and static artificial metasurfaces to be installed/embedded on the
facades of urban buildings, such as coating skins, to enhance the
coverage by reflecting the \emph{EM} wave coming from the \emph{BS}
towards the desired directions within an \emph{AoI}. In order to fulfil
user-defined coverage conditions, while minimizing the cost/complexity,
the design of the \emph{EMS} has been cast as a multi-objective optimization
problem and it has been addressed by means of a binary implementation
of the \emph{NSGA-II} algorithm.

\noindent From a technological and methodological viewpoint, the main
novelties, to the best of the authors' knowledge, of this research
work can be summarized as follows:

\begin{itemize}
\item the introduction for the first time of a novel cost-effective solution,
to be possibly implemented through cheap printed technology, for the
large scale deployment of artificial metasurfaces to be installed
on the facades of buildings for improving the wireless coverage in
urban scenarios;
\item the suitability of the proposed technological solution in future wireless
networks thanks to its {}``green'' (i.e., passive) and non-invasive
(i.e., low profile and without heavy architectural impact) nature;
\item the development of a customized design strategy to enable an effective/efficient
optimization-based design of (also large) \emph{EMS}s composed by
simple (also non-homogeneous) tiles.
\end{itemize}
\noindent From the numerical assessment, which has been carried out
by considering realistic topological urban scenarios and a millimeter-wave
5G frequency band, the following outcomes can be drawn:

\begin{itemize}
\item the use of tiled \emph{EMS}s always improves the coverage of the \emph{AoI};
\item the (\emph{NSGA-II})-based synthesis approach provides the designer
with a Pareto front of multiple \emph{EMS} solutions, which are trade-offs
between coverage requirements and complexity of the \emph{EMS} layout.
In all the considered scenarios, an \emph{EMS} that fulfils the user-defined
coverage condition (i.e., not only the user connection) has been (generally)
found without using the whole area available on the facade of the
building;
\item \noindent the number and the positions of the tiles of the \emph{EMS}
layout depend on the relative position between the \emph{BS} and the
\emph{AoI}. Moreover, the dimension (e.g., small, medium, large) and
the distribution (e.g., uniform or non-uniform) of the tiles composing
the \emph{EMS} are other \emph{DoF}s, which can be exploited to fit
the coverage conditions as well as other architectural constraints
(e.g., misaligned windows and open window shutters).
\end{itemize}
Future research activities, beyond the scope of this paper, will integrate
the design of the tiles layout within the proposed iterative optimization
loop in order to take into account non-ideal reflections and polarization
losses. Moreover, the presence of multiple \emph{BS}s and \emph{AoI}s
will be dealt with towards the definition of a tool for network planning.

\section*{\noindent Acknowledgements}

\noindent This work benefited from the networking activities carried
out within the Project {}``CYBER-PHYSICAL ELECTROMAGNETIC VISION:
Context-Aware Electromagnetic Sensing and Smart Reaction (EMvisioning)''
(Grant no. 2017HZJXSZ){}`` funded by the Italian Ministry of Education,
University, and Research under the PRIN2017 Program (CUP: E64I19002530001).
Moreover, it benefited from the networking activities carried out
within the Project {}``SPEED'' (Grant No. 61721001) funded by National
Science Foundation of China under the Chang-Jiang Visiting Professorship
Program, the Project 'Inversion Design Method of Structural Factors
of Conformal Load-bearing Antenna Structure based on Desired EM Performance
Interval' (Grant no. 2017HZJXSZ) funded by the National Natural Science
Foundation of China, and the Project 'Research on Uncertainty Factors
and Propagation Mechanism of Conformal Loab-bearing Antenna Structure'
(Grant No. 2021JZD-003) funded by the Department of Science and Technology
of Shaanxi Province within the Program Natural Science Basic Research
Plan in Shaanxi Province. A. Massa wishes to thank E. Vico for her
never-ending inspiration, support, guidance, and help.

\newpage
\section*{FIGURE CAPTIONS}

\begin{itemize}
\item \textbf{Figure 1.} \emph{Problem geometry}. Illustrative sketches
of the wireless communication scenario: (\emph{a}) top view and (\emph{b})
detailed zoom.
\item \textbf{Figure 2.} \emph{Problem geometry.} Graphical representation
of the \emph{EMS} local coordinate system.
\item \textbf{Figure 3.} \emph{Numerical Validation}. Layout of (\emph{a})
the \emph{ANSYS HFSS} model of the single-tile \emph{EMS} and plot
of (\emph{b})(\emph{c}) the angular distribution of the power, $\mathcal{P}_{\Re}\left(\mathbf{r}\right)$,
reflected from the \emph{EMS} on a sphere at a distance of $5$ {[}m{]}
and computed with (\emph{b}) \emph{ANSYS HFSS} or using (\emph{c})
the closed-form relationship (\ref{eq:_Danufane}).
\item \textbf{Figure 4.} \emph{Numerical Validation} - \emph{Simple Skin
Layout} - \emph{Orthogonal Incidence} ($f=27$ {[}GHz{]}, $\mathcal{S}=15$
{[}$\textnormal{m}^{2}${]}, $L=0.5$ {[}m{]}, $N=60$, $\Delta\Omega=500$
{[}$\textnormal{m}^{2}${]}, $\mathcal{P}_{th}=-70$ {[}dB{]}). \emph{}Sketch
\emph{}of (\emph{a}) the scenario and of (\emph{b}) the admissible
surface $\mathcal{S}$ along with its tile discretization.
\item \textbf{Figure 5.} \emph{Numerical Validation} - \emph{Simple Skin
Layout} - \emph{Orthogonal Incidence} ($f=27$ {[}GHz{]}, $\mathcal{S}=15$
{[}$\textnormal{m}^{2}${]}, $L=0.5$ {[}m{]}, $N=60$, $\Delta\Omega=500$
{[}$\textnormal{m}^{2}${]}, $\mathcal{P}_{th}=-70$ {[}dB{]}). Iterative
($i=100$, $i=500$, and $i=I$) evolution of the population of the
$P$ ($P=2\times N$) trial solutions, $\left\{ \mathbf{T}_{i}^{\left(p\right)};\, p=1,...,P\right\} $,
in the space of the design objectives and Pareto front at convergence
($i=I$), \{$\mathbf{T}_{opt}^{\left(o\right)}$; $o=1,...,O$\}.
\item \textbf{Figure 6.} \emph{Numerical Validation} - \emph{Simple Skin
Layout} - \emph{Orthogonal Incidence} ($f=27$ {[}GHz{]}, $\mathcal{S}=15$
{[}$\textnormal{m}^{2}${]}, $L=0.5$ {[}m{]}, $N=60$, $\Delta\Omega=500$
{[}$\textnormal{m}^{2}${]}, $\mathcal{P}_{th}=-70$ {[}dB{]}). Plot
of (\emph{a}) the \emph{EMS} layout of the $O$-th ($O=12$) solution,
$\mathcal{S}_{opt}^{\left(O\right)}$, of the Pareto front in Fig.
5 and map of the spatial distribution of the power, $\mathcal{P}_{\Re}\left(\mathbf{r}\right)$,
reflected from the \emph{EMS} in (\emph{b}) the region $\Psi$ and
within (\emph{c}) the AoI $\Omega$ ($\Omega\subset\Psi$).
\item \textbf{Figure 7.} \emph{Numerical Validation} - \emph{Simple Skin
Layout} - \emph{Orthogonal Incidence} ($f=27$ {[}GHz{]}, $\mathcal{S}=15$
{[}$\textnormal{m}^{2}${]}, $L=0.5$ {[}m{]}, $N=60$, $\Delta\Omega=500$
{[}$\textnormal{m}^{2}${]}, $\mathcal{P}_{th}=-70$ {[}dB{]}, $\mathcal{P}_{bls}\approx-100$
{[}dB{]}). Plot of (\emph{a})(\emph{b}) the \emph{EMS} layout and
of (\emph{c})-(\emph{f}) the corresponding spatial distributions of
the power, $\mathcal{P}_{\Re}\left(\mathbf{r}\right)$, reflected
from the \emph{EMS} along with (\emph{g})(\emph{h}) the coverage/connectivity
maps for (\emph{a})(\emph{c})(\emph{e})(\emph{g}) the $o=1$ and (\emph{b})(\emph{d})(\emph{f})(\emph{h})
the $o=4$ solutions of the Pareto front in Fig. 5.
\item \textbf{Figure 8.} \emph{Numerical Validation} - \emph{Simple Skin
Layout} - \emph{Oblique Incidence} ($f=27$ {[}GHz{]}, $\mathcal{S}=15$
{[}$\textnormal{m}^{2}${]}, $L=0.5$ {[}m{]}, $N=60$, $\Delta\Omega=500$
{[}$\textnormal{m}^{2}${]}, $\mathcal{P}_{th}=-70$ {[}dB{]}). \emph{}Sketch
\emph{}of (\emph{a}) the scenario and plot of (\emph{b}) the Pareto
front at convergence ($i=I$), \{$\mathbf{T}_{opt}^{\left(o\right)}$;
$o=1,...,O$\}.
\item \textbf{Figure 9.} \emph{Numerical Validation} - \emph{Simple Skin
Layout} - \emph{Oblique Incidence} ($f=27$ {[}GHz{]}, $\mathcal{S}=15$
{[}$\textnormal{m}^{2}${]}, $L=0.5$ {[}m{]}, $N=60$, $\Delta\Omega=500$
{[}$\textnormal{m}^{2}${]}, $\mathcal{P}_{th}=-70$ {[}dB{]}). Plot
of (\emph{a}) the \emph{EMS} layout of the $O$-th ($O=14$) solution,
$\mathcal{S}_{opt}^{\left(O\right)}$, of the Pareto front in Fig.
5 and map of the spatial distribution of the power, $\mathcal{P}_{\Re}\left(\mathbf{r}\right)$,
reflected from the \emph{EMS} in (\emph{b}) the region $\Psi$ and
within (\emph{c}) the AoI $\Omega$ ($\Omega\subset\Psi$).
\item \textbf{Figure 10.} \emph{Numerical Validation} - \emph{Simple Skin
Layout} - \emph{Varying Tiles Size} ($f=27$ {[}GHz{]}, $\mathcal{S}=15$
{[}$\textnormal{m}^{2}${]}, $N=\left\{ 240,\,15\right\} $, $\mathcal{P}_{th}=-70$
{[}dB{]}). Sketches \emph{}of the scenario and of the admissible surface
$\mathcal{S}$ along with its discretization when using tiles with
side-length (\emph{a}) $L=1.0$ {[}m{]} and (\emph{b}) $L=0.25$ {[}m{]}.
\item \textbf{Figure 11.} \emph{Numerical Validation} - \emph{Simple Skin
Layout} - \emph{Varying Tiles Size} ($f=27$ {[}GHz{]}, $\mathcal{S}=15$
{[}$\textnormal{m}^{2}${]}, $\mathcal{P}_{th}=-70$ {[}dB{]}). Plot
of the Pareto fronts at convergence ($i=I$), \{$\mathbf{T}_{opt}^{\left(o\right)}$;
$o=1,...,O$\}, for different square tile sizes ($L$ being the side-length
of the tile) in correspondence with an \emph{AoI} $\Omega$ of dimension
(\emph{a}) $\Delta\Omega=10\times50=500$ {[}$\textnormal{m}^{2}${]}
and (\emph{b}) $\Delta\Omega=10\times100=1000$ {[}$\textnormal{m}^{2}${]}.
\item \textbf{Figure 12.} \emph{Numerical Validation} - \emph{Simple Skin
Layout} - \emph{Varying Tiles Size} ($f=27$ {[}GHz{]}, $\mathcal{S}=15$
{[}$\textnormal{m}^{2}${]}, $L=\left\{ 0.25,\,0.5,\,1.0\right\} $
{[}m{]}, $N=\left\{ 240,\,60,\,15\right\} $, $\Delta\Omega=1000$
{[}$\textnormal{m}^{2}${]}, $\mathcal{P}_{th}=-70$ {[}dB{]}). Plot
of (\emph{a})-(\emph{c}) the \emph{EMS} layouts of the $O$-th solution,
$\mathcal{S}_{opt}^{\left(O\right)}$, of the Pareto fronts in Fig.
11(\emph{b}) and maps of the spatial distribution of the power, $\mathcal{P}_{\Re}\left(\mathbf{r}\right)$,
reflected from the \emph{EMS} in (\emph{d})-(\emph{f}) the region
$\Psi$ and within (\emph{g})-(\emph{i}) the AoI $\Omega$ ($\Omega\subset\Psi$)
when using tiles with side-length (\emph{a})(\emph{d})(\emph{g}) $L=0.25$
{[}m{]}, (\emph{b})(\emph{e})(\emph{h}) $L=0.25$ {[}m{]}, and (\emph{c})(\emph{f})(\emph{i})
$L=1.0$ {[}m{]}.
\item \textbf{Figure 13.} \emph{Numerical Validation} - \emph{Complex Skin
Layout} - \emph{Orthogonal Incidence} ($f=27$ {[}GHz{]}, $\mathcal{S}=15$
{[}$\textnormal{m}^{2}${]}, $L=0.5$ {[}m{]}, $N=60$, $\Delta\Omega=1000$
{[}$\textnormal{m}^{2}${]}, $\mathcal{P}_{th}=-70$ {[}dB{]}). Sketch
\emph{}of (\emph{a}) the scenario and of the admissible surface $\mathcal{S}$
along with its tile discretization and plot of (\emph{b}) the iterative
($i=100$, $i=500$, and $i=I$) evolution of the population of the
$P$ ($P=2\times N$) trial solutions, $\left\{ \mathbf{T}_{i}^{\left(p\right)};\, p=1,...,P\right\} $,
in the space of the design objectives along with the Pareto front
at convergence ($i=I$), \{$\mathbf{T}_{opt}^{\left(o\right)}$; $o=1,...,O$\}.
\item \textbf{Figure 14.} \emph{Numerical Validation} - \emph{Complex Skin
Layout} - \emph{Orthogonal Incidence} ($f=27$ {[}GHz{]}, $\mathcal{S}=15$
{[}$\textnormal{m}^{2}${]}, $L=0.5$ {[}m{]}, $N=60$, $\Delta\Omega=1000$
{[}$\textnormal{m}^{2}${]}, $\mathcal{P}_{th}=-70$ {[}dB{]}). Plot
of (\emph{a}) the \emph{EMS} layout and of (\emph{b})(\emph{c}) the
corresponding spatial distributions of the power, $\mathcal{P}_{\Re}\left(\mathbf{r}\right)$,
reflected from the \emph{EMS} along with (\emph{d}) the coverage/connectivity
maps for the $O$-th ($O=31$) solution of the Pareto front in Fig.
13(\emph{b}).
\end{itemize}

\section*{TABLE CAPTIONS}

\begin{itemize}
\item \textbf{Table I.} \emph{Numerical Validation}. Statistics of the reflected
power $\mathcal{P}_{\Re}\left(\mathbf{r}\right)$ within the \emph{AoI},
$\Omega$.
\end{itemize}
\newpage
\begin{center}~\vfill\end{center}

\begin{center}\begin{tabular}{c}
\includegraphics[%
  width=0.40\columnwidth]{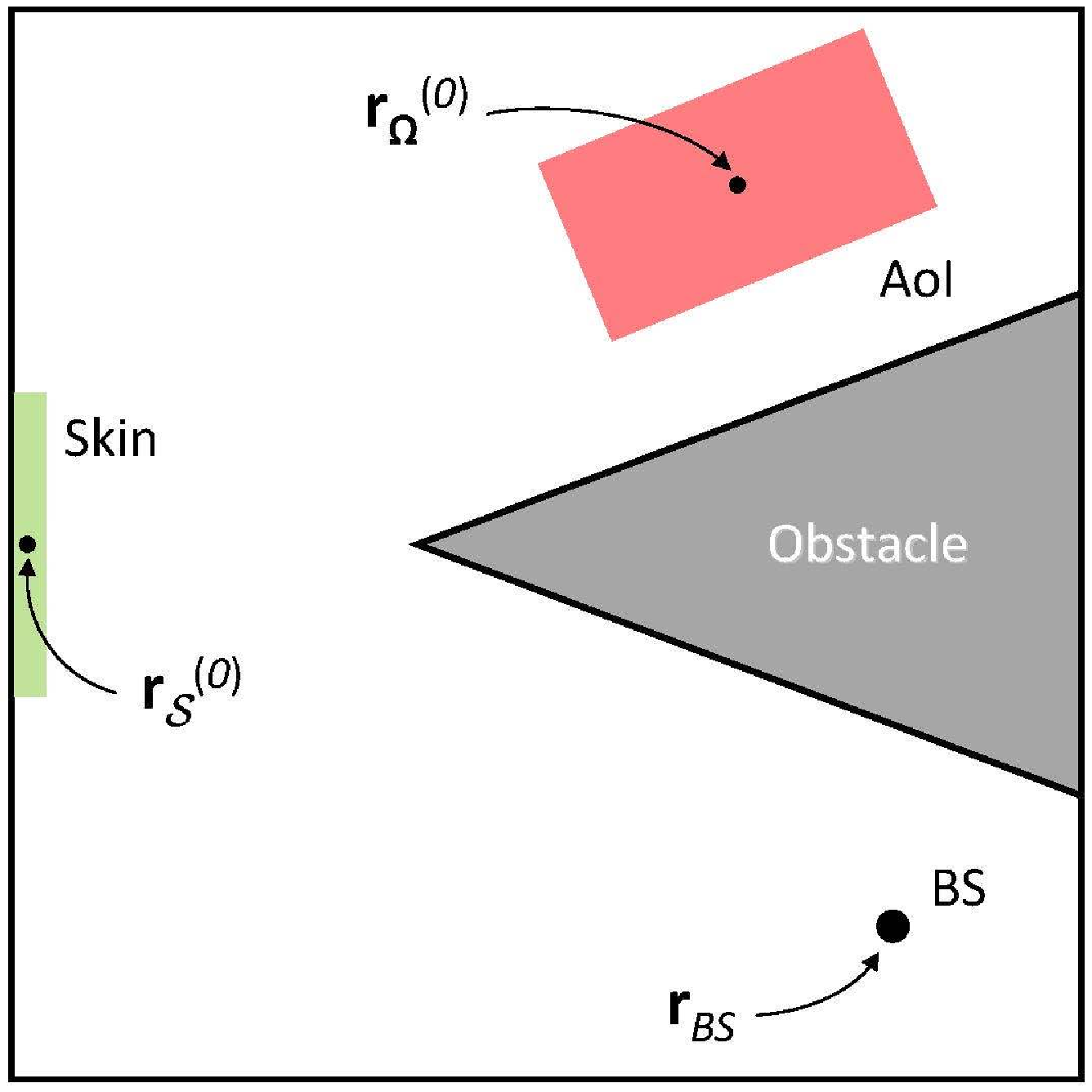}\tabularnewline
(\emph{a})\tabularnewline
\tabularnewline
\includegraphics[%
  width=0.80\columnwidth]{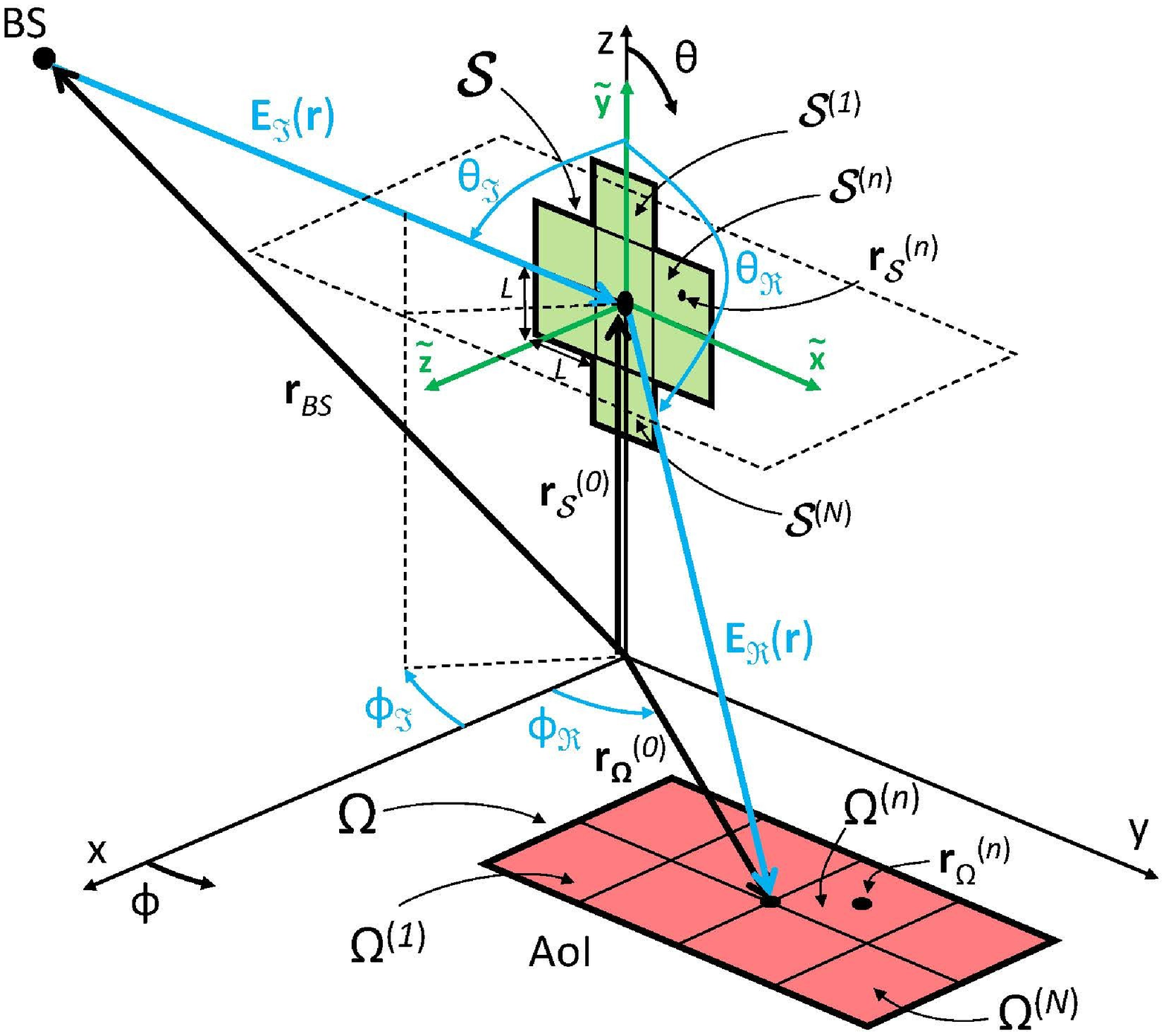}\tabularnewline
(\emph{b})\tabularnewline
\end{tabular}\end{center}

\begin{center}~\vfill\end{center}

\begin{center}\textbf{Fig. 1 - P. Rocca} \textbf{\emph{et al.}}\textbf{,}
\textbf{\emph{{}``}}On the Design of Modular Reflecting EM Skins
...''\end{center}

\newpage
\begin{center}~\vfill\end{center}

\begin{center}\includegraphics[%
  width=0.80\columnwidth]{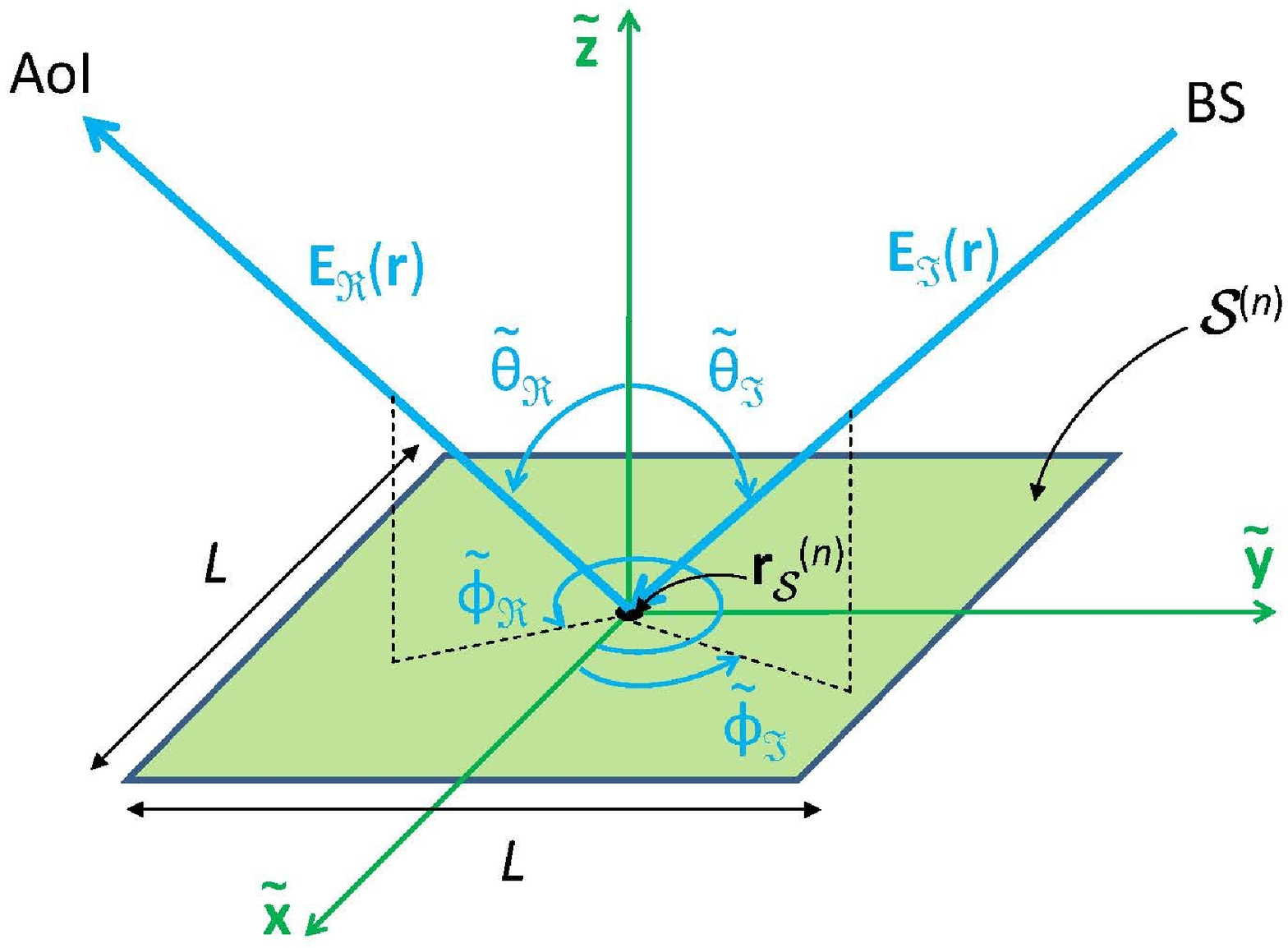}\end{center}

\begin{center}~\vfill\end{center}

\begin{center}\textbf{Fig. 2 - P. Rocca} \textbf{\emph{et al.}}\textbf{,}
\textbf{\emph{{}``}}On the Design of Modular Reflecting EM Skins
...''\end{center}

\newpage
\begin{center}~\vfill\end{center}

\begin{center}\begin{tabular}{c}
\includegraphics[%
  width=0.50\columnwidth]{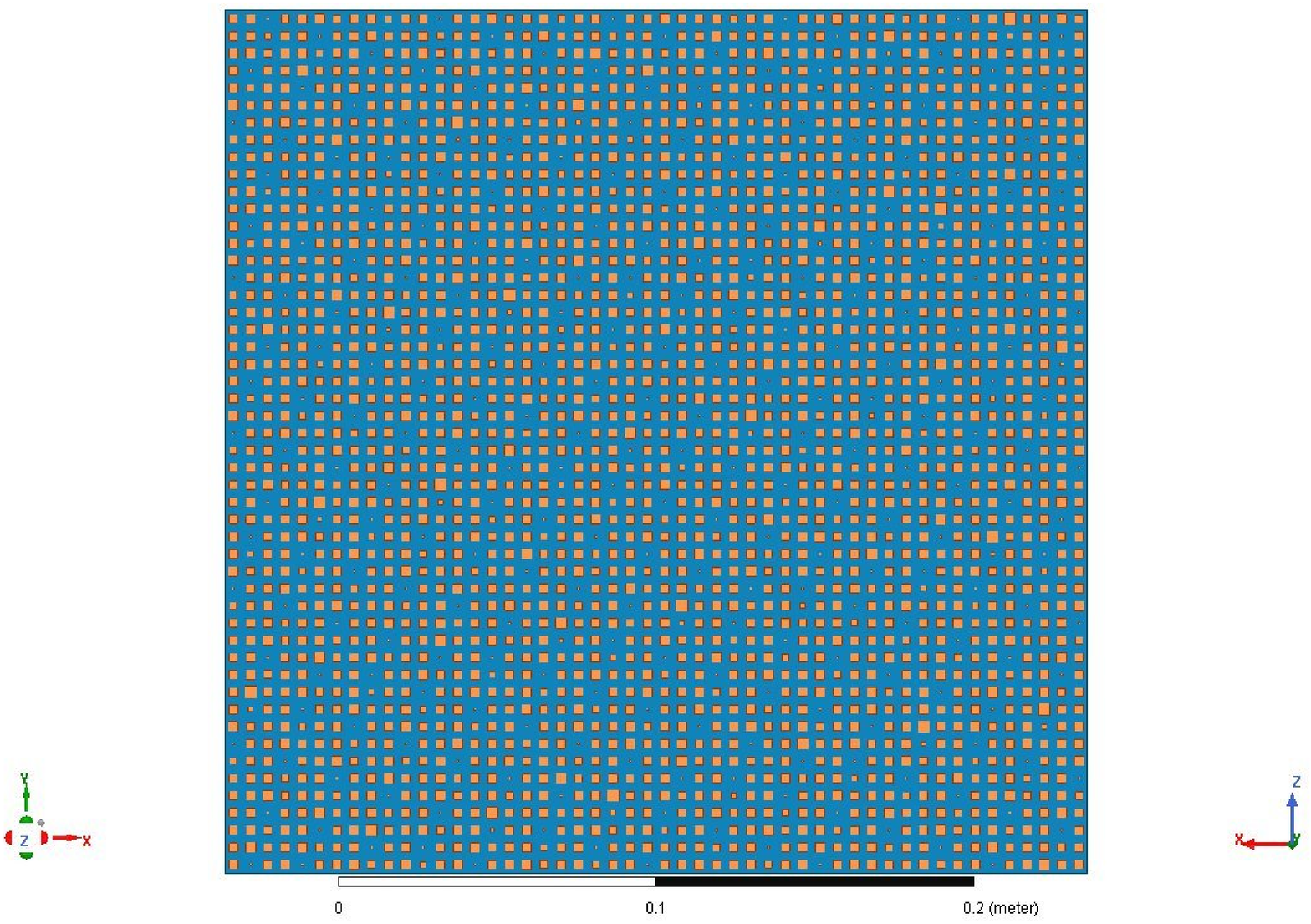}\tabularnewline
(\emph{a})\tabularnewline
\includegraphics[%
  width=0.50\columnwidth]{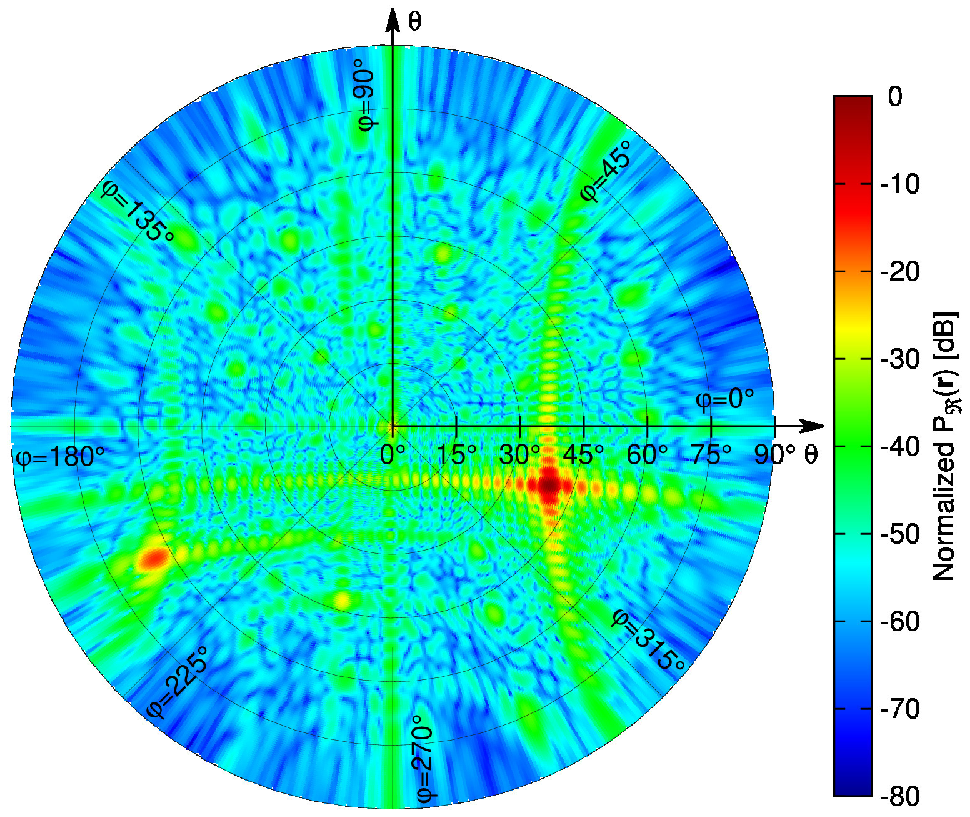}\tabularnewline
(\emph{b})\tabularnewline
\includegraphics[%
  width=0.50\columnwidth]{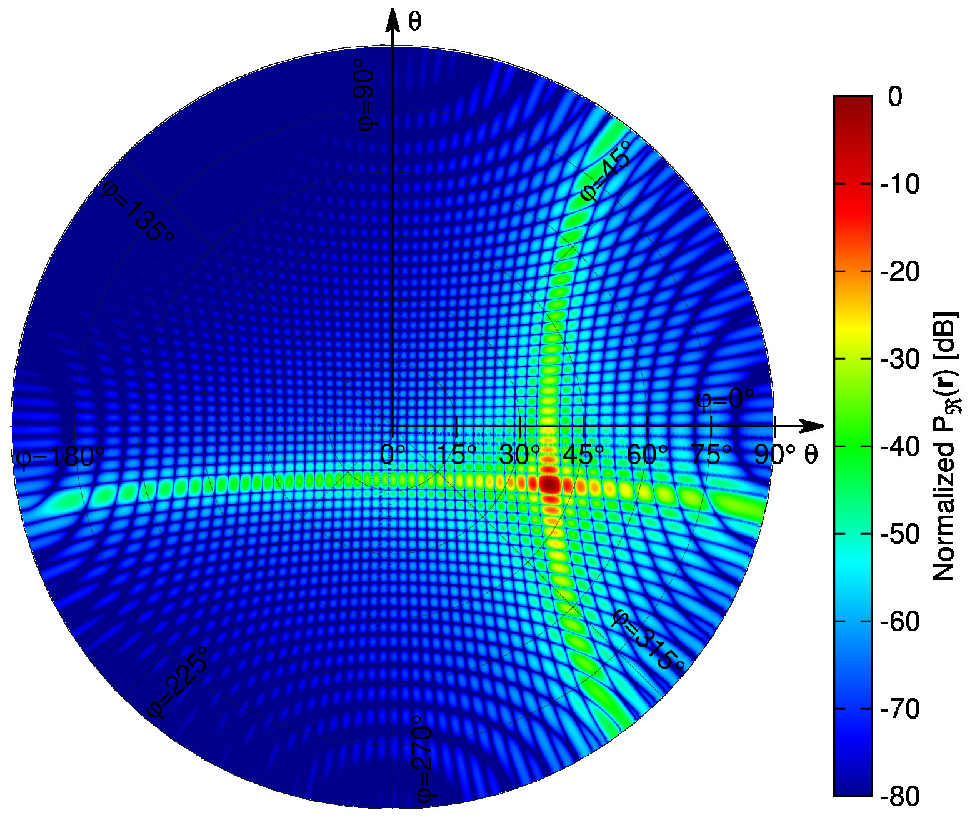}\tabularnewline
(\emph{c})\tabularnewline
\end{tabular}\end{center}

\begin{center}~\vfill\end{center}

\begin{center}\textbf{Fig. 3 - P. Rocca} \textbf{\emph{et al.}}\textbf{,}
\textbf{\emph{{}``}}On the Design of Modular Reflecting EM Skins
...''\end{center}

\newpage
\begin{center}~\vfill\end{center}

\begin{center}\begin{tabular}{c}
\includegraphics[%
  width=0.50\columnwidth]{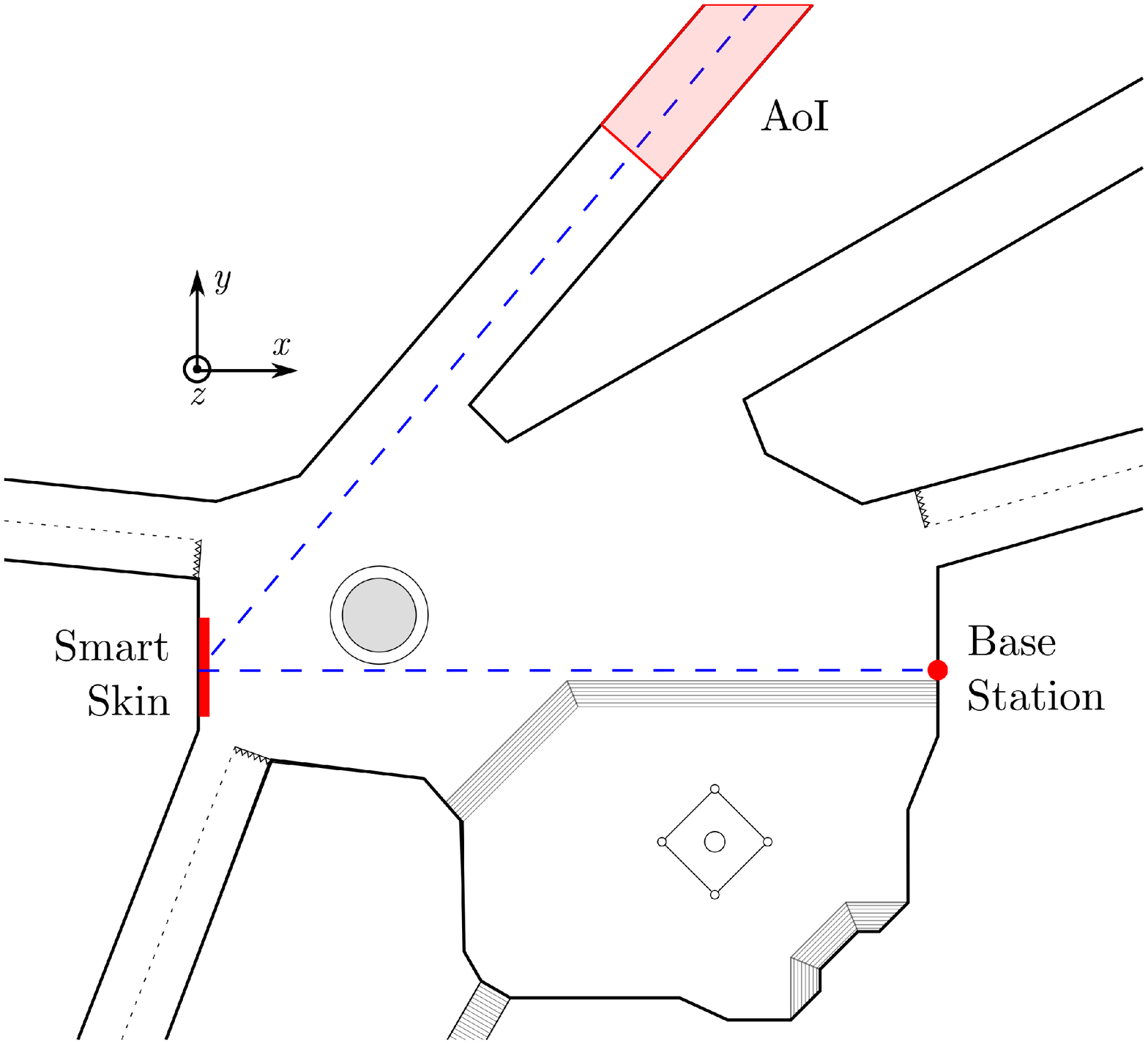}\tabularnewline
(\emph{a})\tabularnewline
\tabularnewline
\includegraphics[%
  width=0.40\columnwidth]{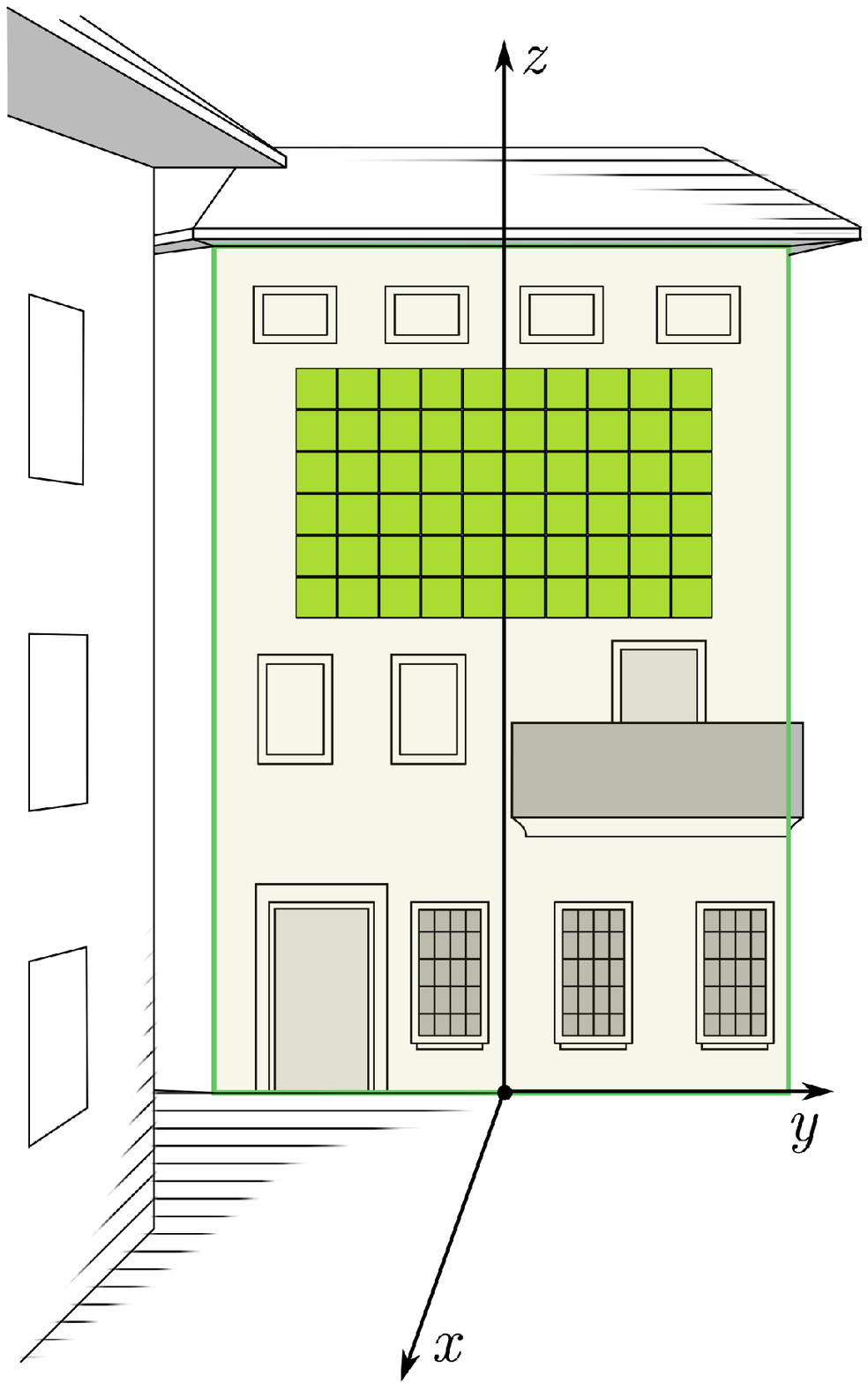}\tabularnewline
(\emph{b})\tabularnewline
\end{tabular}\end{center}

\begin{center}~\vfill\end{center}

\begin{center}\textbf{Fig. 4 - P. Rocca} \textbf{\emph{et al.}}\textbf{,}
\textbf{\emph{{}``}}On the Design of Modular Reflecting EM Skins
...''\end{center}

\newpage
\begin{center}~\vfill\end{center}

\begin{center}\begin{tabular}{c}
\includegraphics[%
  width=0.75\columnwidth]{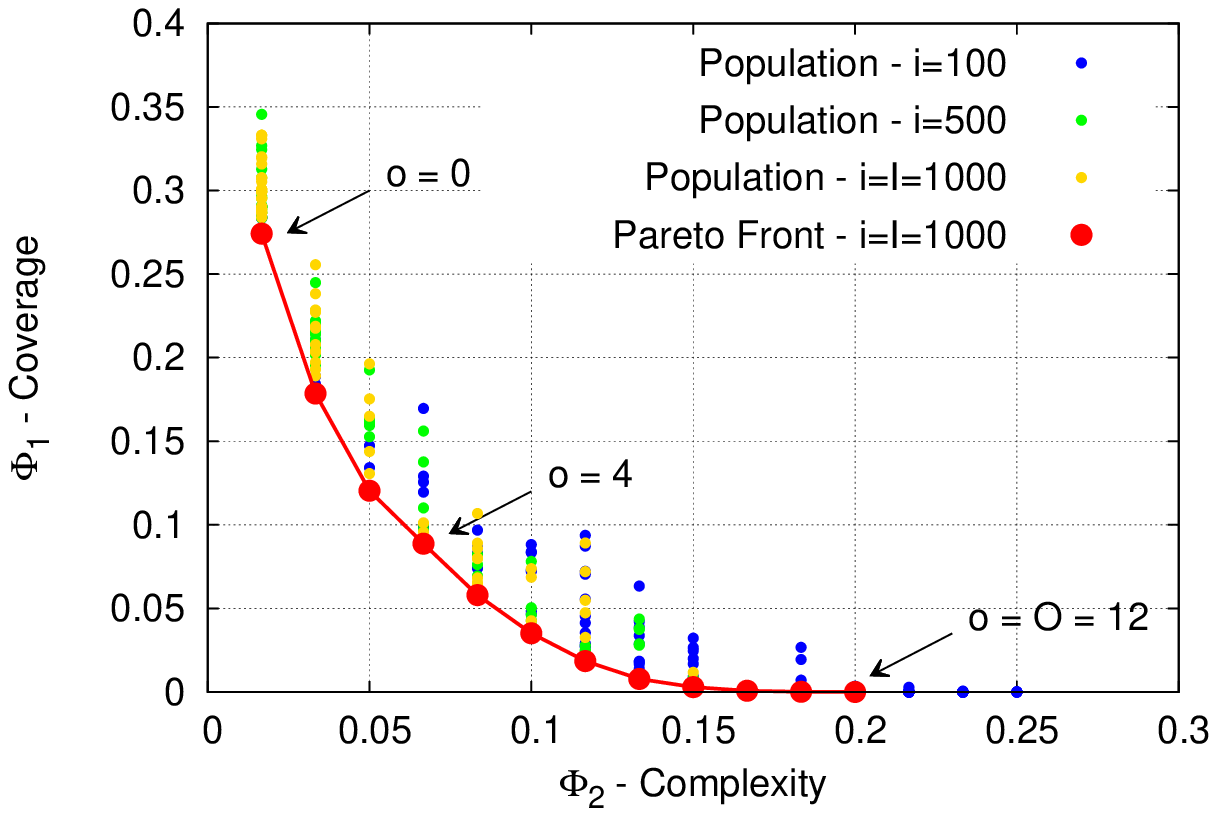}\tabularnewline
\end{tabular}\end{center}

\begin{center}~\vfill\end{center}

\begin{center}\textbf{Fig. 5 - P. Rocca} \textbf{\emph{et al.}}\textbf{,}
\textbf{\emph{{}``}}On the Design of Modular Reflecting EM Skins
...''\end{center}

\newpage
\begin{center}~\vfill\end{center}

\begin{center}\begin{tabular}{c}
\includegraphics[%
  width=0.35\columnwidth]{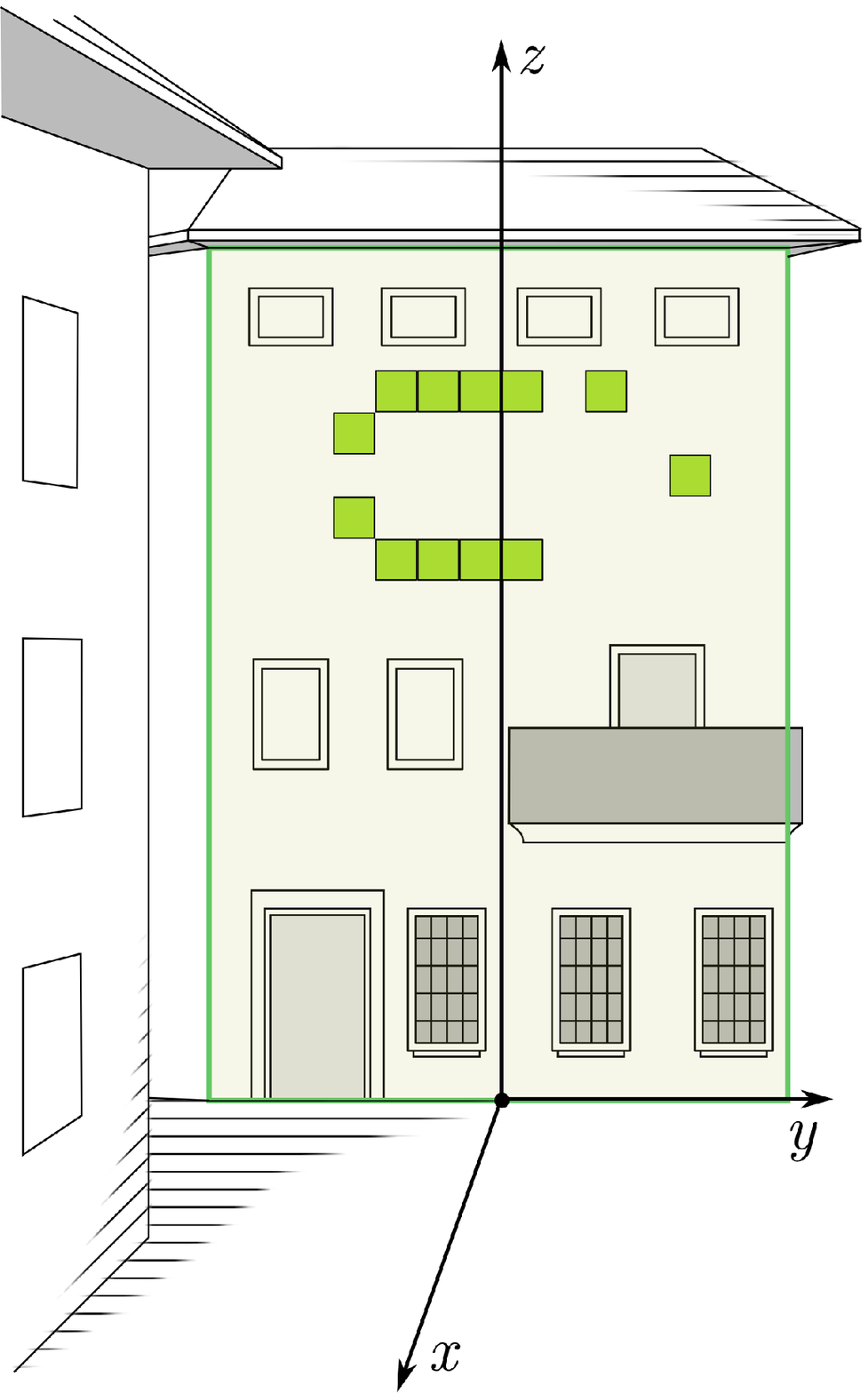}\tabularnewline
(\emph{a})\tabularnewline
\includegraphics[%
  width=0.50\columnwidth]{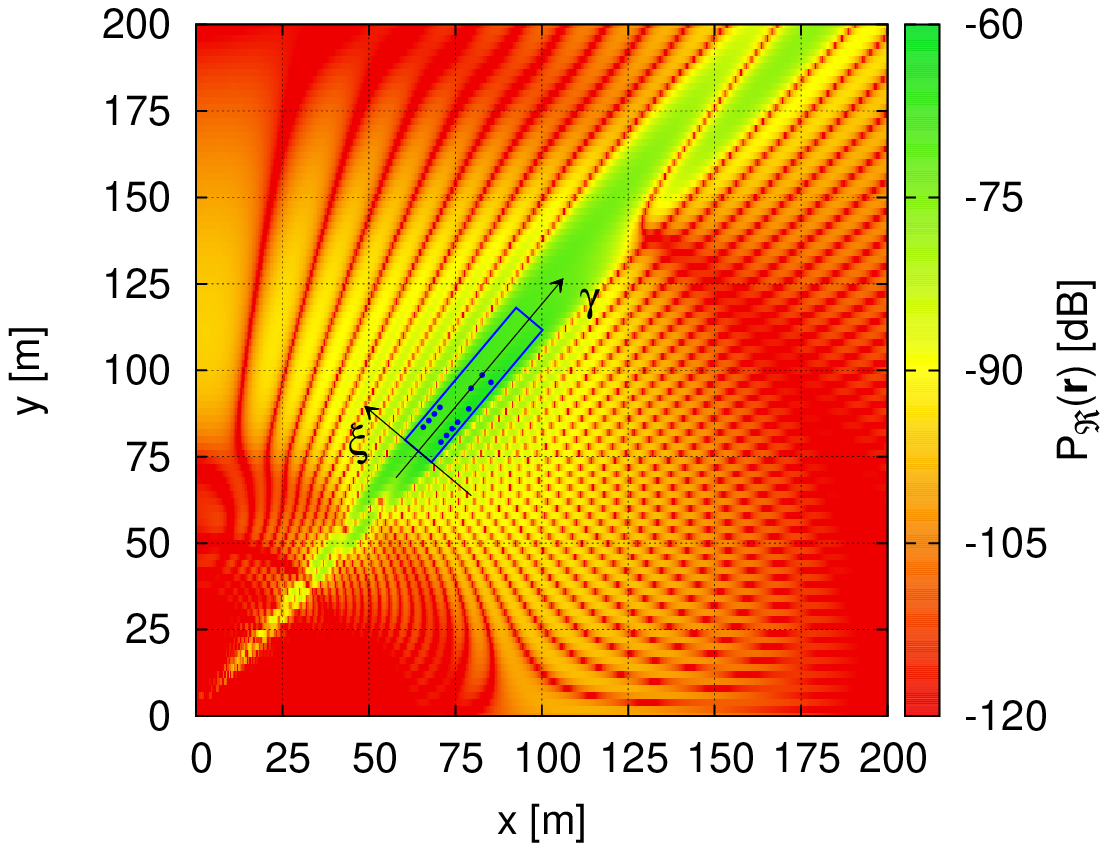}\tabularnewline
(\emph{b})\tabularnewline
\includegraphics[%
  width=0.50\columnwidth]{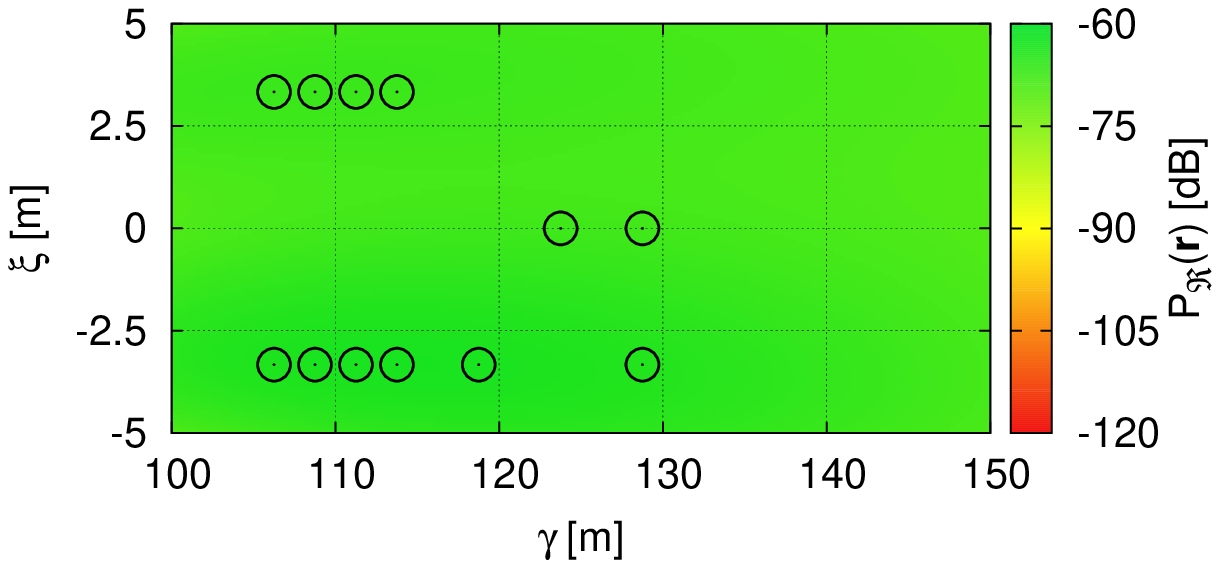}\tabularnewline
(\emph{c})\tabularnewline
\end{tabular}\end{center}

\begin{center}~\vfill\end{center}

\begin{center}\textbf{Fig. 6 - P. Rocca} \textbf{\emph{et al.}}\textbf{,}
\textbf{\emph{{}``}}On the Design of Modular Reflecting EM Skins
...''\end{center}

\newpage
\begin{center}\begin{tabular}{cc}
\includegraphics[%
  width=0.30\columnwidth]{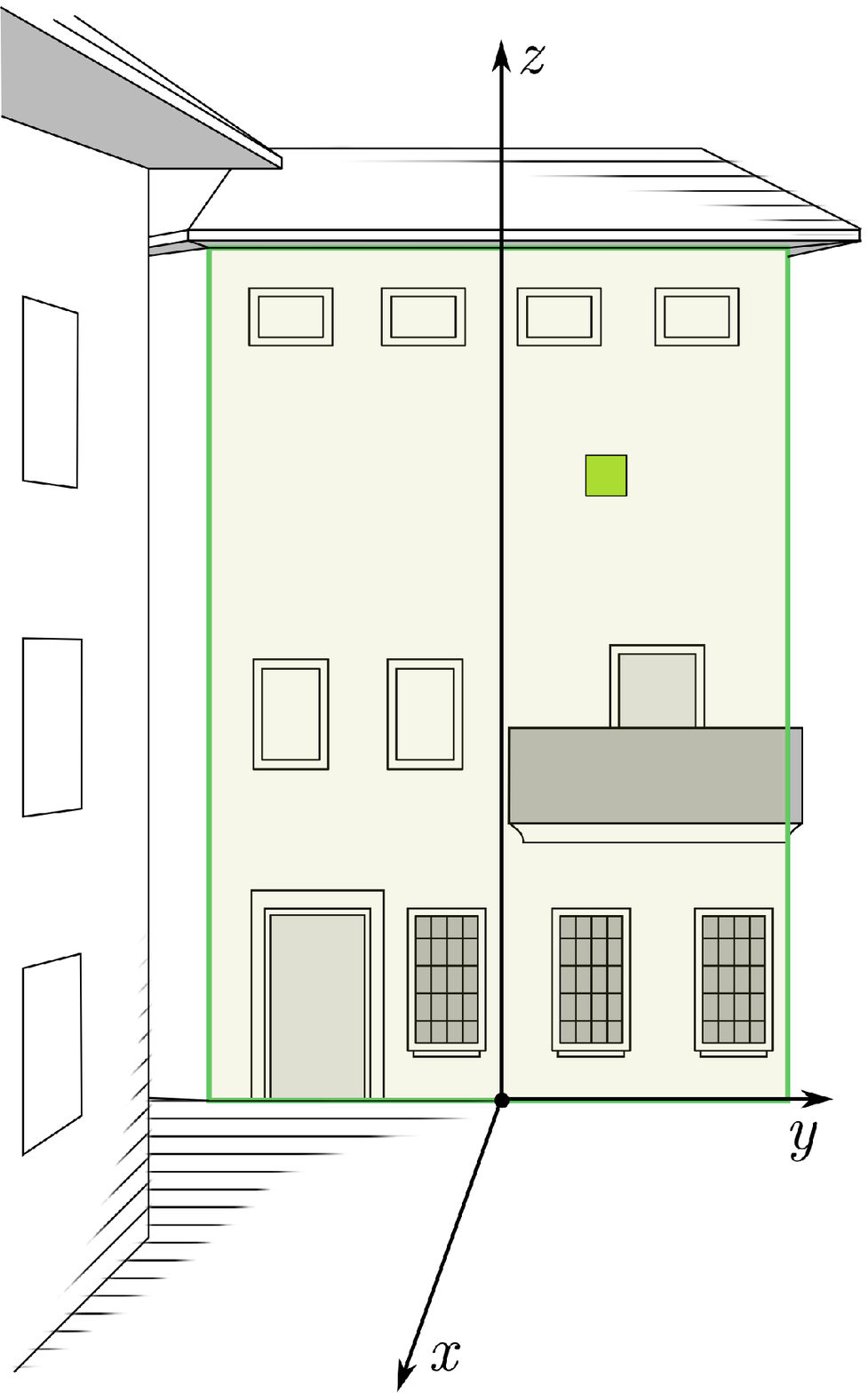}&
\includegraphics[%
  width=0.30\columnwidth]{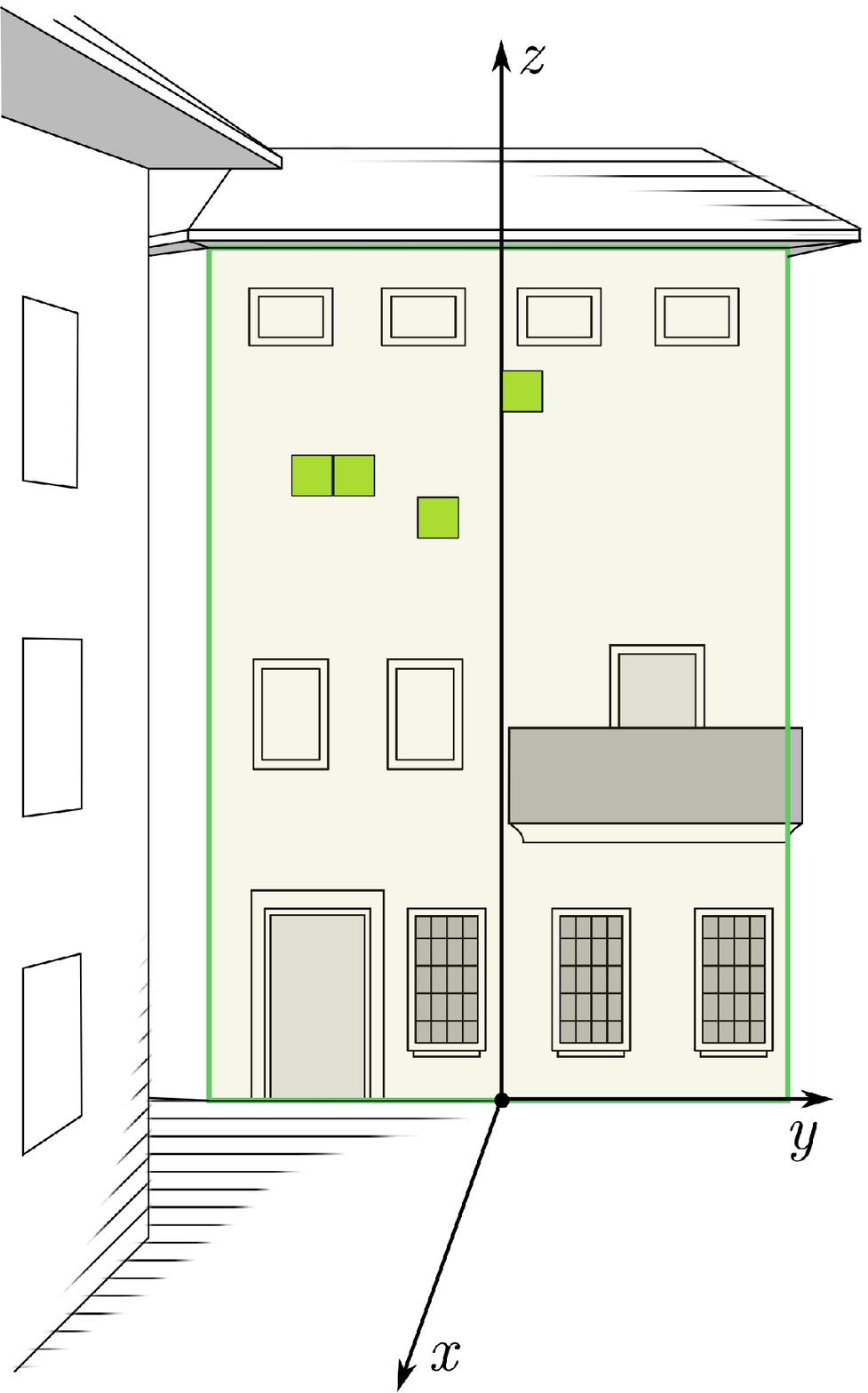}\tabularnewline
(\emph{a})&
(\emph{b})\tabularnewline
\includegraphics[%
  width=0.45\columnwidth]{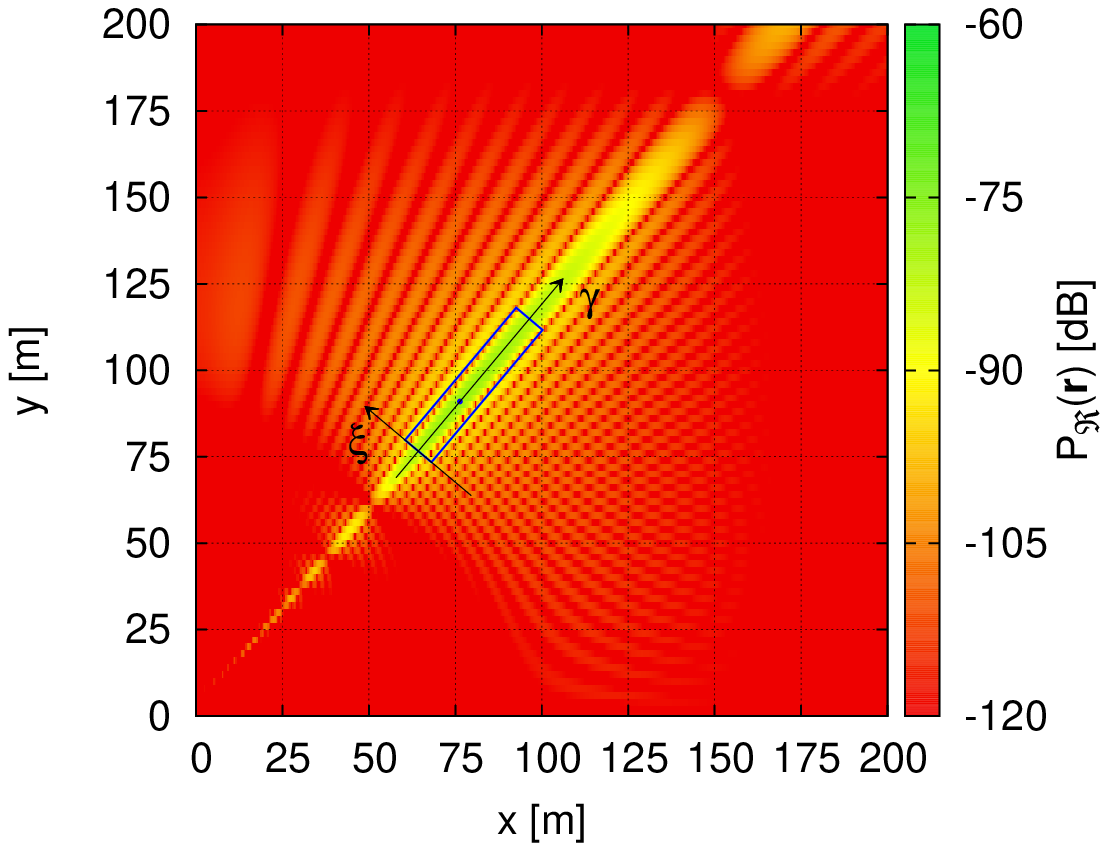}&
\includegraphics[%
  width=0.45\columnwidth]{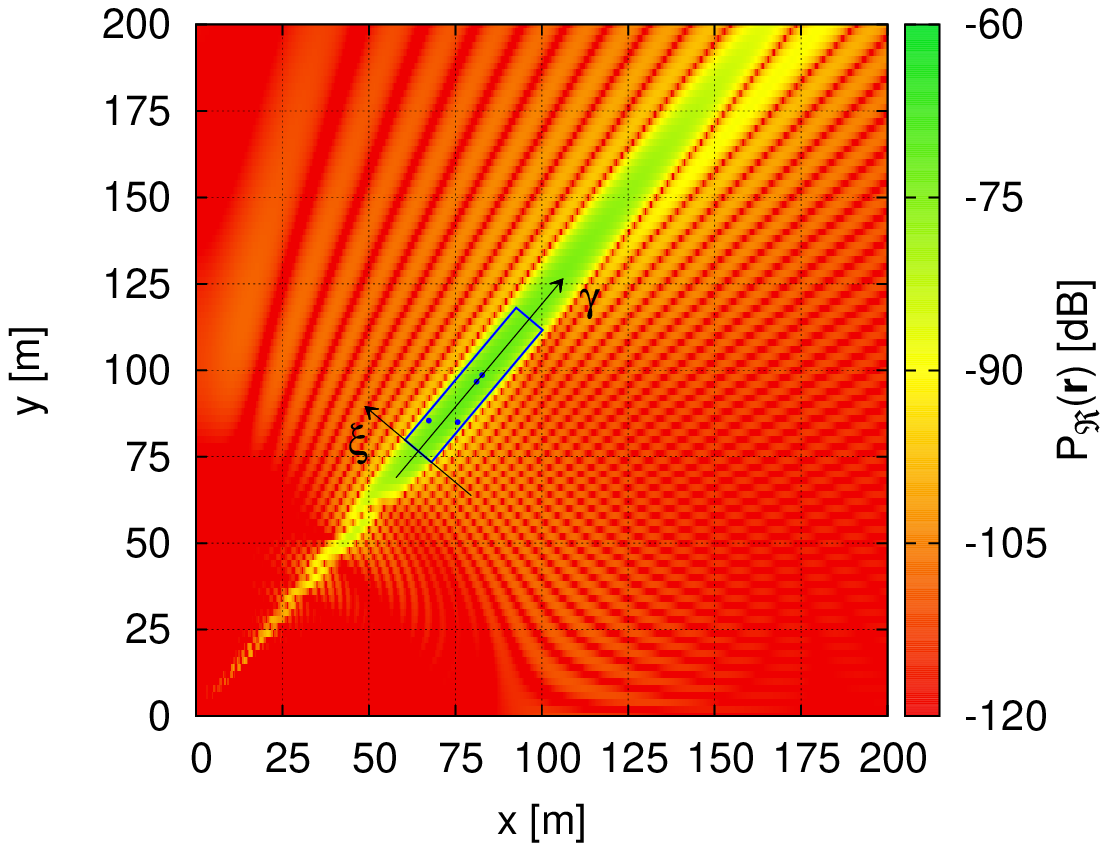}\tabularnewline
(\emph{c})&
(\emph{d})\tabularnewline
\includegraphics[%
  width=0.40\columnwidth]{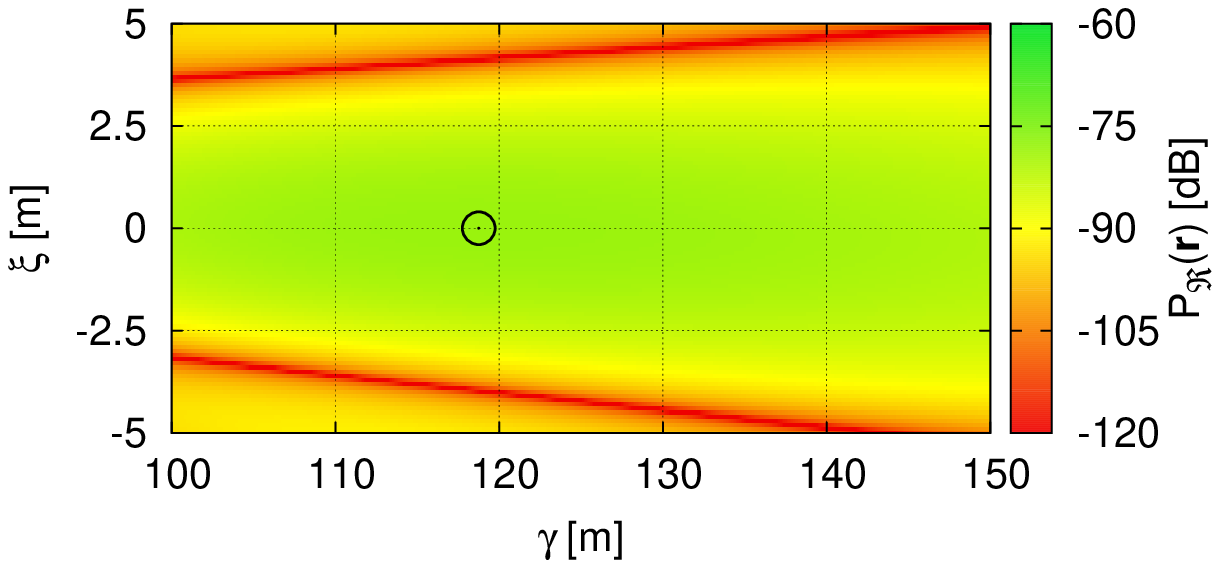}&
\includegraphics[%
  width=0.40\columnwidth]{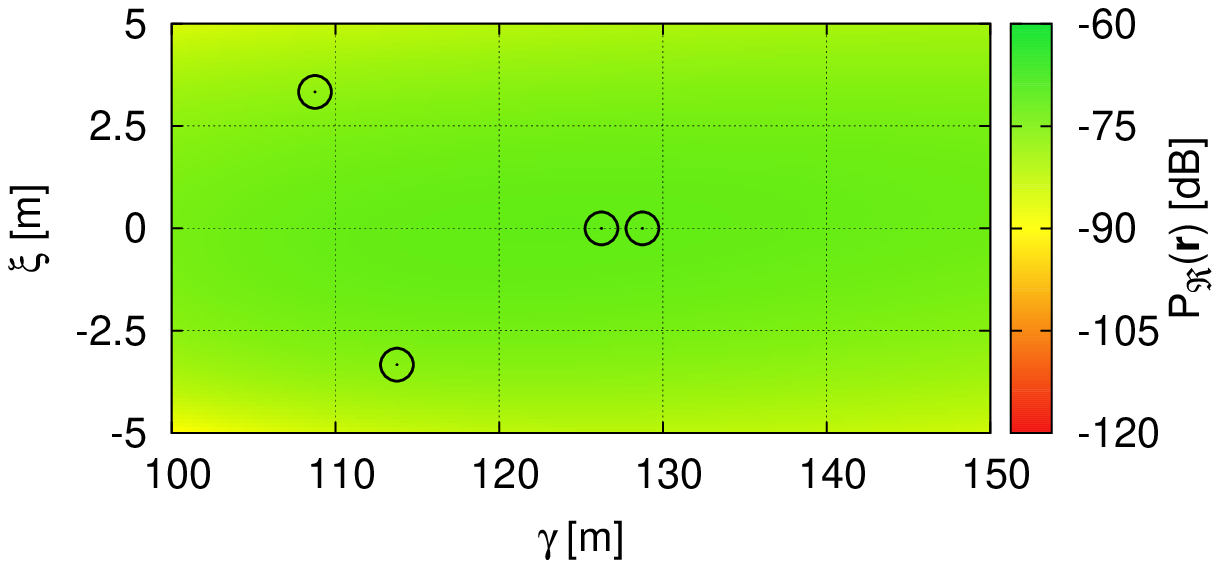}\tabularnewline
(\emph{e})&
(\emph{f})\tabularnewline
\includegraphics[%
  width=0.35\columnwidth]{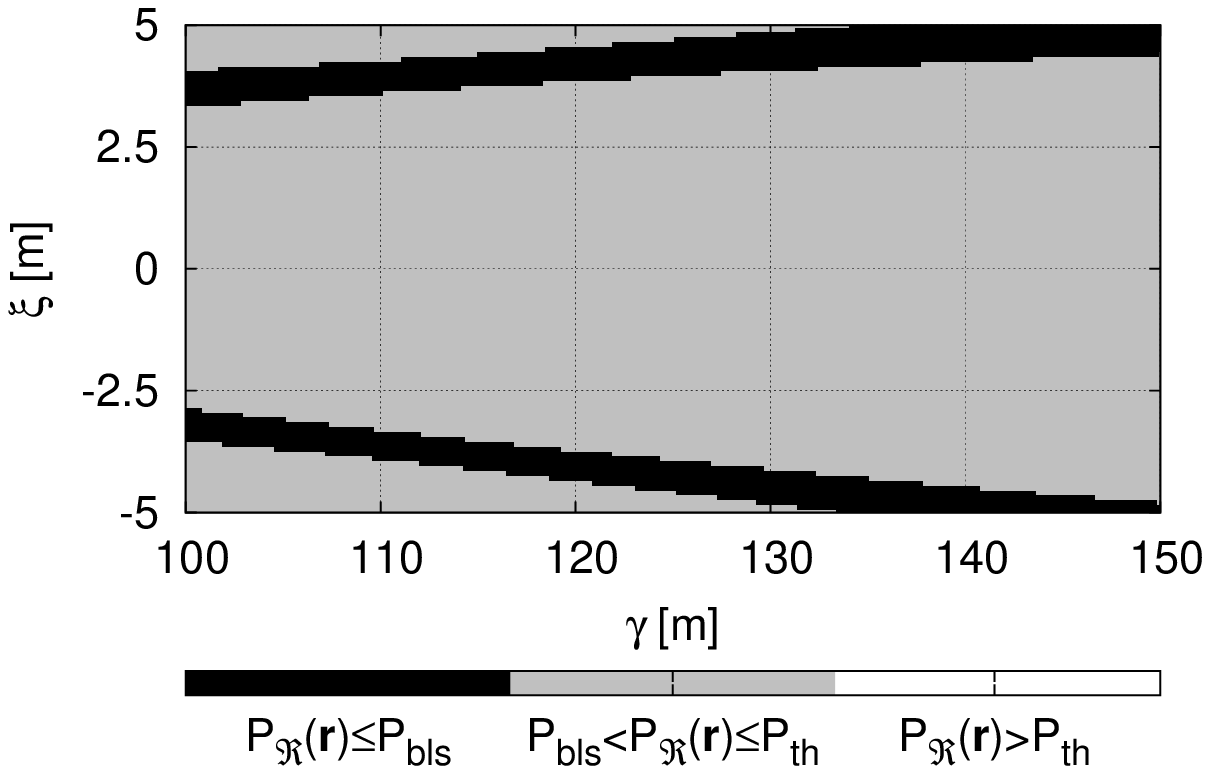}&
\includegraphics[%
  width=0.35\columnwidth]{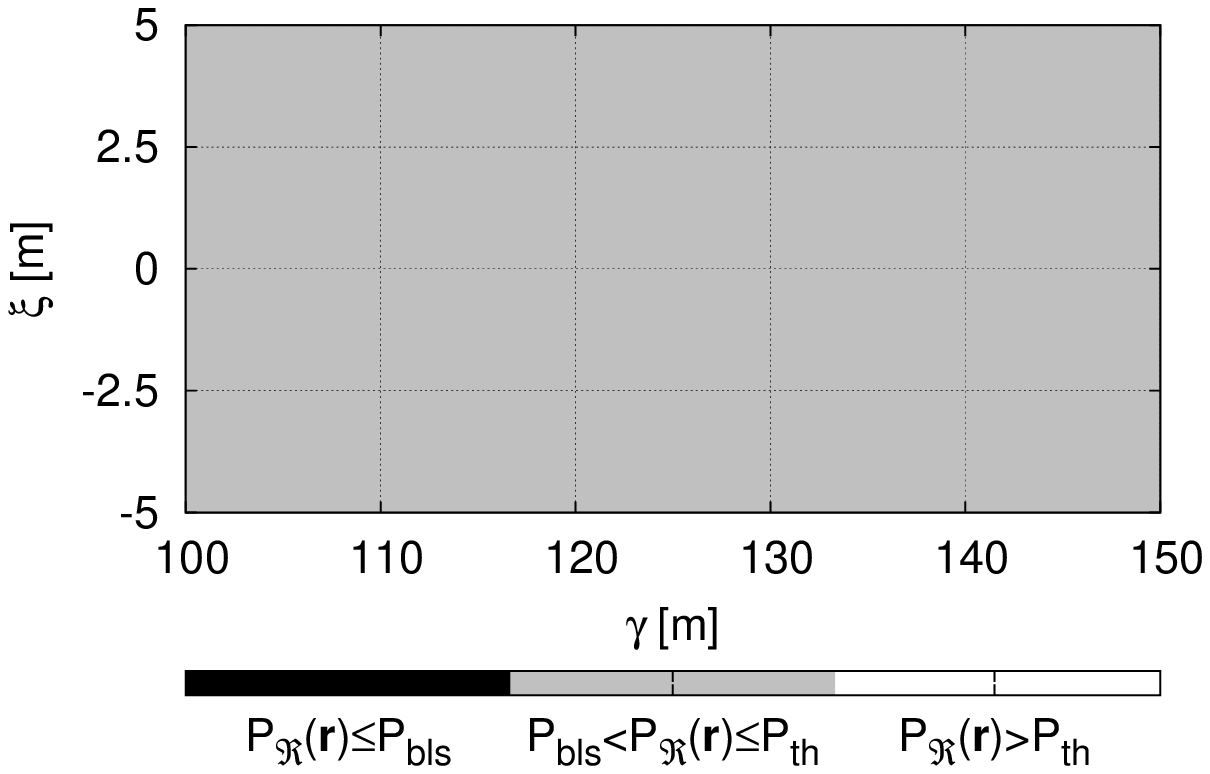}\tabularnewline
(\emph{g})&
(\emph{h})\tabularnewline
\end{tabular}\end{center}

\begin{center}\textbf{Fig. 7 - P. Rocca} \textbf{\emph{et al.}}\textbf{,}
\textbf{\emph{{}``}}On the Design of Modular Reflecting EM Skins
...''\end{center}

\newpage
\begin{center}~\vfill\end{center}

\begin{center}\begin{tabular}{c}
\includegraphics[%
  width=0.60\columnwidth]{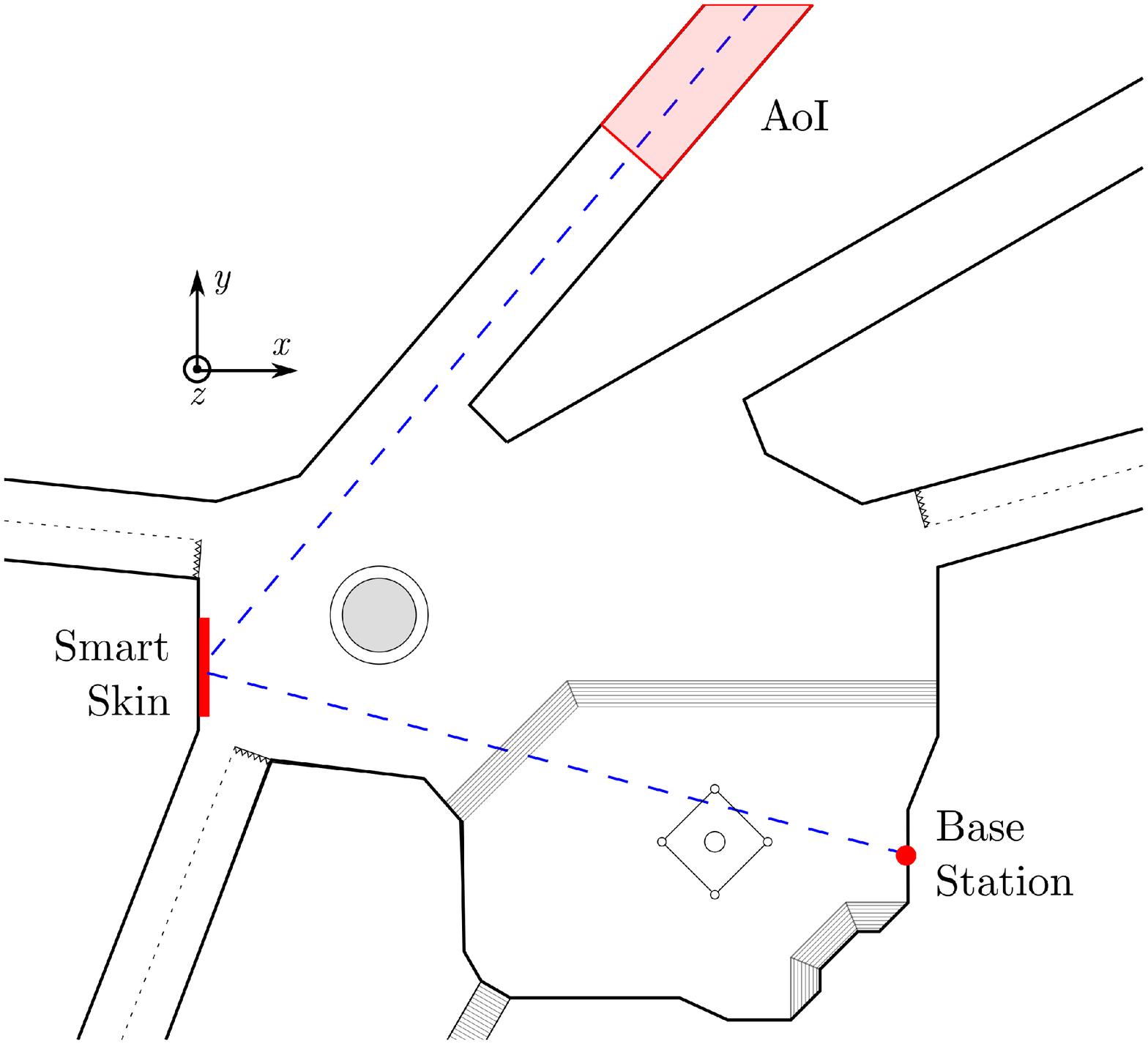}\tabularnewline
(\emph{a})\tabularnewline
\tabularnewline
\includegraphics[%
  width=0.75\columnwidth]{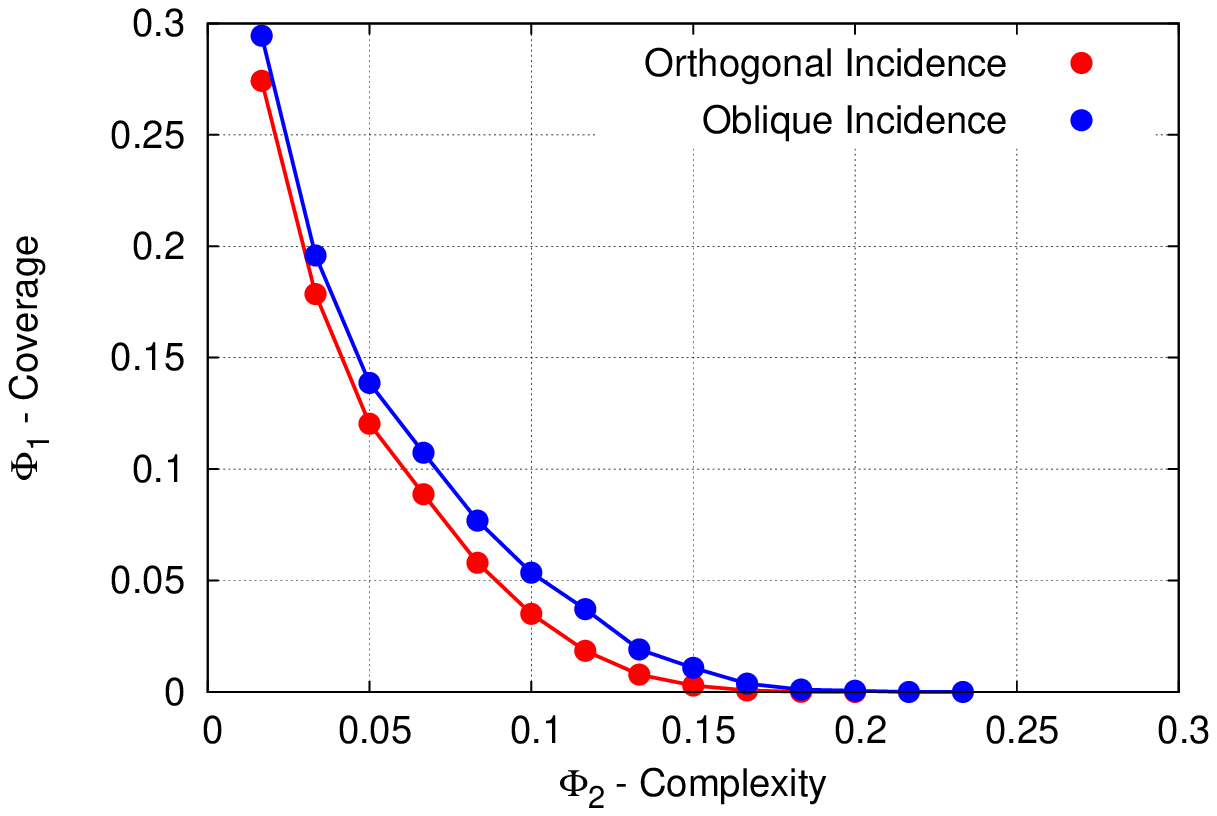}\tabularnewline
(\emph{b})\tabularnewline
\end{tabular}\end{center}

\begin{center}~\vfill\end{center}

\begin{center}\textbf{Fig. 8 - P. Rocca} \textbf{\emph{et al.}}\textbf{,}
\textbf{\emph{{}``}}On the Design of Modular Reflecting EM Skins
...''\end{center}

\newpage
\begin{center}~\vfill\end{center}

\begin{center}\begin{tabular}{c}
\includegraphics[%
  width=0.35\columnwidth]{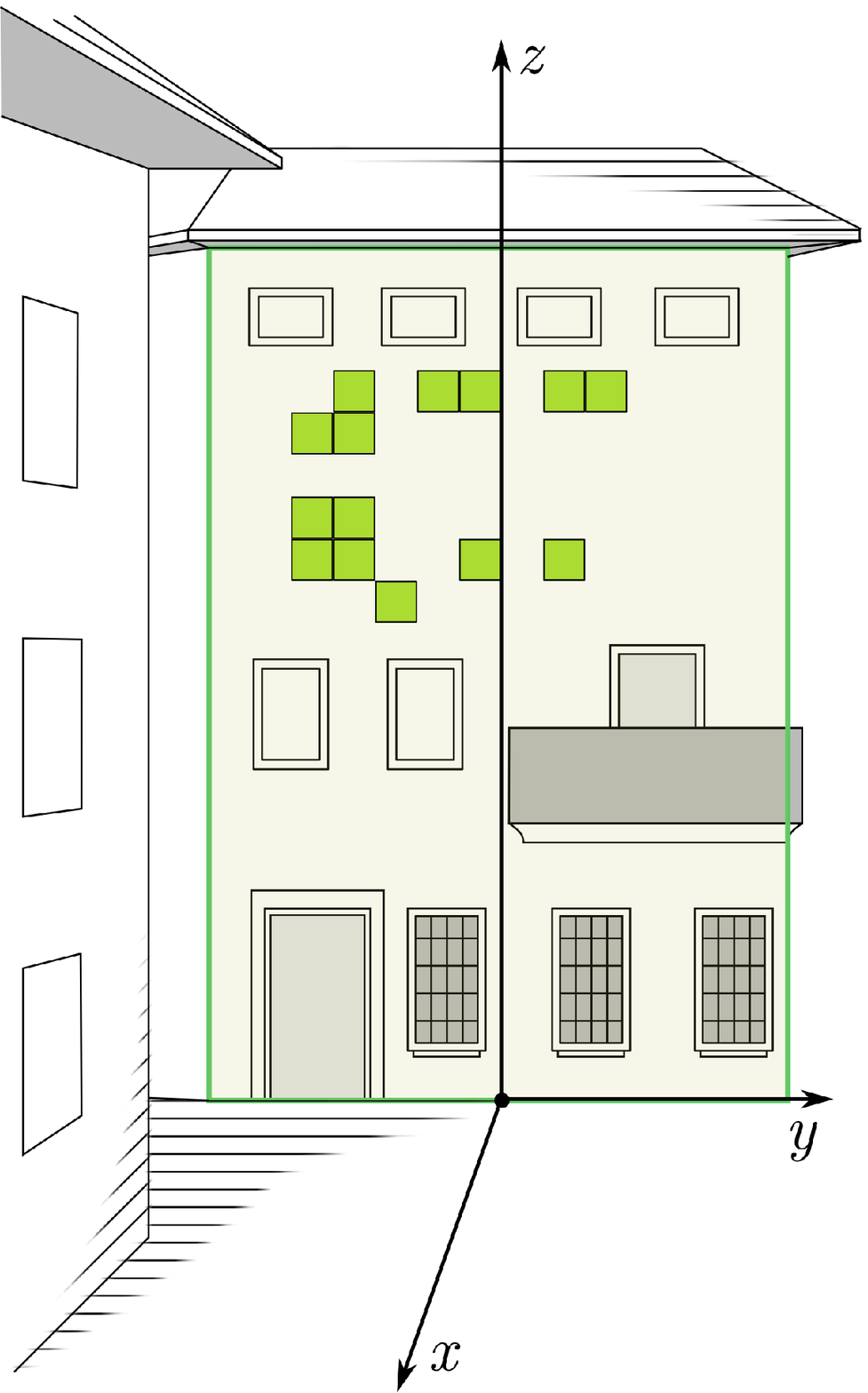}\tabularnewline
(\emph{a})\tabularnewline
\includegraphics[%
  width=0.50\columnwidth]{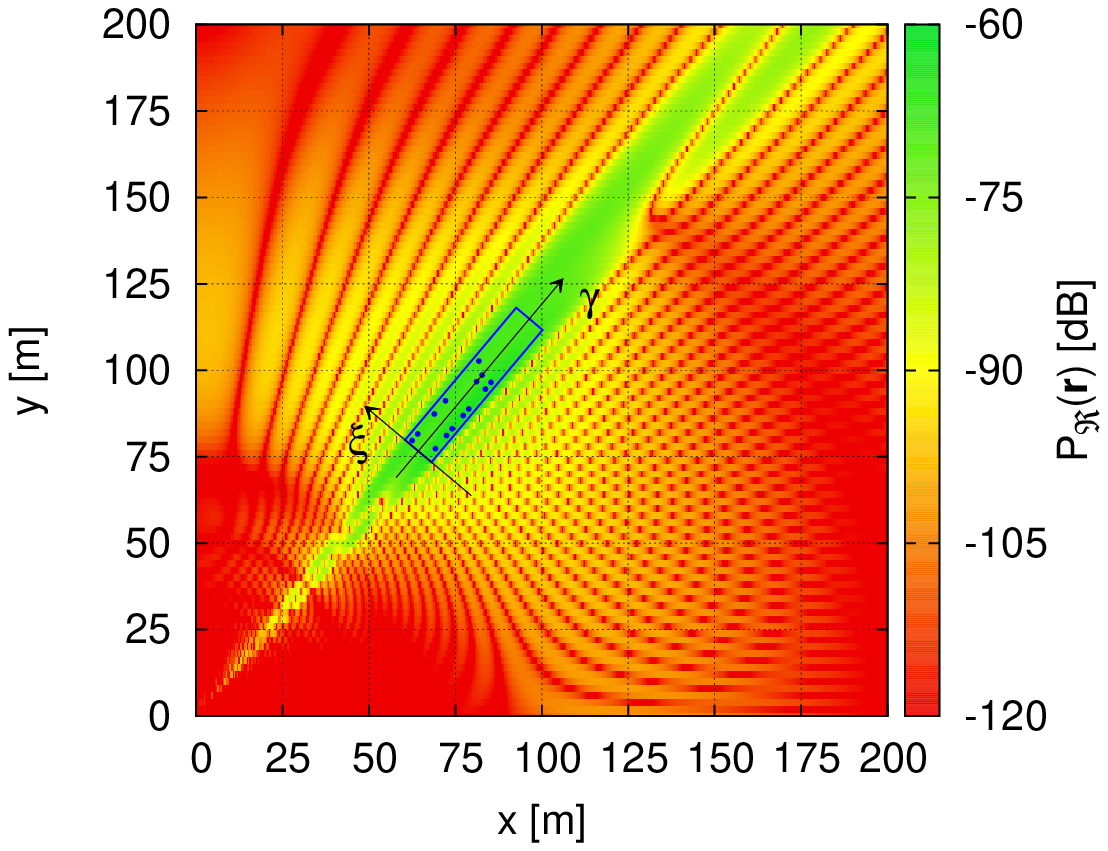}\tabularnewline
(\emph{b})\tabularnewline
\includegraphics[%
  width=0.50\columnwidth]{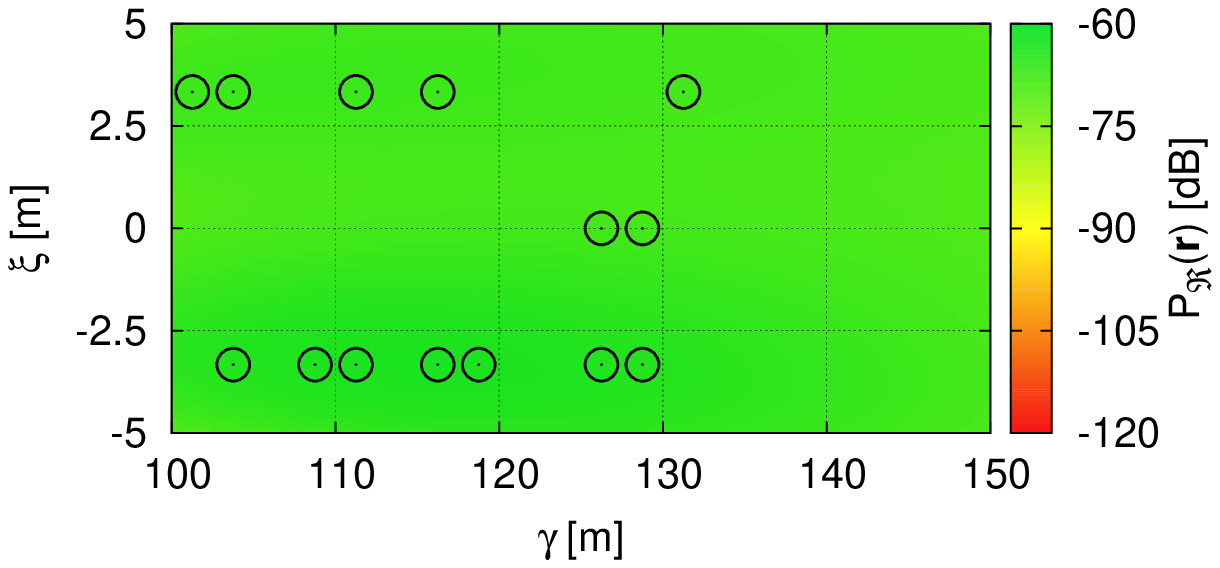}\tabularnewline
(\emph{c})\tabularnewline
\end{tabular}\end{center}

\begin{center}~\vfill\end{center}

\begin{center}\textbf{Fig. 9 - P. Rocca} \textbf{\emph{et al.}}\textbf{,}
\textbf{\emph{{}``}}On the Design of Modular Reflecting EM Skins
...''\end{center}

\newpage
\begin{center}~\vfill\end{center}

\begin{center}\begin{tabular}{c}
\includegraphics[%
  width=0.40\columnwidth]{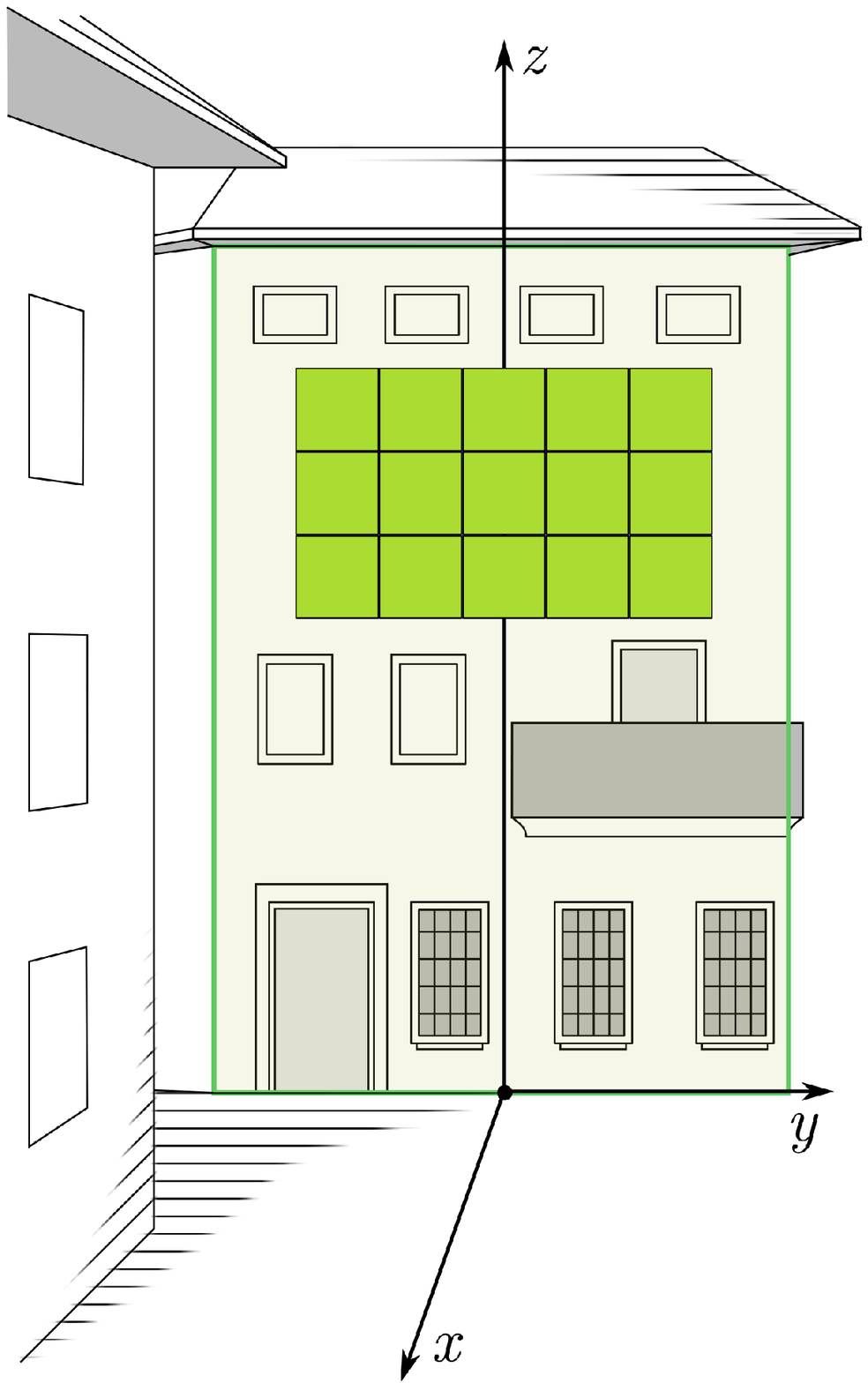}\tabularnewline
(\emph{a})\tabularnewline
\tabularnewline
\includegraphics[%
  width=0.40\columnwidth]{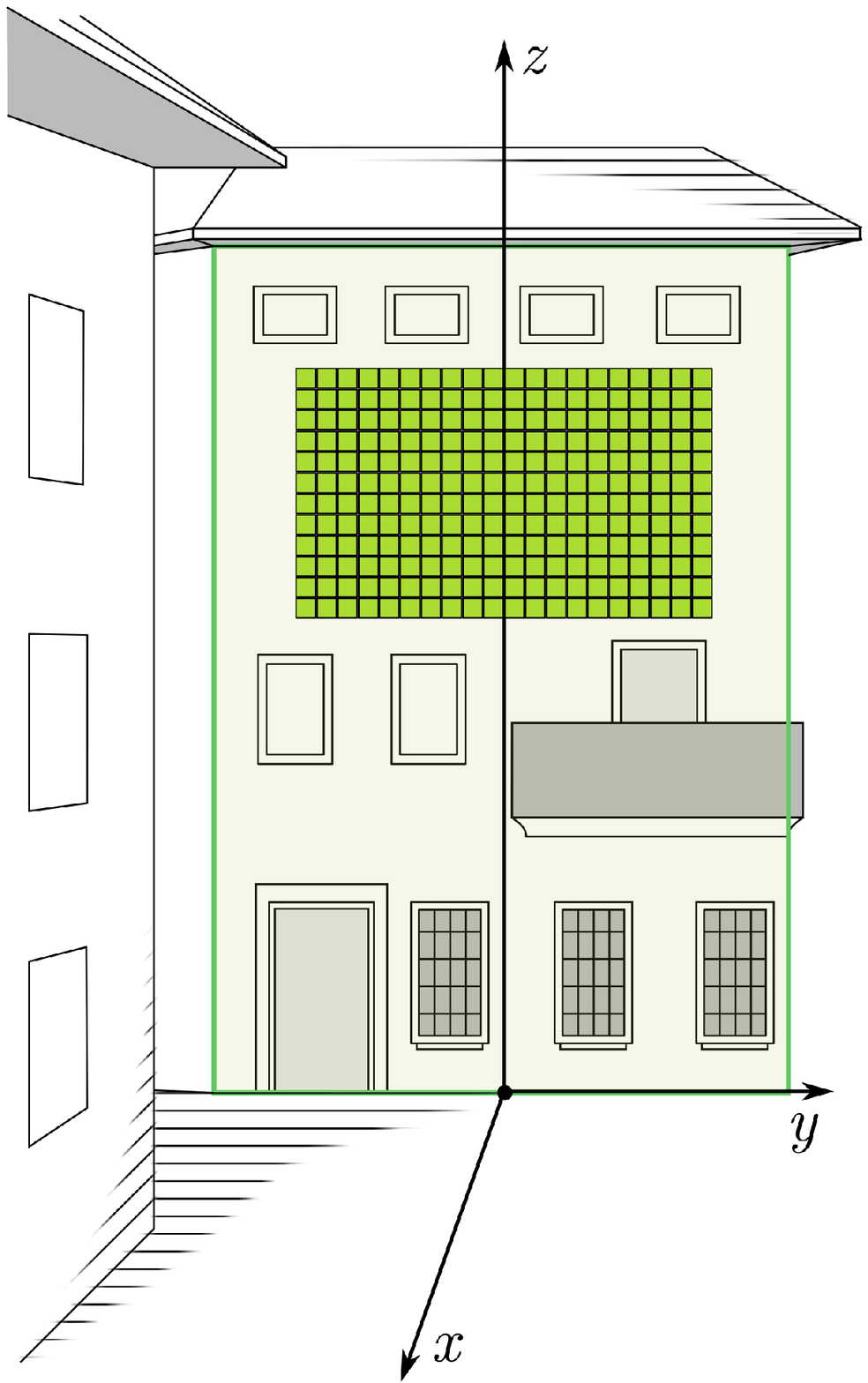}\tabularnewline
(\emph{b})\tabularnewline
\end{tabular}\end{center}

\begin{center}~\vfill\end{center}

\begin{center}\textbf{Fig. 10 - P. Rocca} \textbf{\emph{et al.}}\textbf{,}
\textbf{\emph{{}``}}On the Design of Modular Reflecting EM Skins
...''\end{center}

\newpage
\begin{center}~\vfill \end{center}

\begin{center}\begin{tabular}{c}
\includegraphics[%
  width=0.75\columnwidth]{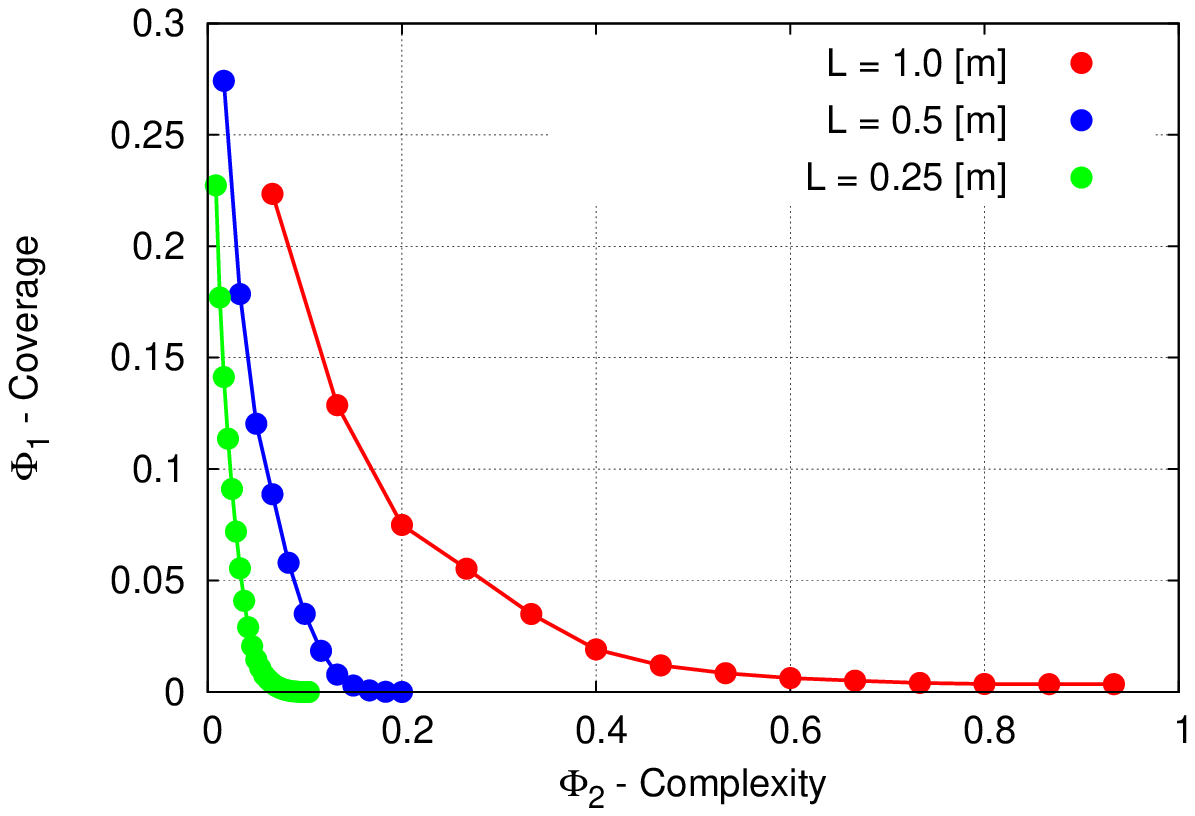}\tabularnewline
(\emph{a})\tabularnewline
\tabularnewline
\includegraphics[%
  width=0.75\columnwidth]{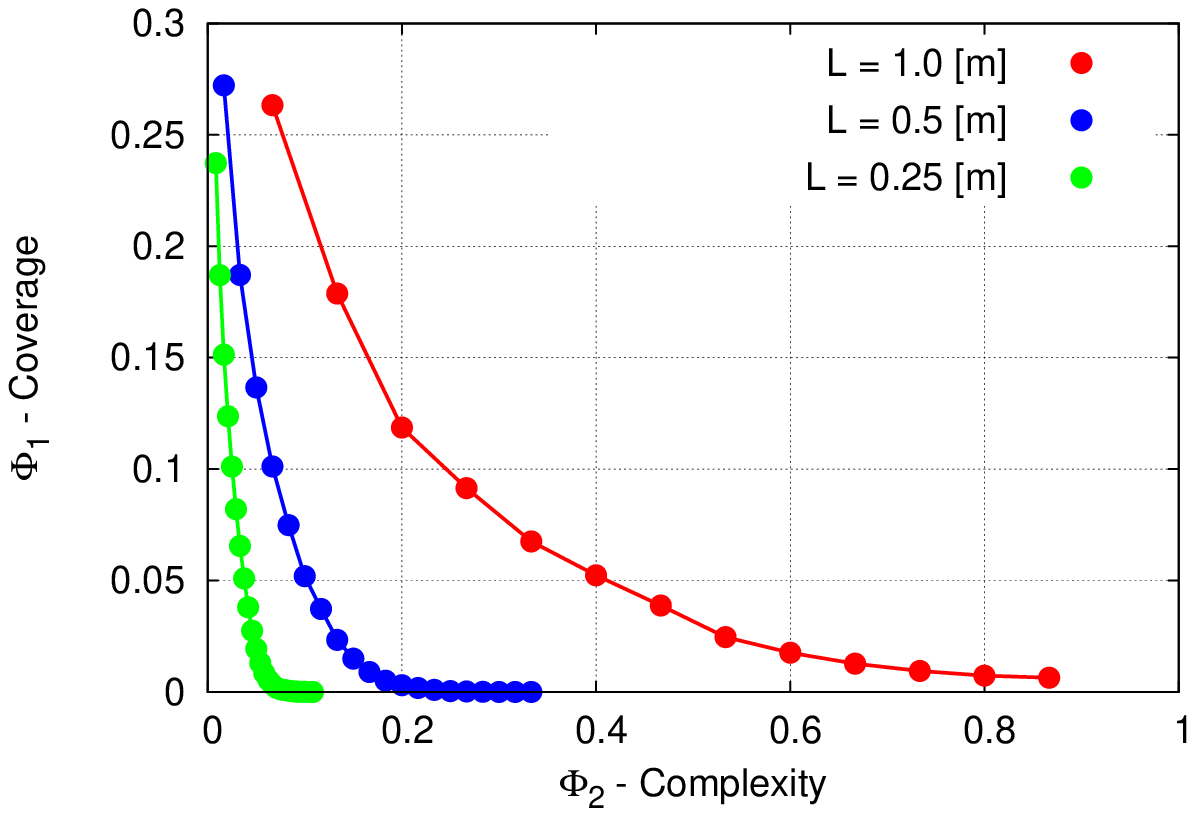}\tabularnewline
(\emph{b})\tabularnewline
\end{tabular}\end{center}

\begin{center}~\vfill\end{center}

\begin{center}\textbf{Fig. 11 - P. Rocca} \textbf{\emph{et al.}}\textbf{,}
\textbf{\emph{{}``}}On the Design of Modular Reflecting EM Skins
...''\end{center}

\newpage
\begin{center}~\vfill\end{center}

\begin{center}\begin{tabular}{ccc}
\includegraphics[%
  width=0.30\columnwidth]{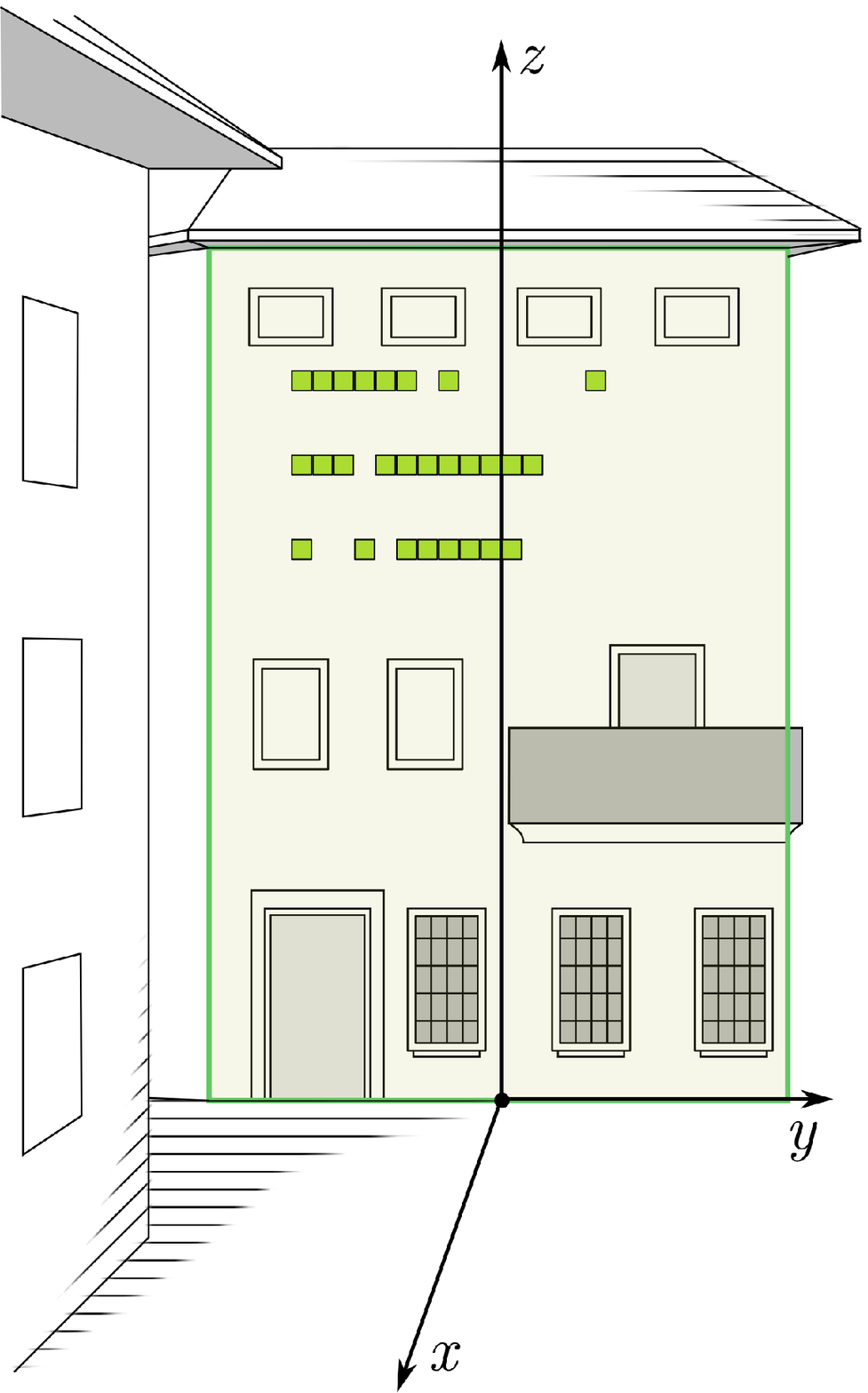}&
\includegraphics[%
  width=0.30\columnwidth]{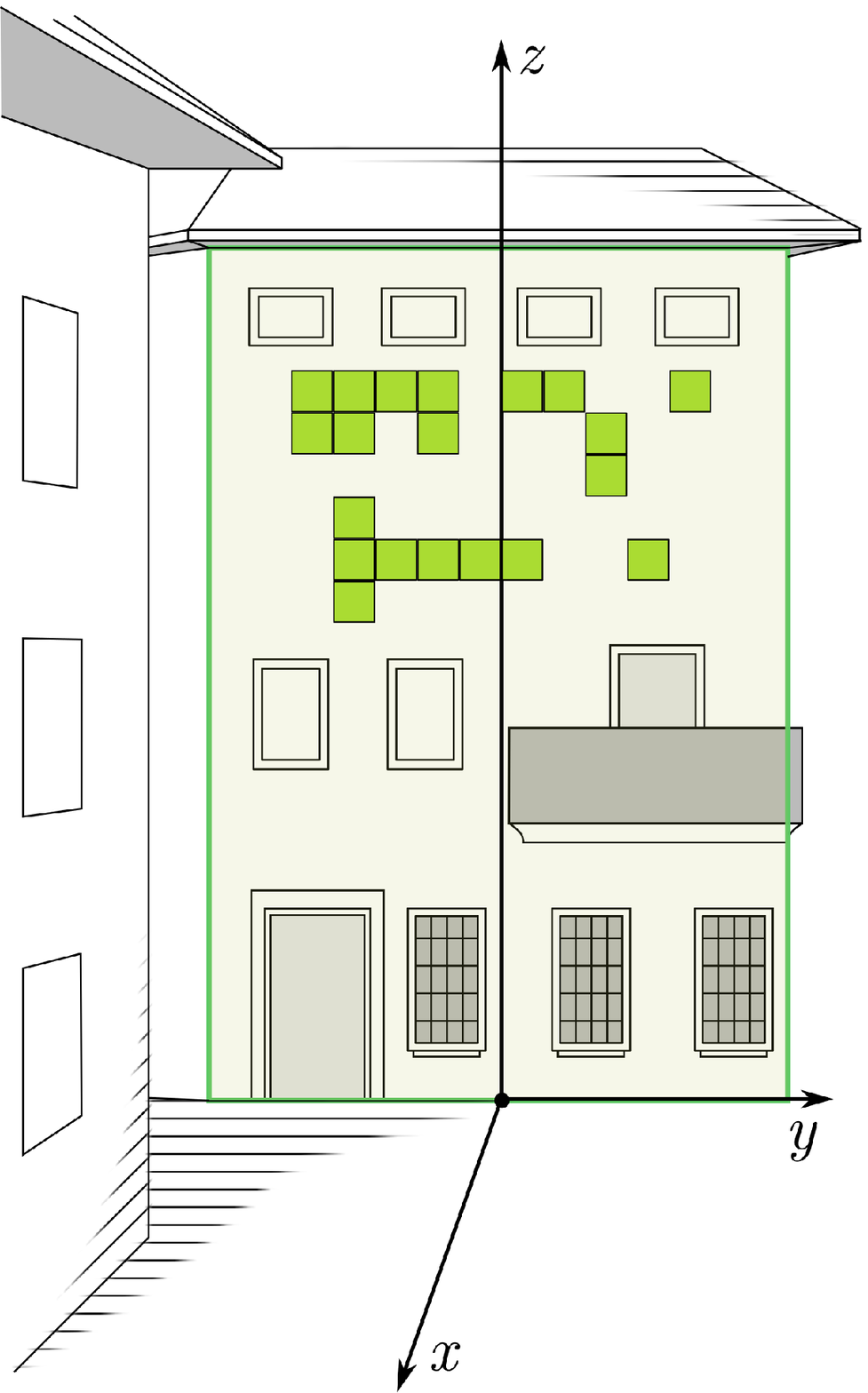}&
\includegraphics[%
  width=0.30\columnwidth]{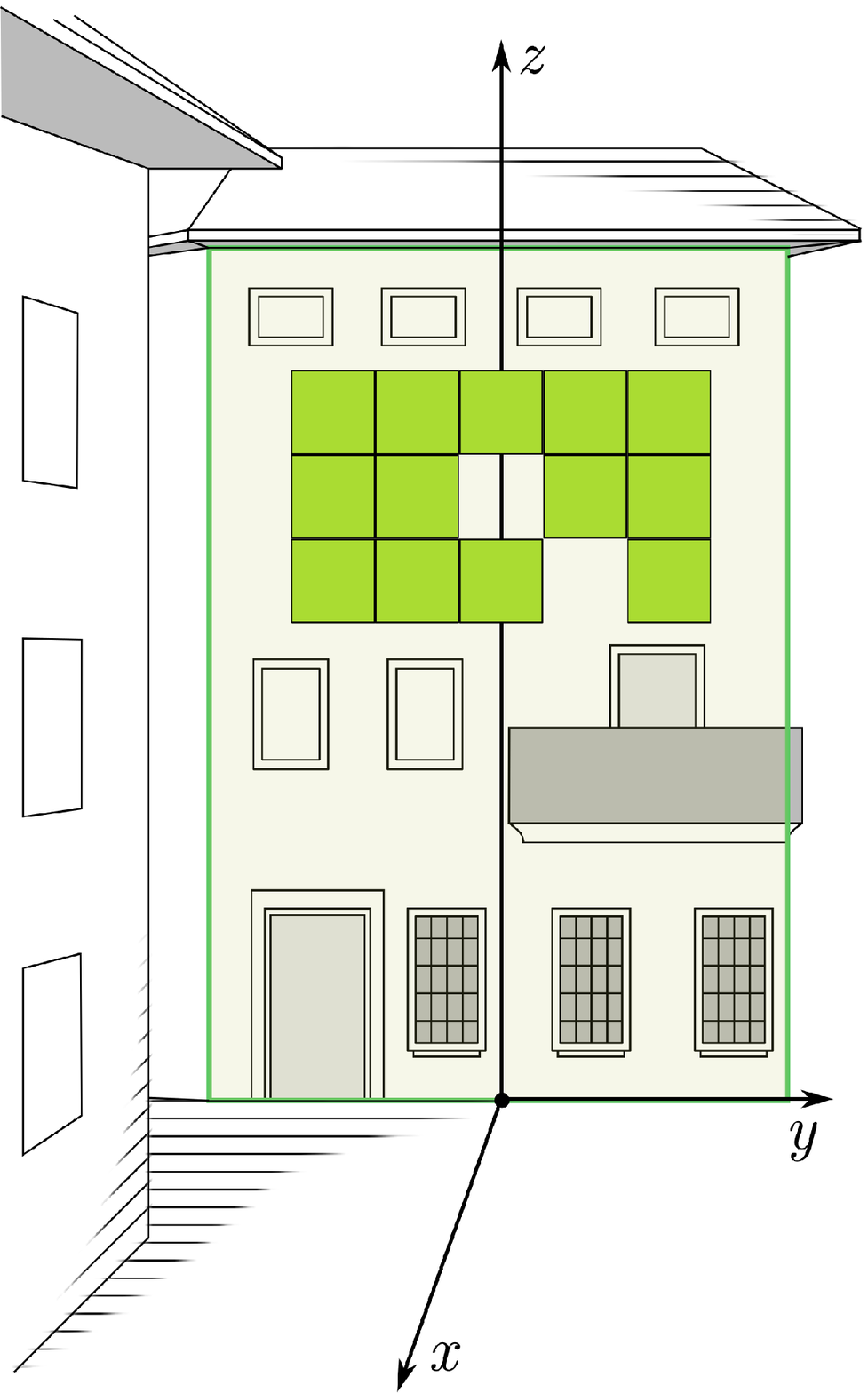}\tabularnewline
(\emph{a})&
(\emph{b})&
(\emph{c})\tabularnewline
&
&
\tabularnewline
\includegraphics[%
  width=0.35\columnwidth]{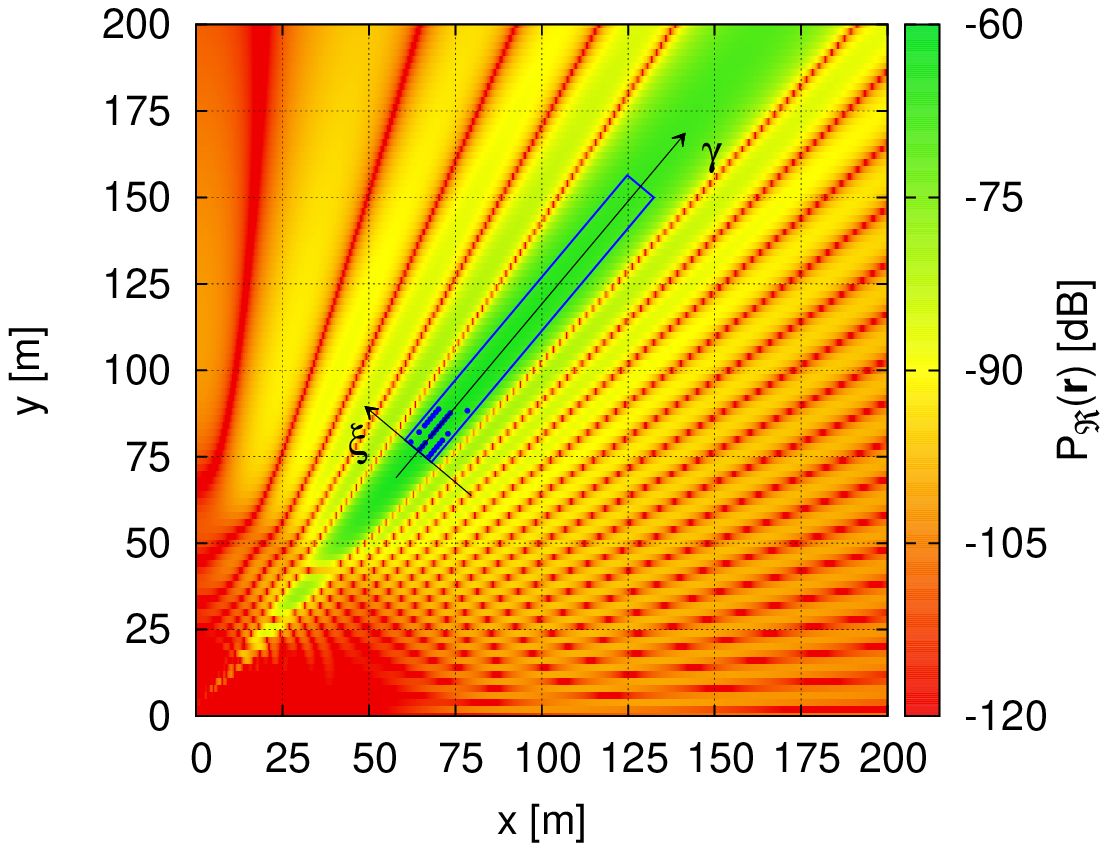}&
\includegraphics[%
  width=0.35\columnwidth]{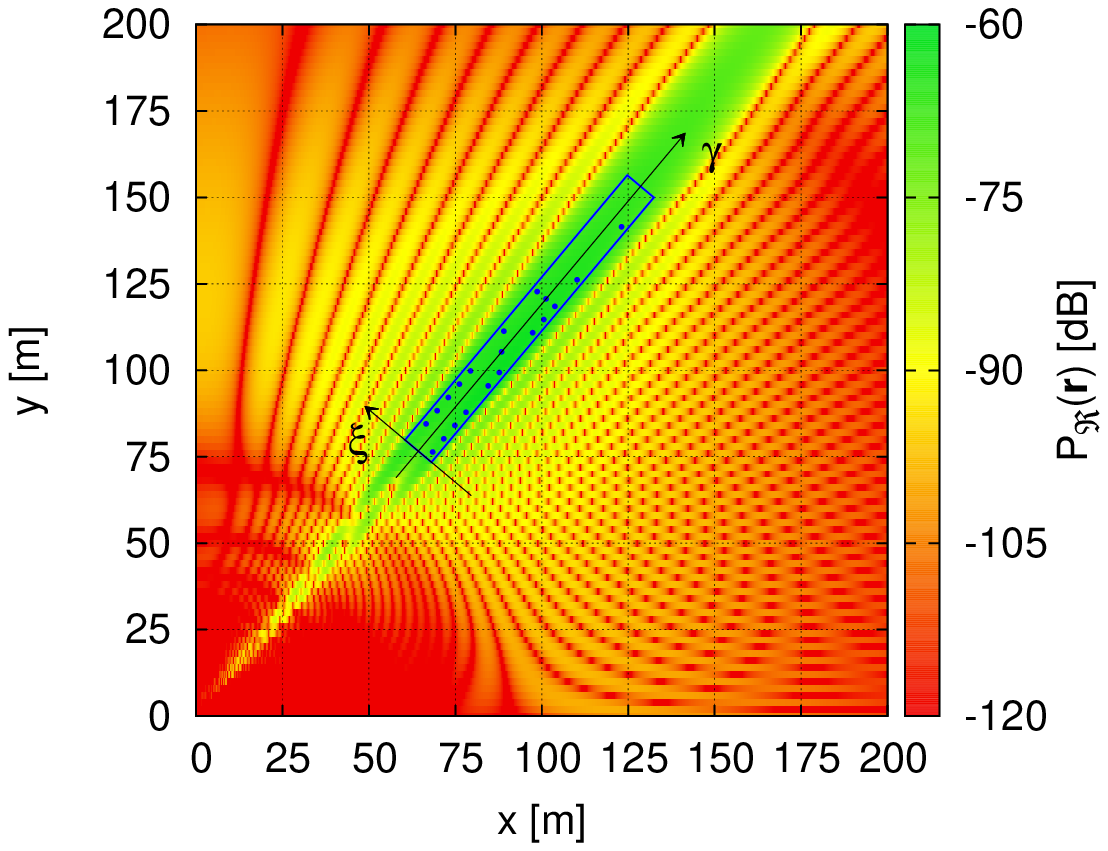}&
\includegraphics[%
  width=0.35\columnwidth]{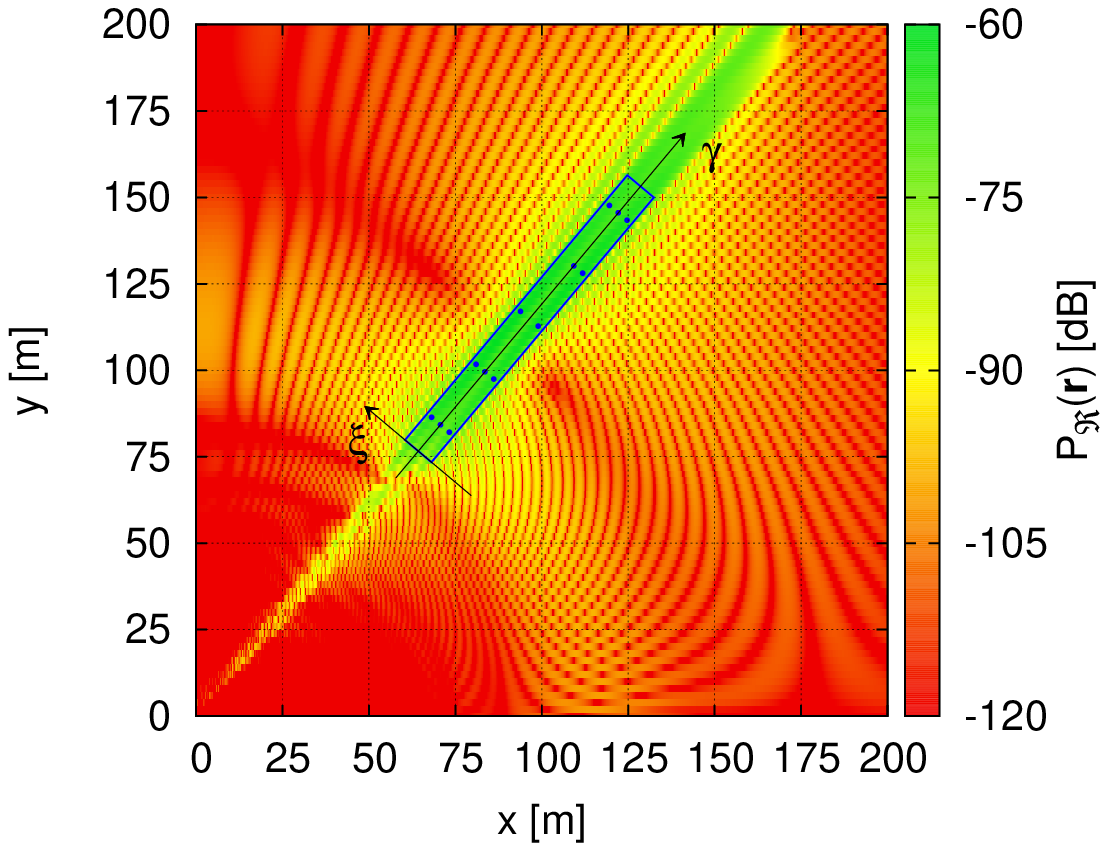}\tabularnewline
(\emph{d})&
(\emph{e})&
(\emph{f})\tabularnewline
&
&
\tabularnewline
\includegraphics[%
  width=0.30\columnwidth]{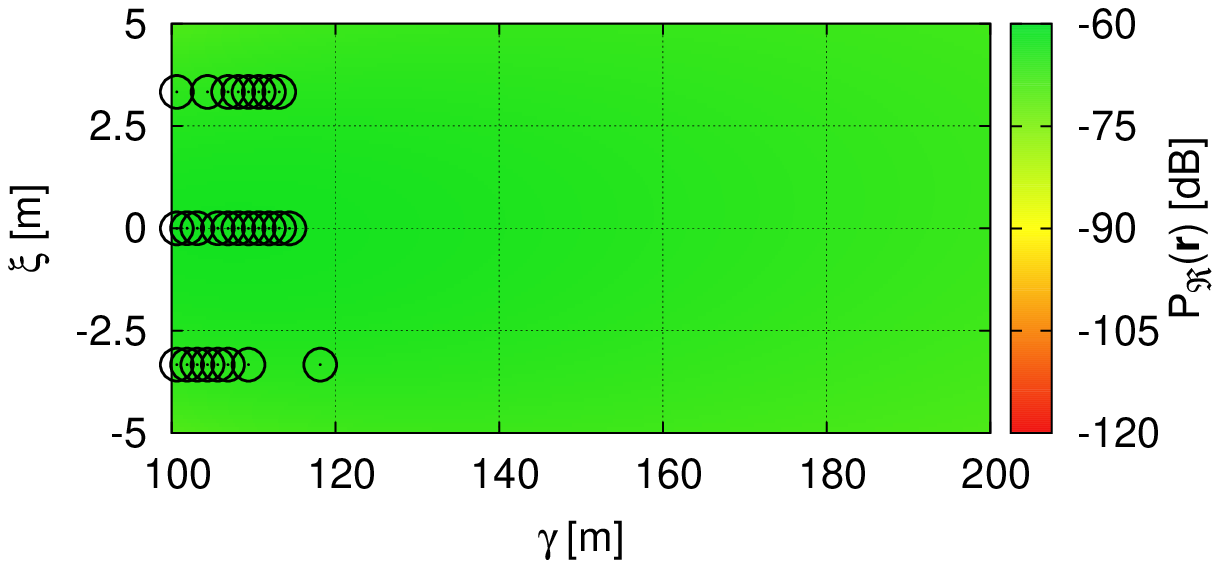}&
\includegraphics[%
  width=0.30\columnwidth]{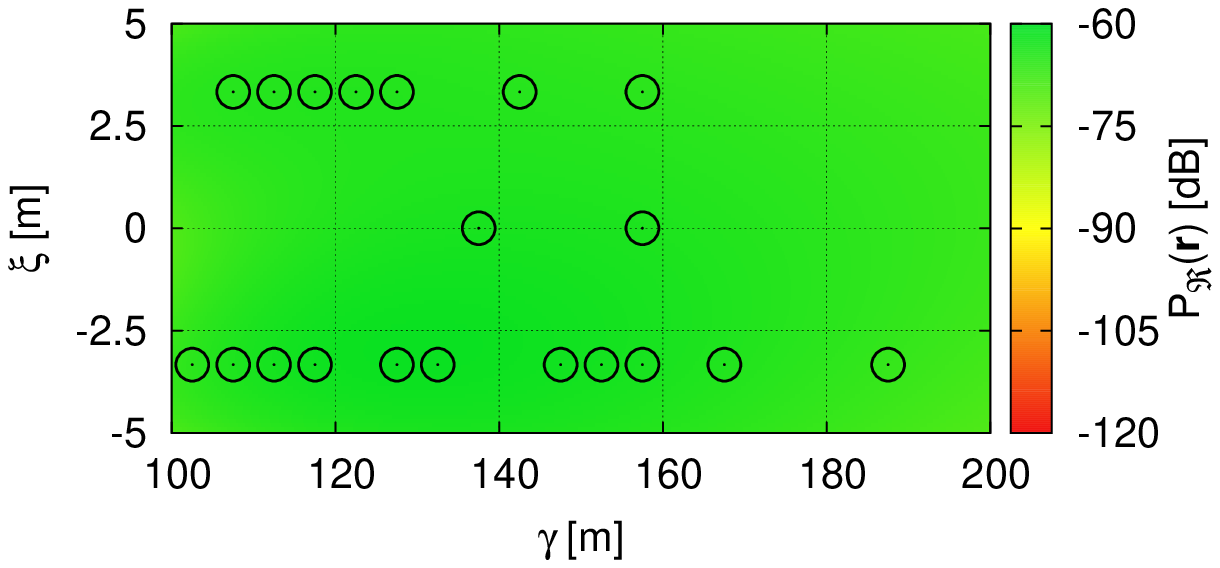}&
\includegraphics[%
  width=0.30\columnwidth]{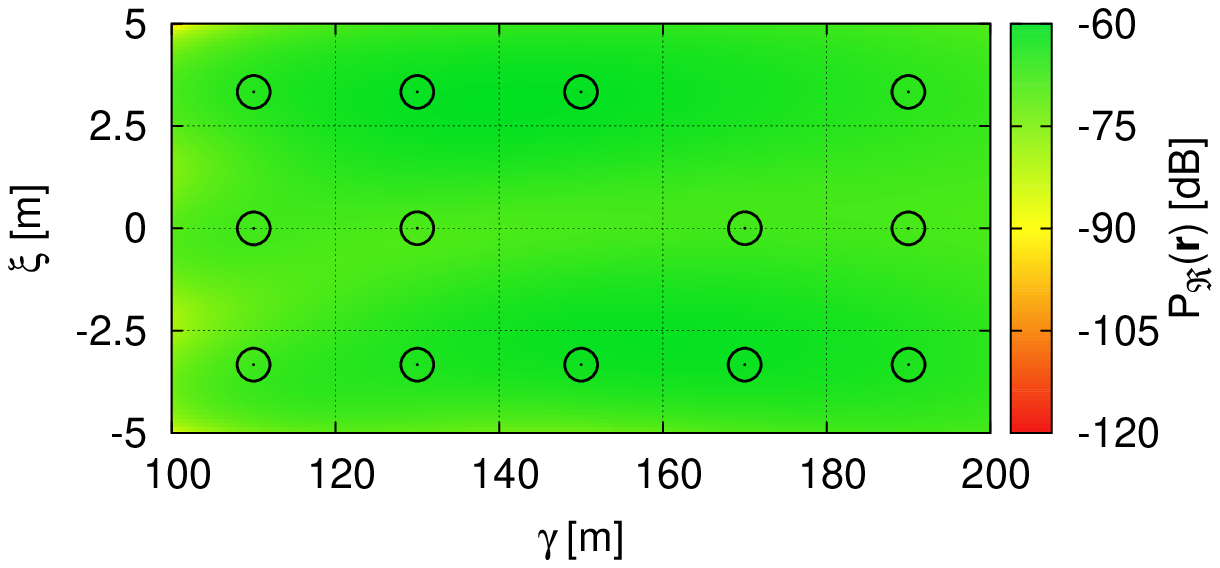}\tabularnewline
(\emph{g})&
(\emph{h})&
(\emph{i})\tabularnewline
\end{tabular}\end{center}

\begin{center}~\vfill\end{center}

\begin{center}\textbf{Fig. 12 - P. Rocca} \textbf{\emph{et al.}}\textbf{,}
\textbf{\emph{{}``}}On the Design of Modular Reflecting EM Skins
...''\end{center}

\newpage
\begin{center}~\vfill\end{center}

\begin{center}\begin{tabular}{c}
\includegraphics[%
  width=0.50\columnwidth]{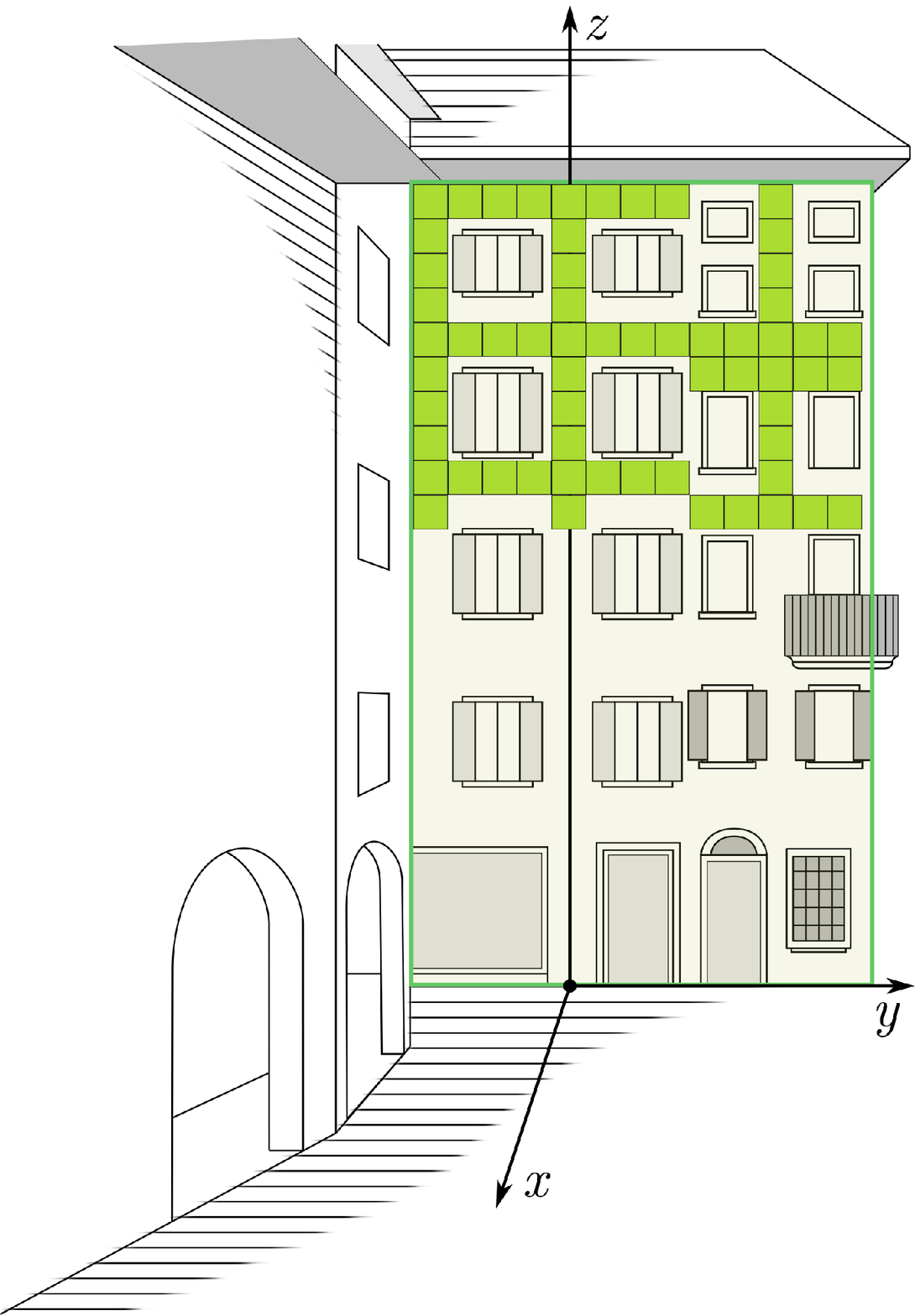}\tabularnewline
(\emph{a})\tabularnewline
\tabularnewline
\includegraphics[%
  width=0.75\columnwidth]{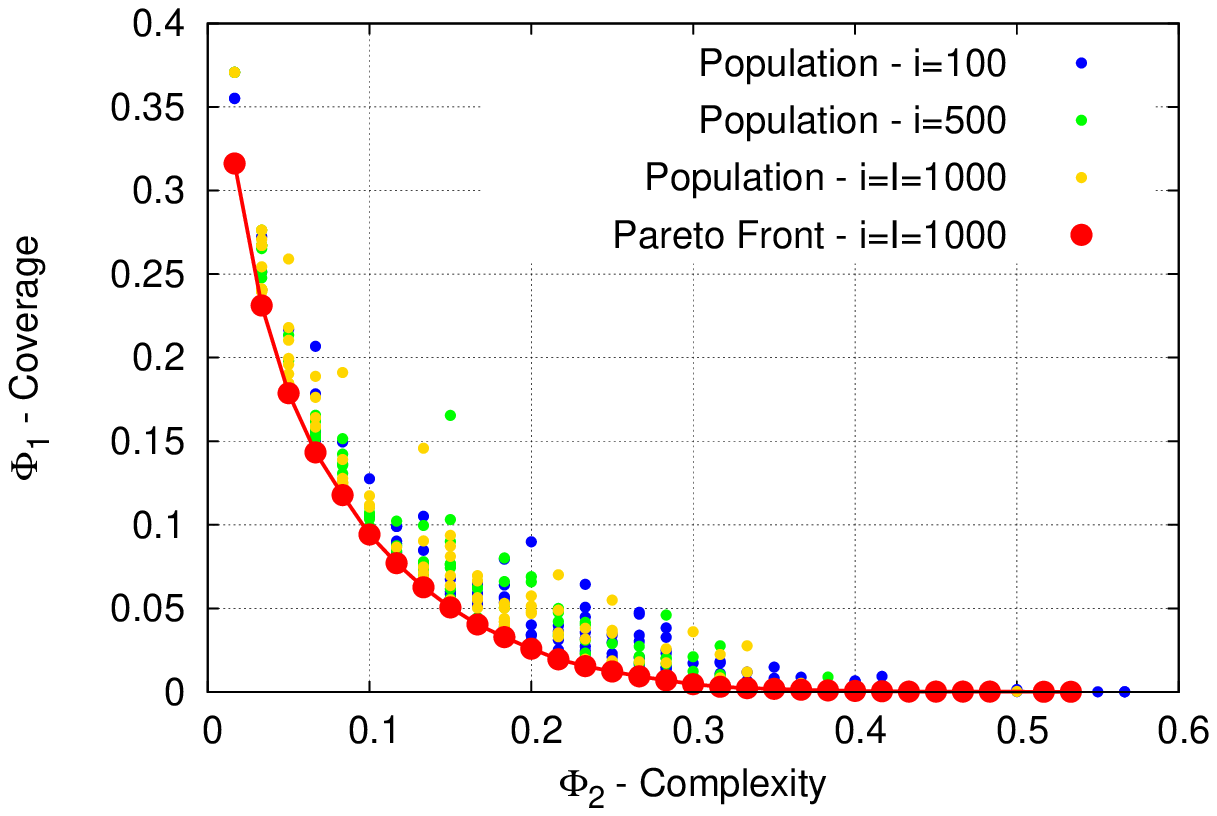}\tabularnewline
(\emph{b})\tabularnewline
\end{tabular}\end{center}

\begin{center}~\vfill\end{center}

\begin{center}\textbf{Fig. 13 - P. Rocca} \textbf{\emph{et al.}}\textbf{,}
\textbf{\emph{{}``}}On the Design of Modular Reflecting EM Skins
...''\end{center}

\newpage
\begin{center}~\vfill\end{center}

\begin{center}\begin{tabular}{>{\centering}m{0.50\columnwidth}>{\centering}m{0.50\columnwidth}}
\includegraphics[%
  width=0.35\columnwidth]{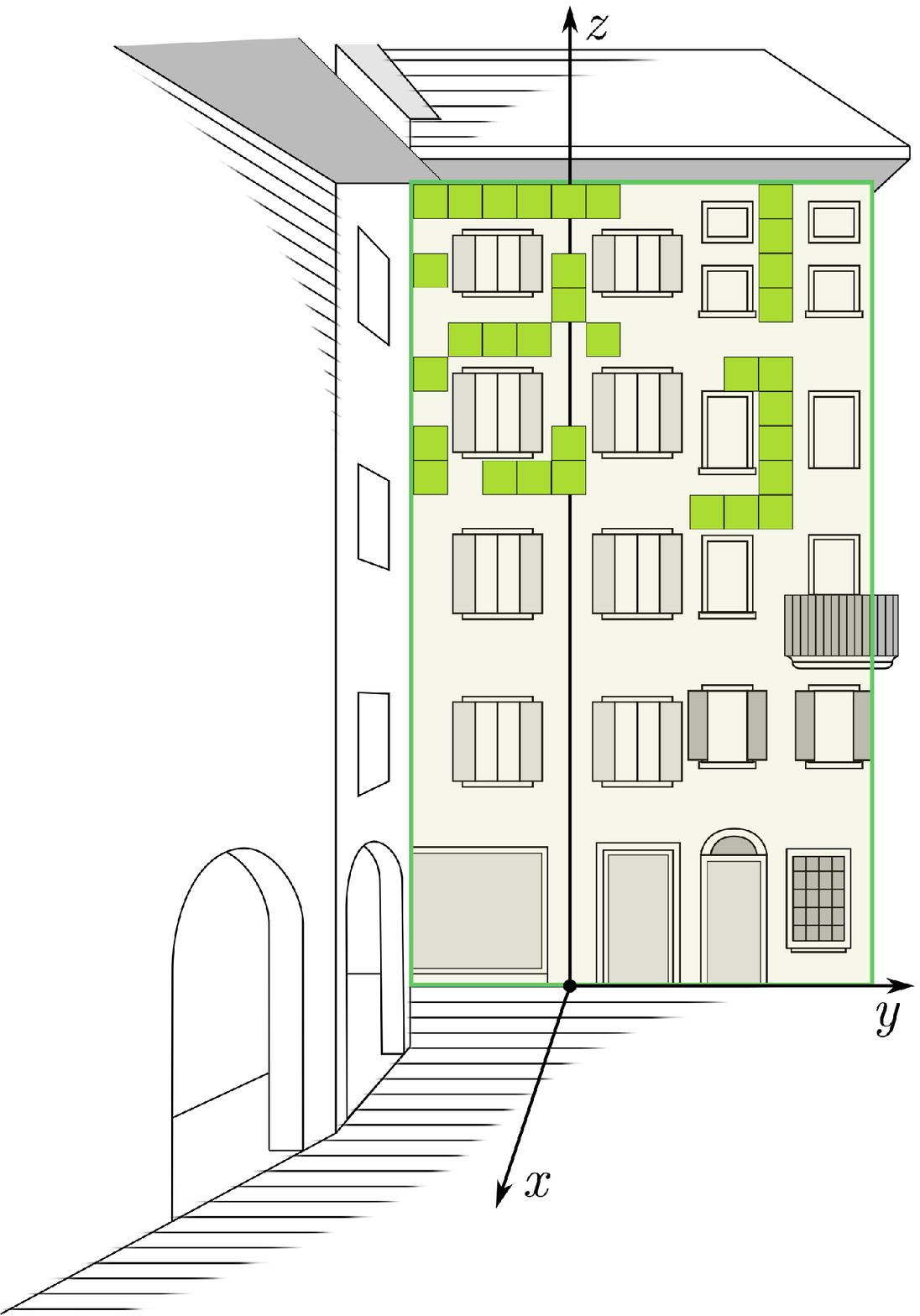}&
\includegraphics[%
  width=0.50\columnwidth]{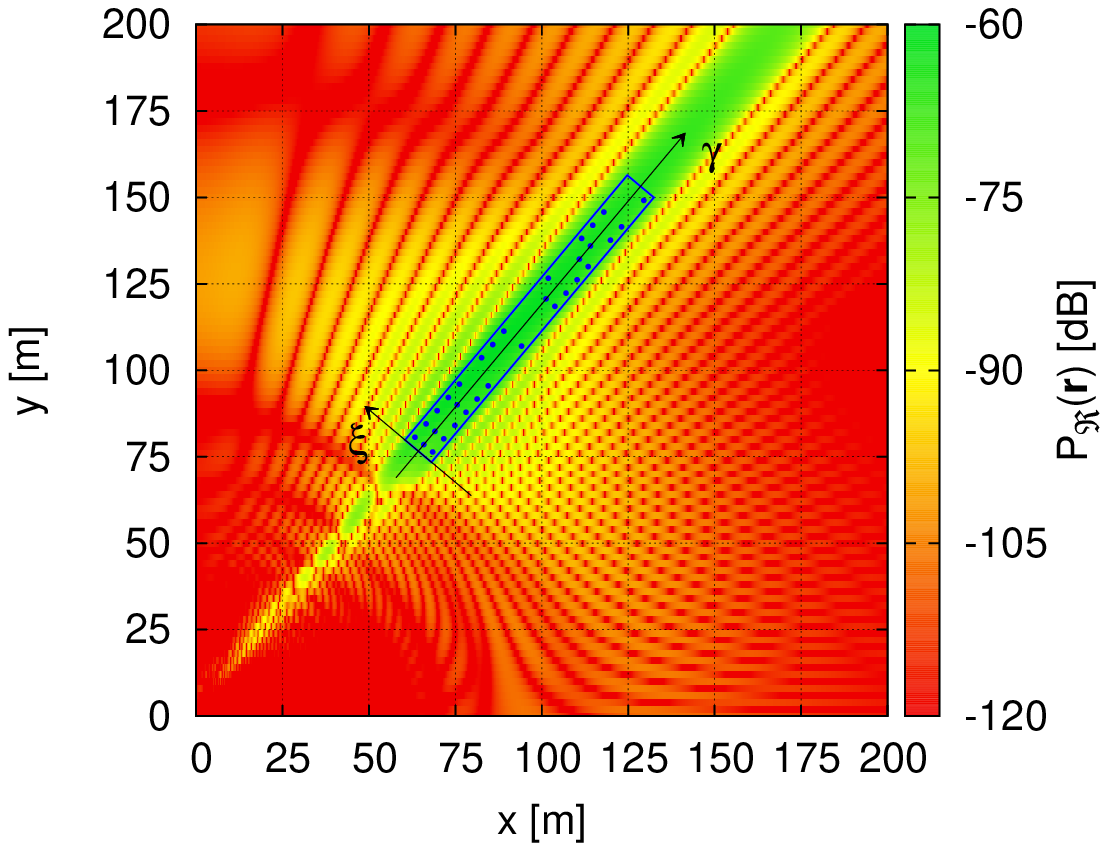}\tabularnewline
(\emph{a})&
(\emph{b})\tabularnewline
&
\tabularnewline
\includegraphics[%
  width=0.50\columnwidth]{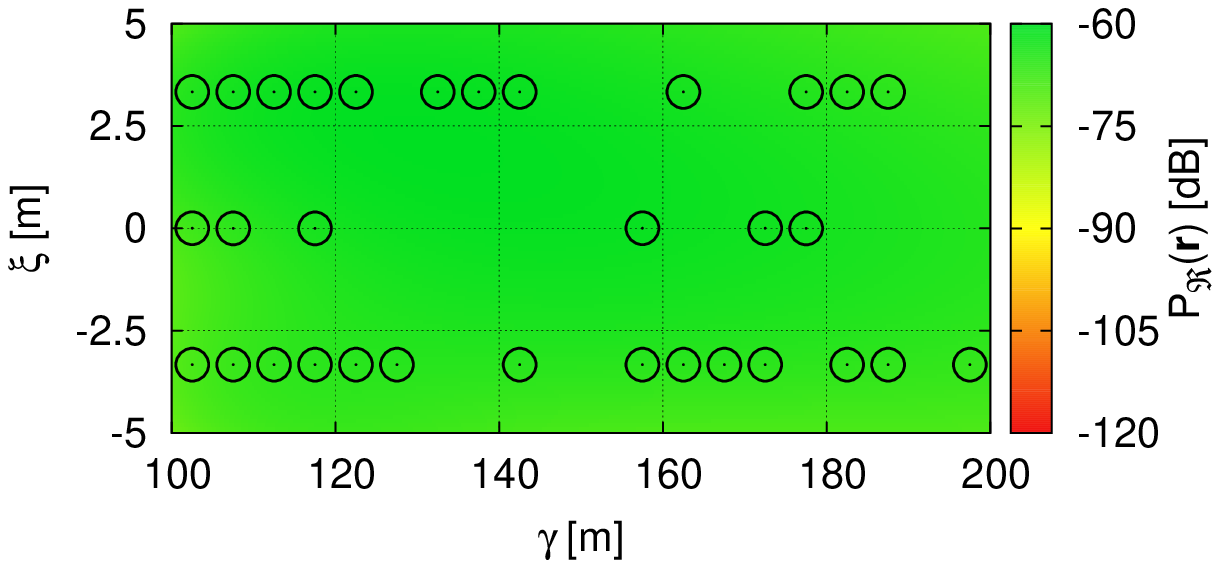}&
\includegraphics[%
  width=0.40\columnwidth]{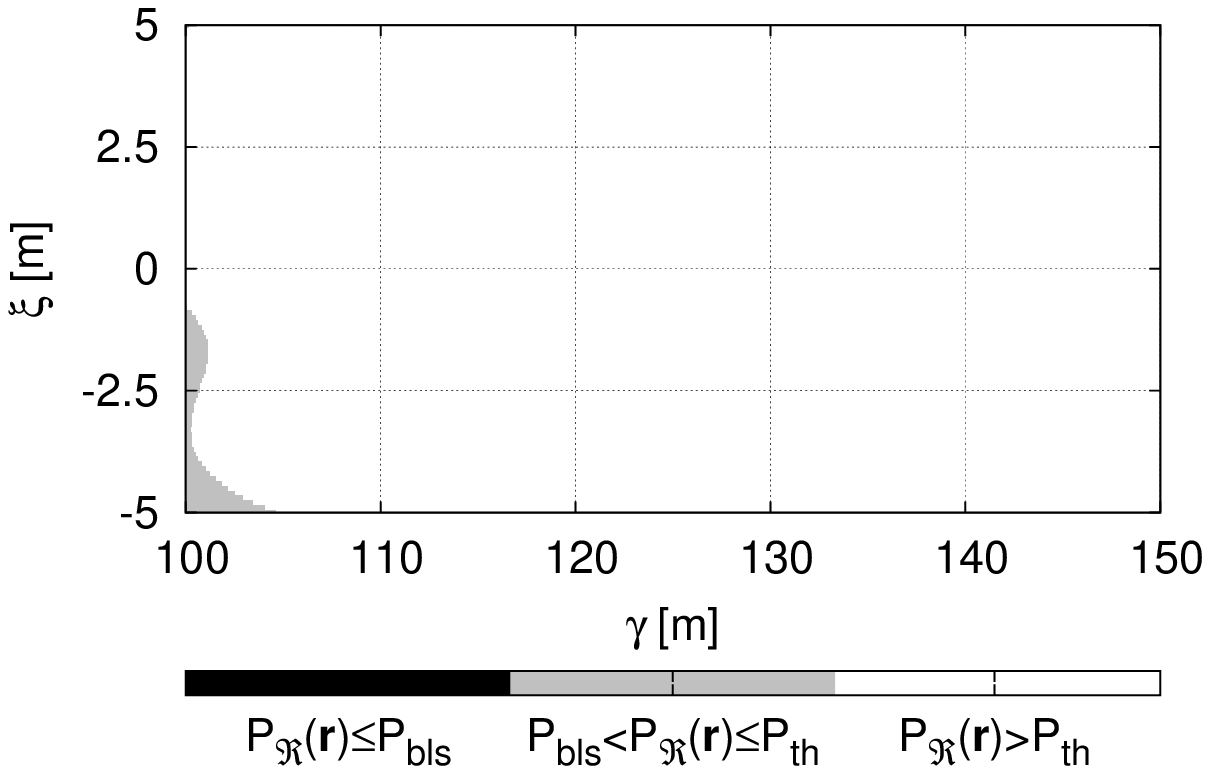}\tabularnewline
(\emph{c})&
(\emph{d})\tabularnewline
\end{tabular}\end{center}

\begin{center}~\vfill\end{center}

\begin{center}\textbf{Fig. 14 - P. Rocca} \textbf{\emph{et al.}}\textbf{,}
\textbf{\emph{{}``}}On the Design of Modular Reflecting EM Skins
...''\end{center}

\newpage
\begin{center}~\vfill\end{center}

\begin{center}\textcolor{black}{}\begin{tabular}{|c|c|c|c|}
\hline 
&
\textcolor{black}{$\min_{\mathbf{r}\in\Omega}\left\{ \left|E_{rx}\left(\mathbf{r}\right)\right|^{2}\right\} $
{[}dB{]}}&
\textcolor{black}{$\max_{\mathbf{r}\in\Omega}\left\{ \left|E_{rx}\left(\mathbf{r}\right)\right|^{2}\right\} $
{[}dB{]}}&
\textcolor{black}{$\mathrm{avg}_{\mathbf{r}\in\Omega}\left\{ \left|E_{rx}\left(\mathbf{r}\right)\right|^{2}\right\} =\Phi_{1}$
{[}dB{]}}\tabularnewline
\hline
\hline 
\textcolor{black}{Fig. 6(}\textcolor{black}{\emph{c}}\textcolor{black}{)}&
\textcolor{black}{$-69.9$}&
\textcolor{black}{$-63.0$}&
\textcolor{black}{$-66.8$}\tabularnewline
\hline 
\textcolor{black}{Fig. 7(}\textcolor{black}{\emph{e}}\textcolor{black}{) }&
\textcolor{black}{$-173.9$}&
\textcolor{black}{$-78.1$}&
\textcolor{black}{$-84.7$}\tabularnewline
\hline 
\textcolor{black}{Fig. 7(}\textcolor{black}{\emph{f}}\textcolor{black}{) }&
\textcolor{black}{$-90.9$}&
\textcolor{black}{$-71.7$}&
\textcolor{black}{$-75.0$}\tabularnewline
\hline
\hline 
\textcolor{black}{Fig. 9(}\textcolor{black}{\emph{c}}\textcolor{black}{) }&
\textcolor{black}{$-69.9$}&
\textcolor{black}{$-63.2$}&
\textcolor{black}{$-66.9$}\tabularnewline
\hline
\hline 
\textcolor{black}{Fig. 12(}\textcolor{black}{\emph{g}}\textcolor{black}{)}&
\textcolor{black}{$-69.7$}&
\textcolor{black}{$-62.5$}&
\textcolor{black}{$-65.1$}\tabularnewline
\hline 
\textcolor{black}{Fig. 12(}\textcolor{black}{\emph{h}}\textcolor{black}{)}&
\textcolor{black}{$-69.7$}&
\textcolor{black}{$-61.2$}&
\textcolor{black}{$-64.4$}\tabularnewline
\hline 
\textcolor{black}{Fig. 12(}\textcolor{black}{\emph{i}}\textcolor{black}{)}&
\textcolor{black}{$-101.5$}&
\textcolor{black}{$-60.3$}&
\textcolor{black}{$-65.1$}\tabularnewline
\hline
\hline 
\textcolor{black}{Fig. 14(}\textcolor{black}{\emph{c}}\textcolor{black}{)}&
\textcolor{black}{$-73.1$}&
\textcolor{black}{$-60.1$}&
\textcolor{black}{$-63.5$}\tabularnewline
\hline
\end{tabular}\end{center}

\begin{center}\textcolor{black}{~\vfill}\end{center}

\begin{center}\textbf{\textcolor{black}{Tab. I - P. Rocca}} \textbf{\textcolor{black}{\emph{et
al.}}}\textbf{\textcolor{black}{,}} \textbf{\textcolor{black}{\emph{{}``}}}\textcolor{black}{On
the Design of Modular Reflecting EM Skins ...''}\end{center}
\end{document}